\documentclass[twoside,12pt]{article}
\usepackage{epsfig}
\usepackage{bits}
\usepackage[centertags]{amsmath}
\usepackage{amsfonts}
\usepackage[sectionbib]{bibunits}
\usepackage{cite}
\usepackage{todonotes}
\usepackage{color}
\usepackage{subfigure}
\usepackage[T1]{fontenc}
\usepackage[utf8]{inputenc}
\usepackage{url}
\usepackage[hyperindex]{hyperref}

\definecolor{nu}{RGB}{0,0,0}
\definecolor{nus}{RGB}{0,0,0}
\definecolor{grn}{RGB}{0,0,0}
\definecolor{nu4}{RGB}{0,0,0}
\definecolor{numu}{RGB}{0,0,0}
\definecolor{nue}{RGB}{0,0,0}

\newcommand{\be}{\begin{equation}}
\newcommand{\ee}{\end{equation}}
\newcommand{\bea}{\begin{eqnarray}}
\newcommand{\eea}{\end{eqnarray}}

\newcommand{\nua}[1]{\ensuremath{\rlap{\kern-2.5pt\ensuremath{\overset{\scriptscriptstyle(-)}{\phantom{\nu_{#1}}}}}{\ensuremath{{\nu}_{#1}}}}}
\begin{document}

\title{ \vspace{1cm} Status of Light Sterile Neutrino Searches}
\author{Sebastian B\"oser$^{1}$, Christian Buck$^{2}$, Carlo Giunti$^{3}$, Julien Lesgourgues$^4$, \\
Livia Ludhova$^{5,6}$, Susanne Mertens$^{7,8}$, Anne Schukraft$^{9}$, Michael Wurm$^{1,}$\footnote{corresponding author: \href{mailto:michael.wurm@uni-mainz.de}{michael.wurm@uni-mainz.de}}
\\~\\
$^1$Institut f\"ur Physik und Exzellenzcluster PRISMA$^+$, \\Johannes Gutenberg-Universit\"at Mainz, 55128 Mainz, Germany \\
$^2$Max-Planck-Institut f\"ur Kernphysik,\\
Saupfercheckweg 1, 69117 Heidelberg, Germany\\
$^{3}$Instituto Nazionale di Fisica Nucleare (INFN), Sezione di Torino, \\
Via P. Giuria 1, I--10125 Torino, Italy \\
$^4$Institut für Theoretische Teilchenphysik und Kosmologie (TTK),\\ RWTH Aachen University, D-52056 Aachen, Germany,\\
$^5$ Institut f\"ur Kernphysik 2, Forschungszentrum J\"ulich, 52425 J\"ulich, Germany\\
$^6$ III. Physikalisches Institut B, \\ RWTH Aachen University, D-52056 Aachen, Germany,\\
$^7$ Max Planck Institute for Physics, Föhringer Ring 6, 80805 München, Germany\\
$^8$ Technische Universität München, Arcisstrasse 21, 80333 München, Germany \\ 
$^9$Fermi National Accelerator Laboratory (FNAL), Batavia, IL 60510, USA}
\maketitle

%
%

\begin{abstract}
\noindent A number of anomalous results in short-baseline oscillation may hint at the existence of one or more light sterile neutrino states in the eV mass range and have triggered a wave of new experimental efforts to search for a definite signature of oscillations between active and sterile neutrino states. 
The present paper aims to provide a comprehensive review on the status of light sterile neutrino searches in mid-2019: we discuss not only the basic experimental approaches and sensitivities of reactor, source, atmospheric, and accelerator neutrino oscillation experiments but also the complementary bounds arising from direct neutrino mass experiments and cosmological observations. 
Moreover, we review current results from global oscillation analyses that include the constraints set by running reactor and atmospheric neutrino experiments. They permit to set tighter bounds on the active-sterile oscillation parameters but as yet are not able to provide a definite conclusion on the existence of eV-scale sterile neutrinos.  
\end{abstract}


\newpage

\tableofcontents

\newpage

%
%

\section{Introduction}
\label{sec:intro}

The Standard Model (SM) of particle physics foresees three (originally massless) neutrinos coupling to the $W$ and $Z$ bosons of weak interactions. While not contained in its minimal version, the SM does as well not constrain the existence of further inactive neutrino states that form singlets under the weak interaction (see the reviews in Refs.~\cite{Volkas:2001zb,Mohapatra:2006gs}). In fact, these sterile neutrinos are featuring naturally in many of the proposed SM extensions, and $-$ provided relatively small mixings $-$ can be realized over a wide range of masses  with the three active neutrino states.

The present paper singles out sterile neutrinos on the eV mass scale. Serious interest in their possible existence was sparked by a number of anomalous results in short-baseline oscillation experiments. Since the discovery of flavor oscillations two decades ago, the knowledge on neutrino mixing parameters has been continuously improving (see the recent global fits in Refs.~\cite{deSalas:2017kay,Capozzi:2018ubv,Esteban:2018azc}). Still, the current data of solar, atmospheric, and long-baseline neutrino oscillation experiments leave some room for additional (few) percent-level mixing amplitudes to sterile states \cite{Parke:2015goa}. (In this case, the $3\times3$ PMNS mixing matrix is not unitary.) Using this gap, a number of existing unexpected rate deficits and appearance results at very short baselines can be elegantly explained by an additional fourth sterile neutrino accompanied by an eV-scale mass eigenstate.

The existing anomalies are currently under investigation by a row of dedicated neutrino oscillation experiments, providing exciting new and partially contradictory data on active-sterile mixing. However, oscillation experiments are not the only approach to search for eV-scale sterile neutrinos: the most recent generation of direct neutrino mass measurements, in particular the KATRIN tritium-decay experiment that has recently come into operation, will be sensitive to the fourth mass state due to a notable deformation close to the spectral endpoint. Moreover, observational cosmology sets stringent limits to the number and mass of relativistic neutrino states in the early Universe and in fact strongly disfavors an additional eV-scale neutrino, unless its early thermalization is impeded by some secondary mechanism. 

Several reviews have already been published on the phenomenology of light sterile neutrinos since the observation of the LSND anomaly in the 1990ies \cite{Bilenky:1998dt,GonzalezGarcia:2007ib,Abazajian:2012ys,Conrad:2012qt,Palazzo:2013me,Bellini:2013wra,Gariazzo:2015rra,Giunti:2019aiy} and even before (e.g.~\cite{GomezCadenas:1995sj,Goswami:1995yq}). This review aims at presenting the current status of light sterile neutrinos with an emphasis on the detailed discussion of past, present, and future experiments. At the time of writing, the eV-scale sterile neutrino hypothesis is called into question by new oscillation results and stringent cosmological limits. We compare the old and new experimental constraints with the parameter space preferred by the original anomalies, aiming to provide the reader with a comprehensive overview of the current status of the field.

In preparation, Sec.~\ref{sec:signatures} lines out the possible signatures for the existence of eV-mass sterile neutrinos. The experimental hints triggering the renewed interest in light sterile states are described in Sec.~\ref{sec:hints}. The bulk of the paper is formed by an overview of the on-going experimental efforts ordered by their basic methodology: results as well as upcoming experiments are presented both for the field of very short baseline experiments (Sec.~\ref{sec:oscillations}) and absolute neutrino mass measurements (Sec.~\ref{sec:numass}). As synthesis to the prior sections, Sec.~\ref{sec:status} reviews the status of the field in the light of new limits and novel evidences for sterile neutrinos. These are compared to the existing and future cosmological limits on light sterile neutrinos in Sec.~\ref{sec:cosmology}. 

%
%

\section{Impact of sterile neutrinos}
\label{sec:signatures}
The left-handed {\it active} neutrinos flavors $\nu_{e}$, $\nu_{\mu}$, and $\nu_\tau$ are identified by the charged leptons $e$, $\mu$, and $\tau$ that are generated in their corresponding charged-current weak interactions. From the LEP measurement of the decay witdth of the $Z$ boson into invisble
particles $Z\to \nu \bar\nu$ \cite{ALEPH:2005ab}, we know that there are three of these flavor eigenstates. In contrast to the charged leptons $e^{\pm}, \mu^{\pm}$ and $\tau^{\pm}$ which are heavy and therefore well aligned with their mass eigenstates~\cite{Akhmedov:2007fk}, {\it active} neutrinos $\nu_{e},\nu_{\mu}$ and $\nu_\tau$ are light and can be generated as coherent quantum-mechanical superpositions of the mass eigenstates $\nu_1, \nu_2$ and $\nu_3$. This simplest and standard case in which the neutrino mixing matrix $U$ relating mass and flavor states is not diagonal explains the flavor oscillations observed in solar, atmospheric, and long-baseline neutrino experiments.

However, it is possible that there are additional massive neutrinos that have very small mixing with the three active neutrinos and large mixing with new {\it sterile} neutrinos that are right-handed and do not take part in weak interactions. In this review we consider the simplest case of one additional fourth mass eigenstate $\nu_4$ at the eV scale which corresponds to a sterile neutrino $\nu_s$ in the flavor basis. This has consequences not only for neutrino oscillations, but also for direct neutrino mass experiments and cosmology as briefly outlined below.

\subsection{Oscillation searches}
\label{sec:oscsearches}

The well-established phenomenon of neutrino oscillations~\cite{Tanabashi:2018oca} proceeds via a quantum-mechanical interference effect. The flavor eigenstates $\left| \nu_{\alpha}\right>, \alpha \in \left\{ e,\mu,\tau \right\}$ are superpositions of the mass-eigenstates $\left|\nu_i \right>, i \in \left\{1,2,3\right\}$. In the standard picture with three flavors, the mixing is described by a unitary $3\times 3$ matrix $U$
\begin{equation}
   \arraycolsep=1.0pt\def\arraystretch{1.2}
\left(\begin{array}{l}\nu_e \\ \nu_\mu \\ \nu_\tau\end{array}\right) = 
\left(\begin{array}{ccc}
U_{e1} & U_{e2} & U_{e3} \\
U_{\mu1} & U_{\mu2} & U_{\mu 3} \\
U_{\tau1} & U_{\tau2} & U_{\tau3} \\
\end{array}\right)
\left(\begin{array}{l}\nu_1 \\ \nu_2\\ \nu_3\end{array}\right).
\end{equation}
Due to the unitarity requirement $U^\dagger U = 1$ which ensures that the overall normalization of the wavefunction is conserved, the complex elements $U_{\alpha i}$ can also be described in terms of consecutive rotations of the orthogonal mass eigensystem $\left<\nu_i|\nu_j\right> = \delta_{ij}$ into the orthogonal flavor eigensystem $\left<\nu_\alpha|\nu_\beta\right> = \delta_{\alpha\beta}$ by three Euler angles $\theta_{12}, \theta_{13}$ and $\theta_{23}$~(c.f. Fig.~\ref{fig:euler-angles}) and one complex phase $\delta_{13}$
in the standard parameterization
\cite{Tanabashi:2018oca}
\begin{equation}
U = 
\begin{pmatrix}
  1 & 0 & 0\\0 & c_{23} & s_{23}\\0 & - s_{23} & c_{23}
\end{pmatrix} \times
\begin{pmatrix}
  c_{13} & 0 & e^{-i\delta} s_{13}\\0 & 1 & 0\\- e^{-i\delta} s_{13} & 0 & c_{13}
\end{pmatrix} \times
\begin{pmatrix}
  c_{12} & s_{12} & 0\\- s_{12} & c_{12} & 0\\0 & 0 & 1
\end{pmatrix}, \label{U3nu}
\end{equation}
where $ c_{ab} \equiv \cos\theta_{ab} $ and $ s_{ab} \equiv
\sin\theta_{ab} $.
We neglected possible additional Majorana phases that are irrelevant for neutrino oscillations
(see the review on ``Neutrino Masses, Mixing, and Oscillations'' in Ref.~\cite{Tanabashi:2018oca}).

\twofig[htb]
  {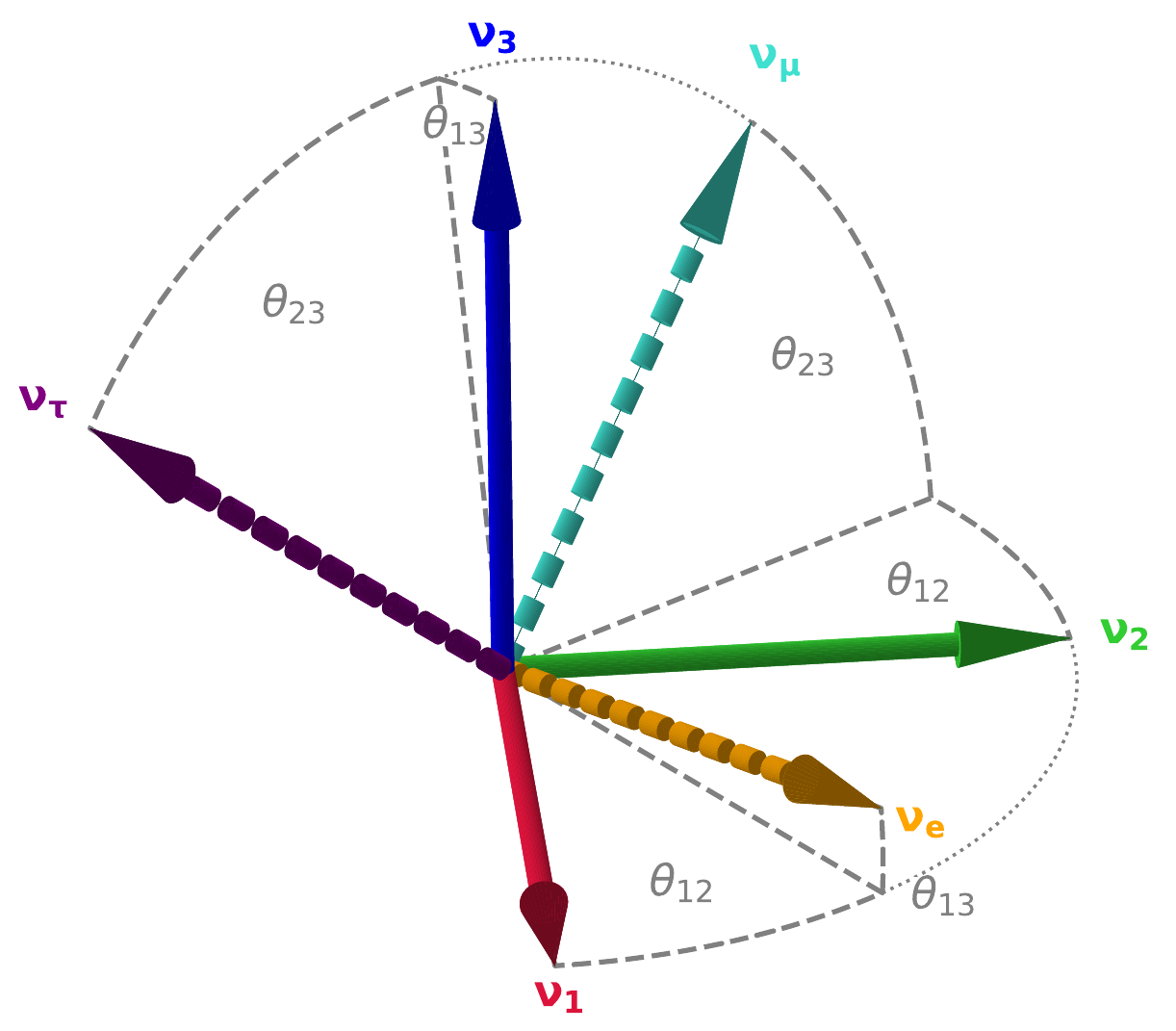}
  {The superposition of neutrino flavor eigenstates (dashed) from the neutrino mass eigenstates (solid) can be described by three mixing angles  $\theta_{12}, \theta_{13}$ and $\theta_{23}$.\label{fig:euler-angles}}
  {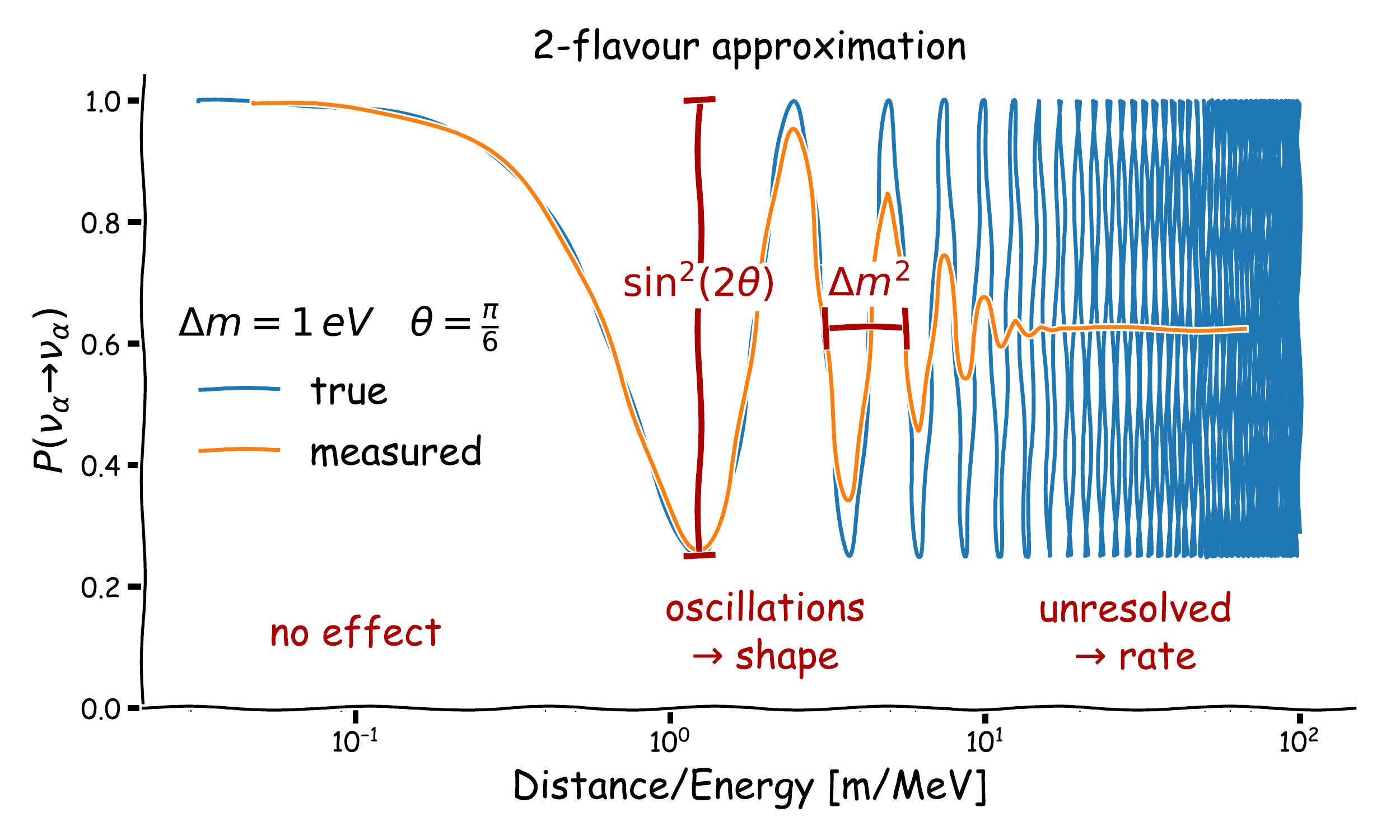}
  {Survival probability $P(\nu_\alpha \to \nu_\alpha)$ as a function of the ratio distance $L$ over energy $E$ in a 2-flavor model. The amplitude of oscillations is governed by the mixing angle while the frequency is determined by the mass splitting. At high values of $L/E$, the rapid oscillations can not be resolved experimentally.\label{fig:nuosc-phenomenology}}

In a weak interaction, a neutrino flavor eigenstate is thus generated as a superposition of mass eigenstates. Due to their different masses, these mass eigenstates evolve with different phase velocities such that the resulting neutrino vector does not remain aligned with a flavor eigenstate. This results in the phenomenon of neutrino oscillations. The transition probability $P(\nu_{\alpha} \to \nu_{\beta})$ for a neutrino of energy E generated in flavor state $\nu_{\alpha}$ to be observed in flavor state $\nu_{\beta}$ after propagating for a distance $L$ can be calculated as 
\begin{equation}
    P(\nu_\alpha \to \nu_\beta) = \left|\left<\nu_{\beta}(L)|\nu_\alpha(0)\right>\right|^2 = \left|\sum_{i=1}^{n} U_{\beta i}U^{*}_{\alpha i} e^{-i\frac{m_{i}^2 L }{2E}}\right|^2 .
    \label{eqn:osc3flav}
\end{equation}

The oscillatory behavior is driven by the phase terms $\exp\left(-i\frac{m_{i}^2 L}{2E}\right)$, which result in interference terms $\exp\left(\frac{\Delta m_{ij}^2 L}{2E}\right)$ with the squared {\it mass splittings} 
\begin{equation}
  \Delta m_{ij}^2 = m_i^2 - m_j^2.
\end{equation} 
For a given distance $L$ and energy $E$ the oscillatory behavior is typically dominated by one of the mass splittings. For many cases, the phenomenology of neutrino oscillations can therefore be well approximated by the two-flavor scenario, in which the oscillation probability reduces to 
\begin{eqnarray}
  P(\nu_\alpha \to \nu_\beta) & \simeq & \left| \delta_{\alpha\beta} - \sin^2(2\theta)\sin^2\left(\frac{\Delta m^2 L}{4E}\right) \right| \nonumber \\
    & = &  \left| \delta_{\alpha\beta} - \sin^2(2\theta)\sin^2\left(1.27 \frac{\Delta m^2/{\rm [eV^2]}\, L/{\rm [km]}}{4E/{\rm [GeV]}}\right) \right|
    \label{eqn:osc2flav}
\end{eqnarray}
Phenomenologically we can thus subdivide the observable oscillation effects into three regimes~(see Fig. \ref{fig:nuosc-phenomenology}):
\begin{itemize}
    \item $4E \gg \Delta m^2 L $: the oscillatory term vanishes and there is no observable effect;
    \item $4E \sim \Delta m^2 L $: the oscillatory behavior is observed as a function of energy (or distance); the {\it frequency} of oscillations is driven by the mass splitting $\Delta m^2$, while the amplitude is governed by the corresponding mixing angle $\sin^2(2\theta)$. 
    \item $4E \ll \Delta m^2 L $: the oscillations become increasingly rapid until the experimental resolution is no longer sufficient. The observable flux of the appearing (or disappearing) neutrinos then averages out at $\left| \delta_{\alpha\beta}-\frac{1}{2}\sin^2(2\theta) \right|$.
\end{itemize}
The mixing angle $\sin^2(2\theta)$ only depends on absolute values of the complex mixing matrix elements $\left|U_{\alpha i}\right|^2$. The oscillation term $\sin^2\left(\frac{\Delta m^2 L}{4E}\right)$ is invariant under the transformation $\Delta m^2 \to -\Delta m^2$. In a 2-flavor scheme and in the absence of other effects such as neutrino matter interactions, the oscillations signatures are therefore insensitive to the sign of the mass splitting and we will hence not differentiate between different ordering of the mass eigenvalues.

In this paper, we consider the so-called (3+1) neutrino mixing scheme
\cite{Goswami:1995yq, Okada:1996kw,Bilenky:1996rw,Bilenky:1999ny,Maltoni:2004ei},
that is the simplest extension of standard $3\nu$ mixing with the addition of
a non-standard massive neutrino at the eV scale\footnote{
More complicated schemes with sterile neutrinos have been also considered in the literature:
(3+2)
\cite{Sorel:2003hf,Karagiorgi:2006jf,Maltoni:2007zf,Karagiorgi:2009nb,Giunti:2011gz,Donini:2012tt,Archidiacono:2012ri,Goswami:2007kv},
(3+3)
\cite{Maltoni:2007zf},
(3+1+1)
\cite{Nelson:2010hz,Fan:2012ca,Kuflik:2012sw,Huang:2013zga,Giunti:2013aea},
and
(1+3+1)
\cite{Kopp:2011qd,Kopp:2013vaa}.
There are also studies of schemes with sterile neutrinos and other non-standard effects:
neutrino non-standard interactions
\cite{Akhmedov:2010vy,Liao:2016reh,Babu:2016fdt,Blennow:2016jkn,Esmaili:2018qzu}
and
radiative sterile neutrino decays
\cite{Gninenko:2011hb,Gninenko:2012rw}.
}.

To embed the fourth mass eigenstate $\color{nu4}\nu_4$, the unitary mixing matrix has to be expanded to a $4\times4$ matrix
\begin{equation}
\arraycolsep=1.0pt\def\arraystretch{1.2}
\left(\begin{array}{l}\nu_e \\ \nu_\mu \\ \nu_\tau \\ {\color{nus}\nu_s} \end{array}\right) = 
\left(\begin{array}{cccc}
U_{e1} & U_{e2} & U_{e3} & {\color{grn} U_{e4}} \\
U_{\mu1} & U_{\mu2} & U_{\mu 3} & {\color{grn} U_{\mu4}} \\
U_{\tau1} & U_{\tau2} & U_{\tau3} & {\color{grn} U_{\tau4}} \\
  {\color{black}U_{s1}} & {\color{black} U_{s2}} & {\color{black} U_{s3}} &  U_{s4}
\end{array}\right)
\left(\begin{array}{l}\nu_1 \\ \nu_2\\ \nu_3 \\ {\color{nu4}\nu_4} \end{array}\right)
\end{equation}

The new state $\color{nus}\nu_s$ is associated with right-handed neutrinos, and hence does not participate in the weak interactions. Lacking any other coupling, the {\it sterile} state $\color{nus}\nu_s$ itself as well as its associated mixing parameters $\color{black}U_{si}$ with the {\it active} neutrinos are generally considered experimentally non-accessible. This is in contrast to the mass eigenstate  $\color{nu4}\nu_4$: if it is fully aligned with the flavor eigenstate $\color{nus}\nu_s$ (i.e. when $U_{s4}=1$ and ${\color{black}U_{si}} = {\color{grn}U_{\alpha 4}} = 0$ for $i \in {1,2,3}$ and $\alpha \in {e,\mu,\tau}$), it can only be observed through gravitational effects and will not affect neutrino oscillations. 

If however like the other states, $\color{nu4}\nu_4$ is not aligned with its corresponding flavor state, the active neutrinos will be generated with a $\color{nu4}\nu_4$ component. Following Eq.~\ref{eqn:osc3flav}, this will affect neutrino oscillation phenomenology as this new mass eigenstate will propagate with yet another phase velocity compared to the other mass eigenstates. This scheme is allowed by the existing experimental
neutrino oscillation data if the non-standard massive neutrino $\nu_{4}$ is mostly sterile,
i.e.
\begin{equation}
|U_{\alpha 4}|^2 \ll 1
\qquad
  (\alpha \in \left\{e,\mu,\tau\right\})
.
\label{smallmix}
\end{equation}
Since $m_4$ is at the eV-scale, $ m_4 \gg m_3,m_2,m_1$ and the corresponding mass-splittings 
\begin{equation}
  \Delta m^2_{4i} \sim {\rm eV}^2 \qquad (\forall i \in \left\{1,2,3\right\})
\end{equation}
are all similar and also greater than the solar and atmospheric mass splittings. Thus, for experimental configurations where $4E \sim \Delta m^2_{4i} L$ and the oscillatory behavior due to sterile neutrinos is observable, the atmospheric and solar mass splittings terms are not effective.

In some cases
(as in Eq.~(\ref{probLBL}))
it is useful to adopt a parameterization of the mixing matrix.
The best choice is to extend the standard three-neutrino parameterization in Eq.~(\ref{U3nu})
by adding the additional rotations generated by the new mixing angles
$\theta_{14}$, $\theta_{24}$, and $\theta_{34}$
on the left of the mixing matrix,
leading to (see, for example, Ref.~\cite{Giunti:2019aiy})
\begin{equation}
U
=
\begin{pmatrix}
c_{12}
c_{13}
c_{14}
&
s_{12}
c_{13}
c_{14}
&
c_{14}
s_{13}
e^{-i\delta_{13}}
&
s_{14}
e^{-i\delta_{14}}
\\[0.3cm] \displaystyle
\cdots
&
\cdots
&
\begin{array}{l} \displaystyle
c_{13} c_{24} s_{23}
\\ \displaystyle
- s_{13} s_{14} s_{24} e^{i(\delta_{14}-\delta_{13})}
\end{array}
&
c_{14}
s_{24}
\\[0.5cm] \displaystyle
\cdots
&
\cdots
&
\begin{array}{l} \displaystyle
c_{13} c_{23} c_{34}
\\ \displaystyle
- ( c_{24} s_{13} s_{14} e^{i(\delta_{14}-\delta_{13})}
\\ \displaystyle
\phantom{-(}
+  c_{13} s_{23} s_{24} ) s_{34} e^{-i\delta_{34}}
\end{array}
&
c_{14}
c_{24}
s_{34}
e^{-i\delta_{34}}
\\[0.9cm] \displaystyle
\cdots
&
\cdots
&
\cdots
&
c_{14}
c_{24}
c_{34}
\end{pmatrix}
,
\label{U4nu}
\end{equation}
where the dots replace the elements with long expressions that are not
relevant for short- and long-baseline neutrino oscillations.
Note that there are two new CP-violating phases
$\delta_{14}$
and
$\delta_{34}$,
and we neglected (as in Eq.~(\ref{U3nu})) possible additional Majorana phases that are irrelevant for neutrino oscillations.
The parameterization (\ref{U4nu})
has the advantage of having the first row, which gives the mixing of $\nu_{e}$, as simple as possible.
Also, the second row, which gives the mixing of $\nu_{\mu}$, is simpler than the third.
This is convenient, because the vast majority of experimental data
concern $\nu_{e}$ and $\nu_{\mu}$ oscillations.
There is also the advantage that the long-baseline probability (\ref{probLBL}) of
$\nu_\mu\to\nu_e$ oscillations is independent of the mixing angle
$\theta_{34}$ and the associated phase $\delta_{34}$.

As active neutrino oscillations are not effective at distance $L$ and energy $E$ where flavour exchange with the new sterile state occur, 
again, an effective 2-flavor approximation (Eq.~\ref{eqn:osc2flav}) can be employed~\cite{Bilenky:1996rw} to describe the oscillation probability:
 
\begin{equation}
  P\left(\nu_{\alpha}\to\nu_{\beta}\right)=
\left|
\delta_{\alpha\beta}
-
  \sin^2 (2\theta_{\alpha\beta})
\sin^{2}\!\left( \frac{\Delta{m}^2_{s}L}{4E} \right)
\right|
,
\label{eqn:probSBL}
\end{equation}
where
$\Delta{m}^2_{s} \sim 1 \, \text{eV}^2$,
and
\begin{equation}
\sin^2(2\theta_{\alpha\beta})
=
4
|U_{\alpha 4}|^2
\left| \delta_{\alpha\beta} -  |U_{\beta 4}|^2 \right|
.
\label{eqn:ampSBL}
\end{equation}
Only the absolute magnitude of the matrix elements appear, so that the oscillatory effect will not differentiate between neutrinos and antineutrinos. As the oscillatory behaviour depends on the ratio $L/E$ it can be accessed experimentally in a large variety of configurations using different neutrino sources such as radioactive elements, reactors, accelerator neutrino beams, and atmospheric neutrinos. Figure~\ref{fig:sterile_osc_overview} gives an overview of the experiments discussed in this paper and the energy and distance ranges they cover. Using Eq.~\ref{eqn:probSBL}, we can further classify searches for sterile neutrinos into two broad categories: {\it disappearance} and {\it appearance} searches.

\setlength{\figwidth}{\textwidth}
\onefig[htbp]
{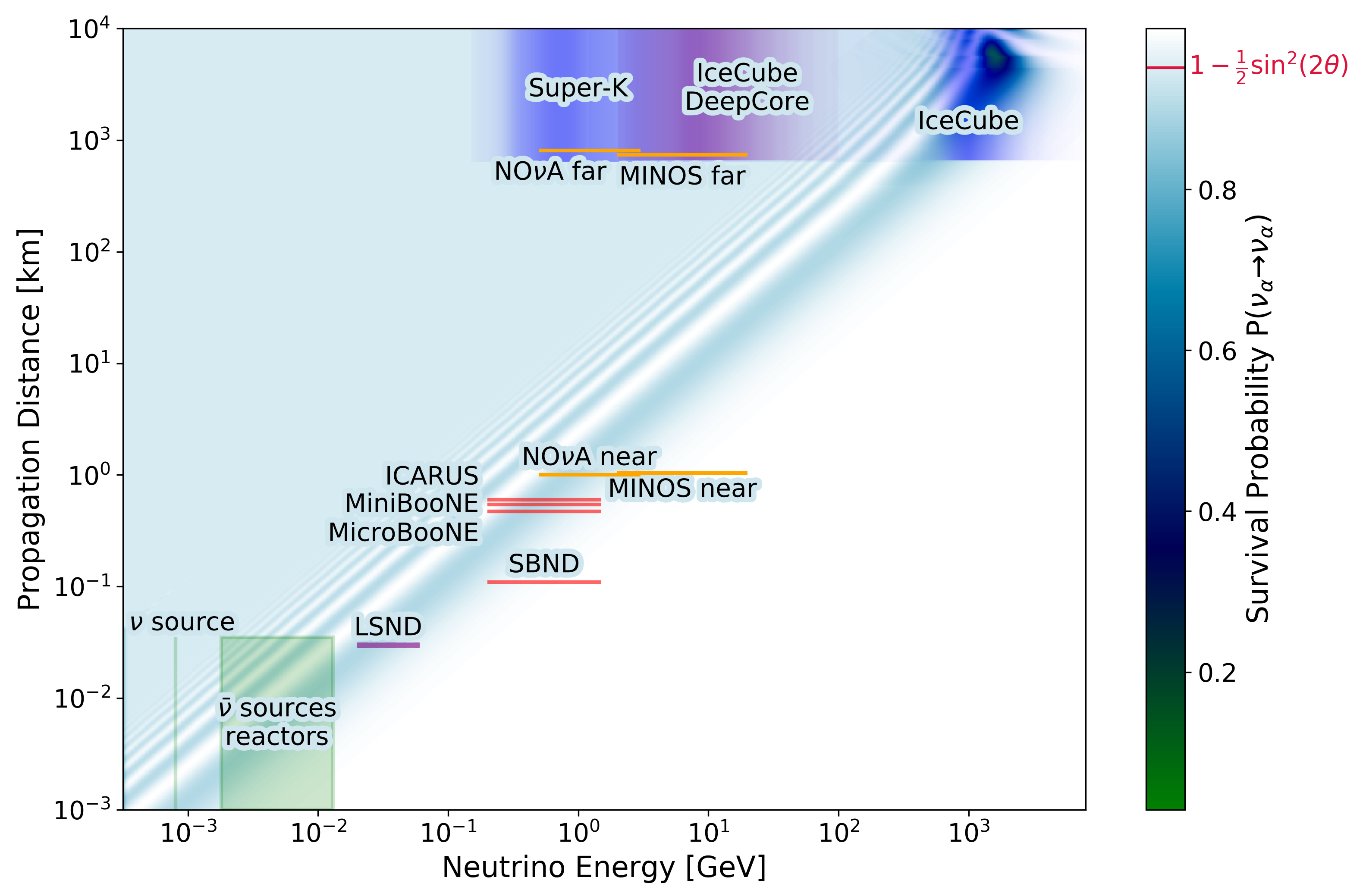}
{Neutrino oscillation probability $P(\nu_\alpha \to \nu_\alpha)$ for a simplified 2-flavor model with one active ($\nu_\mu$-like) and one sterile neutrino mixing with a strength of $\sin^2(2\theta) = 0.1$ at $\Delta m_s=1\un{eV}$. At high energies / small distances no oscillatory behaviour occurs. At low energies / large distances the assumed experimental resolution of $\sigma(L/E) = 30\%$ averages out the oscillation effect to a constant disappearance effect of $P(\nu_\alpha \to \nu_\alpha) = 1-\frac{1}{2}\sin^2 \left(2\theta\right)$. For energies above $E_\nu > 1\un{TeV}$, matter effects become important~(see Sec.~\ref{sec:signature-dissappear}), leading to a strong enhancement of the disappearance for neutrinos crossing the earth. Lines and shaded areas indicate the energy and distance ranges accessible to the various oscillation experiments discussed in this paper: radioactive sources and reactor neutrinos~(green,~Sec.~\ref{sub:reactorexp} and Sec.~\ref{sub:sourceexp}); accelerator neutrinos at Los Alamos~(purple,~Sec.~\ref{sec:lsnd}); the Booster Neutrino Beam (red, MiniBooNE~(Sec.~\ref{sec:miniboone}), the Short Baseline Neutrino program SBN~(Sec.~\ref{sub:sblacc})), and the NuMI beam~(orange,~Sec.~\ref{sub:lblacc}) at Fermilab. The pink and blue shaded areas indicate the sensitive regions to earth-crossing atmospheric neutrinos of Super-Kamiokande and IceCube(DeepCore)~(Chapter~\ref{sub:atmospherics}.)\label{fig:sterile_osc_overview}}
\setlength{\figwidth}{0.6\textwidth}

\subsubsection{Neutrino dissappearance: Active-to-sterile mixing\label{sec:signature-dissappear}}
Active-to-sterile neutrino disappearance searches regard same-flavor oscillation channels $\nu_e \to \nu_e$ and $\nu_\mu \to \nu_\mu$. For these, the oscillation probabilities are driven by the mixing angles

\begin{equation}
  \sin^2(2\theta_{ee})
=
4 |U_{e4}|^2 \left( 1 - |U_{e4}|^2 \right)
=
  \sin^2(2\theta_{14})
\label{see}
\end{equation}
of
$\nu_{e}$
disappearance,
and
the amplitude
\begin{eqnarray}
  \sin^2 (2\theta_{\mu\mu})
  & = & 4 |U_{\mu4}|^2 \left( 1 - |U_{\mu4}|^2 \right) \nonumber \\
  & = & \sin^2(2\theta_{24}) \cos^2(\theta_{14}) + \sin^2(2\theta_{14}) \sin^4(\theta_{24})
\simeq
\sin^2(2\theta_{24})
\label{smm}
\end{eqnarray}
of
$\nu_{\mu}$
disappearance,
where we considered the approximation of small mixing angles given by
the constraint (\ref{smallmix}) to obtain $\cos^2(\theta_{14}) \simeq 1$ and $ \sin^4(\theta_{24})
\simeq 0$.
\medskip\\
For {\bf electron (anti-)neutrino disappearance}, not only powerful radioactive sources but in particular nuclear reactors provide intense MeV neutrino fluxes. However, earlier experiments using radioactive sources were based on chemical extraction schemes and thus they were neither position nor energy sensitive (Sec.~\ref{sub:GA}), while early reactor experiments were situated too far away from the reactor core or had insufficient resolution (Sec.~\ref{sub:RAA}).
In both cases only a general reduction in the observed rate is expected as a consequence of the existence of a sterile neutrino, making the approach rather sensitive to precise knowledge of the source intensity and detection efficiency. The current follow-up experiments have been devised to provide a clear oscillation signature: Placement of radioactive sources close to or even inside a large liquid-scintillator or a segmented gallium detector allows the determination of position and energy of individual neutrino interactions (Sec.~\ref{sub:sourceexp}).
Similarly, a new generation of reactor experiments described in Sec.~\ref{sub:reactorexp} relies on segmented detectors located sufficiently close to the reactor core to probe the oscillatory behavior for baselines of $\sim 10\un{m}$. 
\medskip\\
Artificial muon neutrino beams from accelerators require highest intensities to provide good sensitivity in the {\bf muon neutrino disappearance} channel. The best limits on $\sin^2\theta_{24}$ from disappearance experiments are thus provided by experiments at Fermilab's NuMI beam~(Sec.~\ref{sub:lblacc}). The huge water-/ice-Cherenkov detectors Super-Kamiokande and IceCube exploit the ubiquitous flux of atmospheric muon neutrinos (Sec.~\ref{sub:atmospherics}) yielding comparable sensitivities \cite{Nunokawa:2003ep}. 
\twofig[ht]
{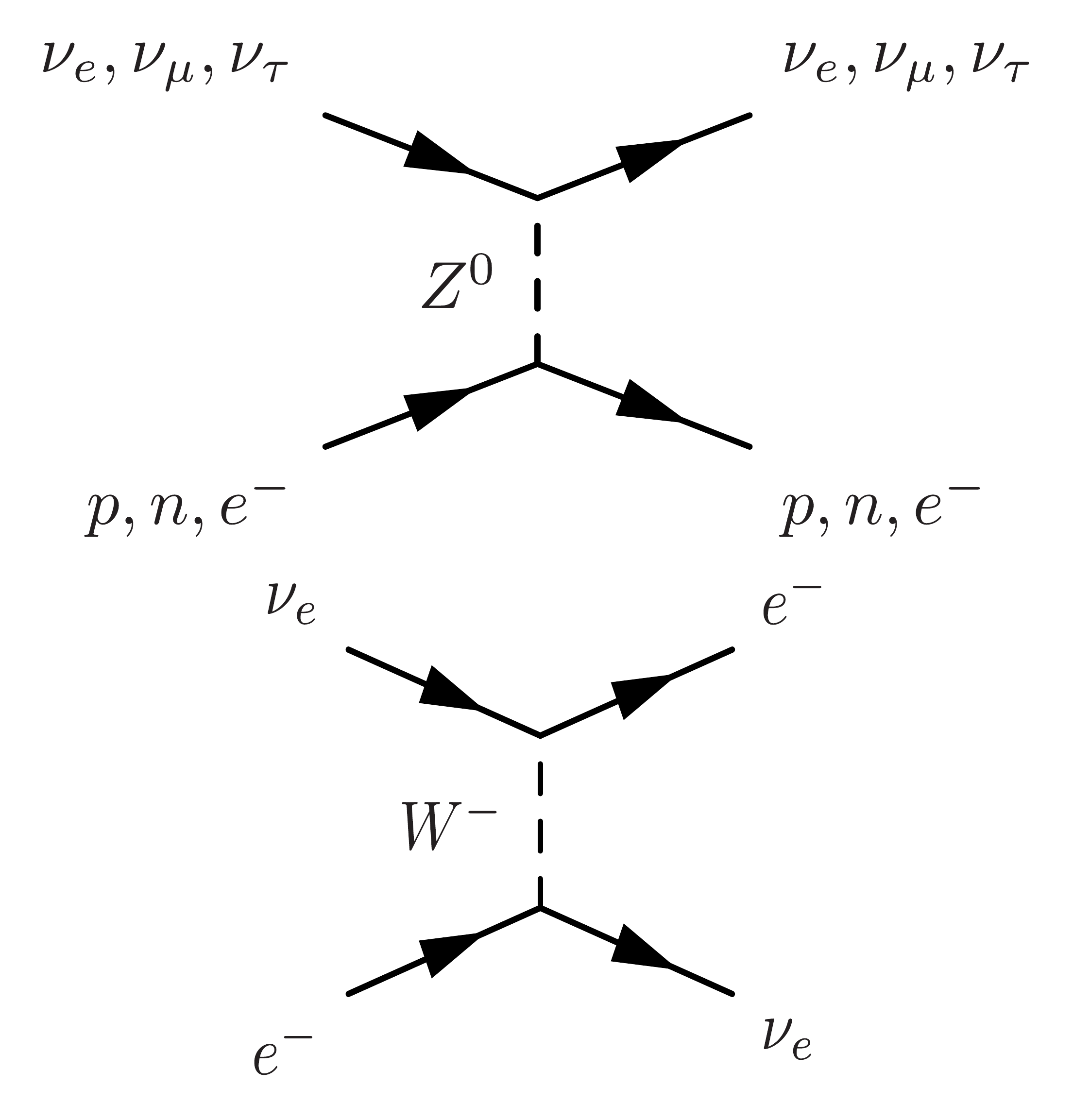}
  {Feynman graphs contributing to the forward elastic scattering of neutrinos in matter.\label{fig:neutrinoforward}}
{atmospheric/CoreResonance}
  {Atmospheric muon neutrino disappearance probability for neutrinos crossing the center of the earth. At $E_\nu \sim 3\un{TeV}$ the resonance condition is reached, strongly enhancing the dissappearence probability for anti-neutrinos.~\cite{Jones:2019nix}\label{fig:resonantosc}}

For Super-Kamiokande and IceCube-DeepCore, which are sensitive up to several tens of GeV, this search is limited to the unresolved reduction regime $4E \ll \Delta m^2_s L$. However, the sensitivity of the IceCube detector extends into the TeV region where $4E \sim \Delta m^2_s L$ for Earth-crossing neutrinos and the oscillatory behavior can be probed. In this setup the uncertainty on the overall normalization and shape of the atmospheric neutrino flux limits the sensitivity in this search. On the other side this experimental configuration profits strongly from an enhancement of the oscillation signature for neutrinos crossing the earth core. All active neutrino flavors can undergo elastic forward scattering with nucleons and electrons in matter in neutral current interactions, with an additional contribution for electron neutrinos from charged current interactions (c.f.~Fig.~\ref{fig:neutrinoforward}). In contrast sterile neutrinos will not interact with the earth matter at all. The additional potential $V(N) = \sqrt{2}G_F N$ provided by the matter number density $N$ only to the active neutrinos affects the phase velocity of the $\nu_\alpha$ waves, resulting in a modified Hamiltonian when the neutrinos traverse matter. In a two-flavor model with one active and one sterile flavor and constant matter density, one can solve the Schrödinger equation explicitly~\cite{Akhmedov:1999uz}. The resulting oscillation probabilities 

\begin{equation}
  P_{\text{MSW}}\left(\bar{\nu}_{\alpha}\to\bar{\nu}_{\beta}\right)=
\left|
\delta_{\alpha\beta}
-
  \frac{1}{\mathcal{C}}\sin^2 (2\theta_{\alpha\beta})
  \sin^{2}\!\left( \frac{\mathcal{C}\Delta{m}^2_{s}L}{4E} \right)
\right|
\label{eqn:probMSW}
\end{equation}
strongly resemble those in vacuum, but with a matter potential modification factor

\begin{equation}
  \mathcal{C}(N) = \left(\cos(2\theta_{\alpha\beta}) - \frac{2V(N)E}{\Delta m_s^2}\right)^2 + \sin^2(2\theta_{\alpha\beta}).
\end{equation}
This has the implication that for $2V(N)E = \cos(2\theta_{\alpha\beta}){\Delta m_s^2}$ the effective mixing angle  $\frac{1}{\mathcal{C}}\sin^2 (2\theta_{\alpha\beta}) = 1$. In other words, the system is in a {\bf resonant mode} where the mixing is maximal independent of how small the values $U_{\alpha4}$ of the mixing matrix are. While the situation is more complex when three active neutrino flavors are considered, the full calculation in Fig.~\ref{fig:resonantosc} shows that the flux of atmospheric anti-muon neutrinos is strongly suppressed when the resonant condition is reached.\footnote{A similar condition can be reached for neutrinos if $\Delta m_s^2 =  m^2_4 - m^2_i < 0$, i.e. if the fourth neutrino is lighter than the active neutrinos. While the simplest of these so called (1+3)-scenarios are ruled out by cosmological limits on $\sum\limits_{i=1,2,3} m_i \ll 1\un{eV}$, these can be reconciled in models with new physics~\cite{Archidiacono:2016kkh,Dasgupta:2013zpn,Gariazzo:2016lsd}.} This strongly enhances the sensitivity of IceCube in the search for anti-muon disappearance into sterile neutrinos as discussed in Sec.~\ref{sub:atmospherics}.

\subsubsection{Neutrino appearance: Active-to-active mixing\label{sec:signature-appear}}
Another important channel to search for sterile neutrinos is via the impact of the fourth mass eigenstate on the active-to-active neutrino oscillations. 
For (accelerator) experiments where the resonance condition $4E \sim \Delta m^2_sL$ is met for eV-scale sterile neutrinos, $4E \gg \Delta m^2_{ij}L$ for the mass splittings $ \Delta m^2_{21}$, $\Delta m^2_{31}$, $\Delta m^2_{32}$ of the active neutrinos. In other words, the 3-flavor oscillation of the active neutrino are not effective for these energies, and no flavor transitions are expected. In the presence of a fourth mass eigenstate, electron neutrinos can still appear from a muon beam (and vice versa) via the mixing matrix elements $|U_{e4}|^2$ and $|U_{\mu4}|^2$, which determine the amplitude
\begin{equation}
  \sin^2(2\theta_{e\mu}) =
  \sin^2(2\theta_{\mu e}) =
  4 |U_{e4}|^2 |U_{\mu4}|^2 =
  \sin^2(2\theta_{14}) \sin^2(\theta_{24})
\label{eqn:Pemu}
\end{equation}
of
$\nu_{\mu}\leftrightarrow\nu_{e}$
transitions via the fourth mass eigenstate.

As in the absence of a sterile neutrino, no electron neutrinos are expected from a muon beam this is a powerful probe to search for a fourth flavor, as is detailed in Sec.~\ref{sub:sblacc}. Moreover, these searches provide a most important verification, as they link the electron- and muon-neutrino disappearance searches, which depend on only $|U_{e4}|^2$ or only $|U_{\mu4}|^2$ respectively. Any sterile signature in the electron neutrino disappearance channels must therefore be compatible with the limits set by the combination of muon neutrino disappearance and electron-neutrino appearance searches, as we discuss in detail in Sec.~\ref{sub:global}.

\subsection{Neutrino mass experiments}

\subsubsection{Direct neutrino mass measurements}
\label{sub:Direct}

\newcommand{\dd}{\mathrm{d}}
Direct neutrino mass experiments solely rely on the kinematics of single beta decay. Close to the kinematic endpoint of the decay, a small amount of energy is taken by the rest mass of the neutrino reducing the maximal kinetic energy of the beta electron. The corresponding signature of the neutrino mass is a reduction of the endpoint and more importantly a characteristic distortion of the spectral shape in the close vicinity of this endpoint.

As the emitted electron-flavor neutrino is a quantum mechanical superposition of neutrino mass eigenstates, also the beta-decay spectrum is in fact a superposition of spectra with different endpoints corresponding to the neutrino mass eigenstates. Due to the tiny mass splitting of the active neutrinos this superposition cannot be resolved by any current direct neutrino mass experiment, therefore the incoherent sum of the neutrino masses $m_{\beta}^2=\sum_i|U_{ei}|^2m_i$ is measured.

\onefig[htb]{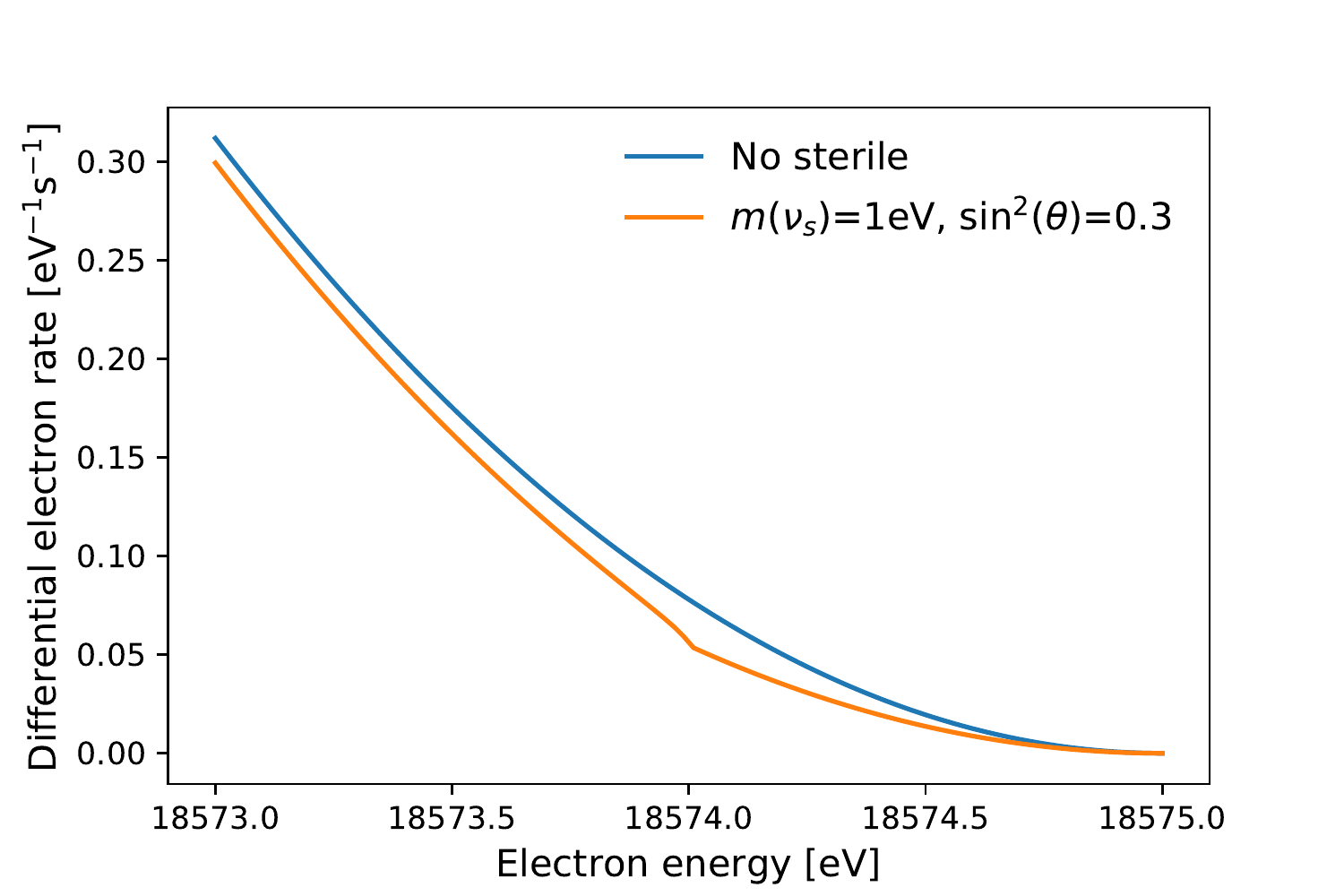}{Signal of sterile neutrino with a mass of 1eV and a mixing angle of $\sin^2\theta_{14}=0.3$ \label{fig:kink}}

If the electron-flavor neutrino contains a small admixture of a fourth neutrino mass eigenstate $\nu_4$, with a mass $m_4$ significantly larger than the absolute neutrino mass scale, the beta decay spectrum $\frac{\dd\Gamma}{\dd E}$ contains a new decay branch with an endpoint reduced by $m_4$, and an amplitude governed by the mixing $|U_{e4}|^2 = \sin^2\theta_{14}$:
\begin{equation}
\frac{\dd\Gamma}{\dd E} = \cos^2\theta_{14} \, \frac{\dd\Gamma}{\dd E} \left( m_\beta^2 \right) + \sin^2\theta_{14} \, \frac{\dd\Gamma}{\dd E} \left( m_s^2 \right)
\end{equation}

Due to the large mass splitting between $m_{\beta}$ and $m_4$, $m_4$ cannot be considered as part of the effective electron neutrino mass. In contrast, the imprint of a forth neutrino mass eigenstate would be a characteristic kink-like signature at $m_4$ below the endpoint, as displayed in Fig.~\ref{fig:kink}. This phenomenon has been extensively discussed in literature, for example in~\cite{Kobzarev:1980nk, McKellar:1980cn, Shrock:1980vy, deGouvea:2006gz, Farzan:2002zq, Farzan:2001cj, Bilenky:2001xq}.

An experimental search for a sterile neutrino signature in beta-decays is highly complementary to neutrino oscillation experiments. First of all, the observable a beta-decay experiment is the squared mass of the sterile neutrino $m^2_4$ itself, whereas an oscillation experiment is sensitive to the squared mass difference $\Delta m^2_{41}$ of the active and sterile neutrino. Moreover, short-baseline experiments are typically sensitive to rather small sterile neutrino masses, as too large masses would imply very small oscillation lengths, which could not be resolved by current experiment. In a beta-decay experiment, in contrast, the larger the sterile neutrino mass the further away from the endpoint the signal occurs. Thus, for larger sterile neutrino mass the signal rate and hence the statistical sensitivity is increased. On the other hand, a number of systematic uncertainties increase in a region further away from the kinematic endpoint.

\subsubsection{Neutrinoless double-beta decay}
\label{sub:0nbb}

The search for neutrinoless double-beta decay ($0\nu\beta\beta$) is generally regarded as the most straight-forward experimental access to determine the Dirac or Majorana nature of the neutrino mass (see the recent review in Ref.\ \cite{Bilenky:2014uka}). Next to phase space and nuclear matrix elements, the expected decay rate crucially depends on the effective Majorana mass $|m_{\beta\beta}|$. The predicted ranges for $|m_{\beta\beta}|$ change dramatically in case of the existence of light sterile neutrinos \cite{Goswami:2005ng, Goswami:2007kv, Barry:2011wb, Li:2011ss, Giunti:2012tn, Rodejohann:2012xd, Girardi:2013zra, Pascoli:2013fiz, Meroni:2014tba, Abada:2014nwa, Giunti:2015kza}. In a (3+1) scenario, the Majorana mass can be written to
\begin{equation}
    |m_{\beta\beta}|=|\mu_1+\mu_2e^{i\alpha_2}+\mu_3e^{i\alpha_3}+\mu_4e^{i\alpha_4}|,
\end{equation}
where the $\mu_k=|U_{ek}|^2m_k$ describe the partial contribution of the massive Majorana neutrino $\nu_k$ with mass $m_k$, $\alpha_{k}$ are the Majorana phases and $U_{ek}$ are the elements of the 4$\times$4-mixing matrix (Eq.~\ref{U4nu}) \cite{Giunti:2015kza}. Since the new mass value $m_4$ associated with the sterile neutrino is likely to be considerably larger than the values $m_{1,2,3}$ mostly associated with active neutrinos, the effective value of $ |m_{\beta\beta}|$ can vary considerable depending on the relative alignment of the Majorana phases $\alpha_{k}$.

\begin{figure}[htbp]
\centering
\includegraphics[width=0.9\textwidth]{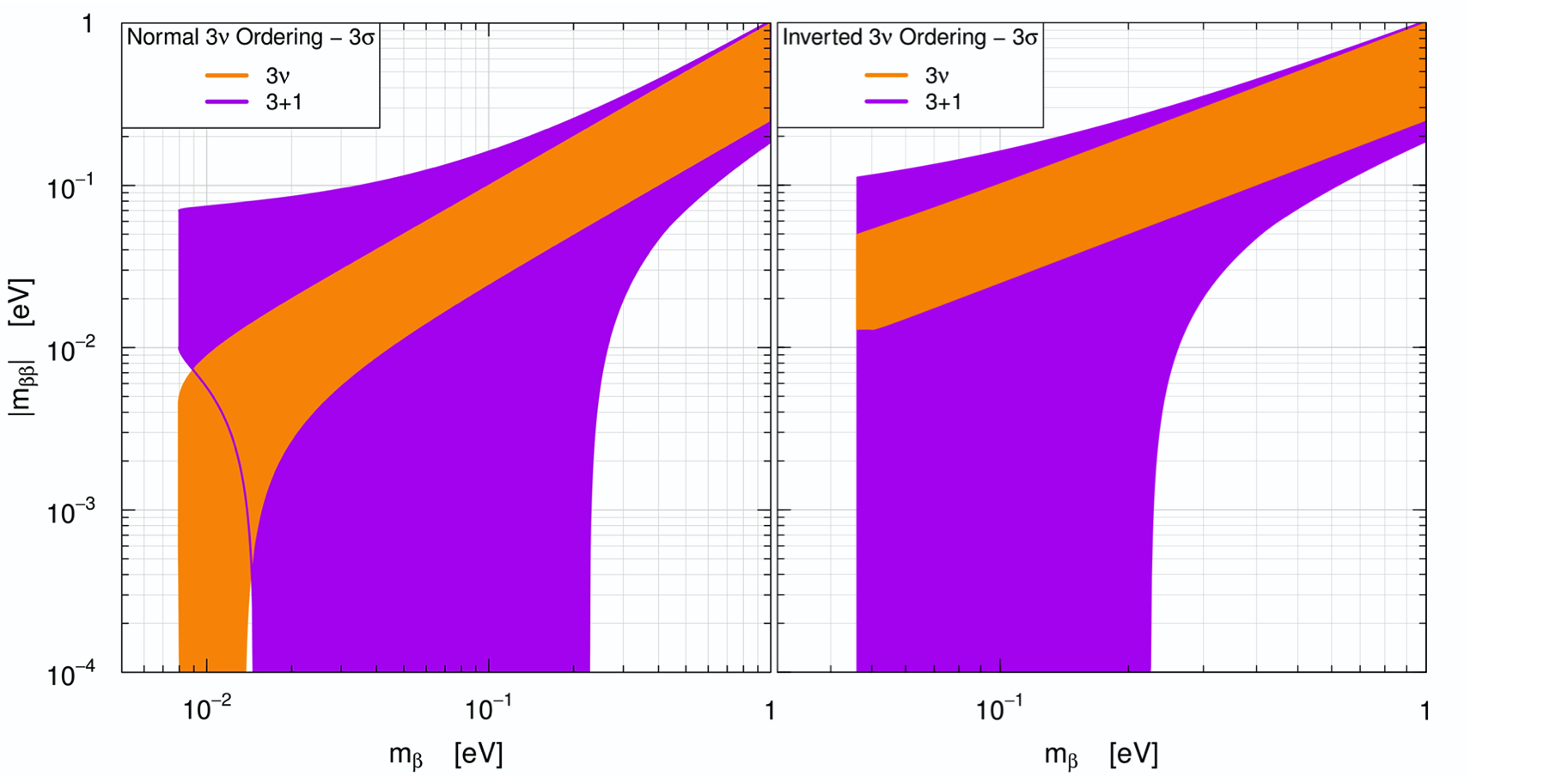}
\caption{Predicted ranges for the Majorana neutrino mass $|m_{\beta\beta}|$ as a function of the effective electron neutrino mass $m_\beta$: While left and right panels distinguish normal and inverted neutrino mass ordering, the different-color bands illustrate the different allowed ranges in case of 3-flavor and (3+1) scenarios \cite{Giunti:2015kza}.}
\label{fig:0nbb}
\end{figure}

Predictions for the possible range of $|m_{\beta\beta}|$ can be displayed as a function of the effective electron neutrino mass $m_\beta$ (sec.\ \ref{sub:Direct}). Since the coherent sum depends on the allowed ranges of the oscillation parameters $\theta_{ij}$ and $\Delta m^2_{ji}$ as well as the experimentally undetermined Majorana phases $\alpha_k$, each value of $m_\beta$ can be associated with a broad range of allowed $|m_{\beta\beta}|$ values. Figure \ref{fig:0nbb} contrasts the allowed parameter spaces for the standard 3-flavor and an extended (3+1) scenario \cite{Giunti:2015kza}. For both normal and inverted mass ordering (of the three active neutrinos), the accessible parameter range is greatly increased. Noteworthy, the range of $m_\beta$ for which total cancellation of the terms in $|m_{\beta\beta}|$ becomes possible for normal ordering is shifted to larger values, while $-$ unlike for the 3-flavor case $-$ complete cancellation becomes possible as well for inverted ordering. 

The increased variability of $|m_{\beta\beta}|$ is of no direct consequence for on-going $0\nu\beta\beta$ experiments. However, if $0\nu\beta\beta$ were observed returning a value for $|m_{\beta\beta}|$,
$m_\beta$ or the sum of $m_{1,2,3}$ were determined by direct mass searches or cosmology, and the mass ordering were measured by oscillation experiments, the combined information could be used to determine the existence of a light sterile neutrino \cite{Giunti:2015kza}.  

\subsection{Effects on cosmology}
\label{sec:cosmoeffects}
\newcommand{\Neff}{N_\mathrm{eff}}
\newcommand{\meff}{m^\mathrm{eff}_{\nu,\mathrm{sterile}}}
\newcommand{\twosigma}{95\%CL}
\newcommand{\LCDM}{$\Lambda$CDM}
\newcommand{\mixing}{\sin^2 (2 \theta_{14})}

\subsubsection{Cosmological background of sterile neutrinos}
\label{sec:cosmoeffects:production}

In standard cosmology, active neutrinos are produced after inflation and remain in thermal equilibrium with most other particles as long as the temperature in the thermal bath exceeds their decoupling temperature $T_\mathrm{dec} \sim 1$~MeV. Active-sterile oscillations before and after that time can produce a population of nearly sterile mass eigenstates, with an efficiency depending mainly on active-sterile mixing angles and squared mass differences. In the simplest (3+1) models, $\nu_4$ neutrinos are produced by non-resonant oscillations in the early universe. They acquire a Fermi-Dirac distribution similar to that of active neutrinos, possibly rescaled by a normalization factor $\chi < 1$ in the case of incomplete thermalization. Once oscillations freeze out, $\chi$ remains constant over time and determines the relic number density of $\nu_4$. This production mechanism is often named after Dodelson and Widrow~\cite{Dodelson:1993je}.

The first calculations of the relic density of sterile neutrinos were performed in pioneering papers by~Barbieri \& Dolgov \cite{Barbieri:1989ti} and Kainulainen~\cite{Kainulainen:1990ds}, who pointed out the importance of taking into account the coherent interactions of neutrinos with the primeval plasma. An updated calculation of the relic density of $\nu_4$ as a function of the mixing angles  $\sin^2 (2 \theta_{i4})$, of the square mass difference $\Delta m^2_{41}$, and of the mass hierarchy between the four mass eigenstates has been presented in~\cite{Gariazzo:2019gyi} (see also~\cite{Hannestad:2012ky,Mirizzi:2013gnd}). Like all previous studies, this work confirms that a (3+1) scenario with $(|\Delta m^2_{41}|, \mixing) \sim (1 \, \mathrm{eV}, 0.1)$ leads by far to a full thermalization of $\nu_4$, and thus to $\chi=1$. The same conclusion is actually reached for any value of $(|\Delta m^2_{41}|, \mixing)$ compatible with DANSS+NEOS~\cite{Gariazzo:2018mwd}, assuming either normal or inverted hierarchy.

More complicated production mechanisms will be mentioned in Sec.~\ref{sec:cosmo:reconcile}. They usually aim at preventing thermalization by delaying or suppressing active-sterile oscillations in the early universe. This can be achieved by invoking a large leptonic asymmetry or non-standard interactions in the neutrino sector.

The population of $\nu_4$ neutrinos can play various roles in cosmological mechanisms that are tested with observations, like Big Bang Nucleosynthesis (BBN, Sec.~\ref{sec:cosmoeffects:BBN}), the formation of Cosmic Microwave Background (CMB, Sec.~\ref{sec:cosmoeffects:CMB}) anisotropies, and that of the Large Scale Structure (LSS, Sec.~\ref{sec:cosmoeffects:LSS}) of the universe.

\subsubsection{Big Bang Nucleosynthesis}
\label{sec:cosmoeffects:BBN}

Big Bang Nucleosynthesis takes place when the expansion of the universe is dominated by the energy density of relativistic species 
\begin{equation}
\rho_R = \frac{\pi^2}{15} \left( 1 + \Neff \frac{7}{8} \left(\frac{4}{11}\right)^{4/3} \right) T_\gamma^4~,
\end{equation}
where $T_\gamma$ is the photon temperature, and the effective neutrino number $\Neff$ represents the density of relativistic relics other than photons expressed in units of one family of instantaneously decoupled neutrinos. In the minimal cosmological model, $\Neff$ is given by the number of active neutrino families plus small corrections due to the fact that neutrinos decouple gradually: the most recent and precise calculations give $\Neff=3.045$~\cite{deSalas:2016ztq}.

Predictions for the abundance of light elements produced at BBN mainly depend on: {\it (i)} the exact value of the temperature of Deuterium formation, which is roughly of the order of 0.07~MeV, and {\it (ii)} the ratio of neutron-to-proton number densities at that temperature. In turn, these two quantities are affected by the number density of baryons relative to photons in our universe, by $\Neff$ (which enters the relation between temperature and proper time), and by the chemical potential $\mu_{\nu_e}$ of electron neutrinos (which take part in $\beta$-decay).  

In the simplest (3+1) scenarios, the only effect of the light sterile neutrinos on BBN comes from an increase of $\Neff$ by $\Delta N_4$, defined as the energy density of the state $\nu_4$ during the BBN epoch expressed in units of one family of  instantaneously decoupled neutrino. An increase in $\Neff$ enhances the primordial abundance of helium and deuterium relative to hydrogen.  Thus, for these scenarios, BBN simply constrains the efficiency of active-sterile neutrino oscillations in the early universe. 

We will see in  Sec.~\ref{sec:cosmo:density} that a fully thermalized population of $\nu_4$ (with  $\Delta N_4 \simeq 1$ and $\Neff\simeq4$) is in significant tension with the measured helium and deuterium abundances, and in even stronger tension with CMB data\footnote{Until 2013, cosmological data were less precise and still compatible with $\Neff\simeq4$. This value was even slightly preferred over $\Neff\simeq3$ around 2011, see e.g.~\cite{Komatsu:2010fb} and the related comments in Sec.~\ref{sec:cosmo:density}.}. This motivated the investigation of more complicated models in which the production of $\nu_4$ is suppressed and $\Neff$ is closer to three. Some of these models can have a more subtle effect on BBN than a modification of $\Neff$. For instance, if one assumed a large primordial leptonic asymmetry residing in any of the four flavor states, neutrino oscillations in the early universe would redistribute the asymmetry between these states and inevitably alter the chemical potential $\mu_{\nu_e}$ at BBN. This may affect the efficiency of $\beta$-decay before Deuterium formation and spoil the final outcome of primordial abundance calculations~\cite{Saviano:2013ktj}.

\subsubsection{Cosmic Microwave Background}
\label{sec:cosmoeffects:CMB}

Depending on their mass $m_4$ and phase-space distribution, $\nu_4$ neutrinos could become non-relativistic before or after photon decoupling. For instance, a population of fully thermalized or Dodelson-Widrow neutrinos $\nu_4$ with a mass in the range $0.57~\mathrm{eV} < m_4 < 1.5~\mathrm{eV}$ would become non-relativistic at a time comprised between radiation-to-matter equality and photon decoupling. For smaller masses, it would still be relativistic at photon decoupling.

A population of $\nu_4$ neutrinos could affect the spectrum of primary CMB anisotropies in temperature and polarization through: its contribution to the relativistic or non-relativistic background density before photon decoupling; the gravitational interactions between $\nu_4$ and photon overdensities; and the contribution of $\nu_4$ particles to the expansion rate between photon decoupling and today. These effects are complicated and intricate. We provide here a very simplified description and refer to reference~\cite{Lesgourgues:2018ncw} for details.

Through their contribution to the background density, sterile neutrinos could change the overall amplitude of the peaks in the CMB spectrum (in the same way as a shift in the time of equality between matter and radiation), as well as the position of the peaks (through a shift in the angular diameter distance to the last scattering surface). Depending on the sterile neutrino model, these effects may or may not be degenerate with other cosmological parameters affecting the background history, such as the Hubble rate $H_0$.

Gravitational interactions between photon and sterile neutrino overdensities can also have very relevant effects, like an enhancement of the ``gravity boost'' that increases the amplitude of acoustic oscillations, raising all the peaks, and of the ``neutrino drag'' that shifts the CMB peaks to slightly larger angular scales.  However these effects would be very different if the sterile neutrinos were not free-streaming but self-interacting, a case that we will encounter in scenarios with ``secret interactions'' in the neutrino sector. Then, sterile neutrinos would have more contrasted density fluctuations, leading to even more ``gravity boost'', and a sound speed smaller than $c$ even when they are relativistic, leading to less ``neutrino drag''.

Our image of the last scattering surface is distorted by weak lensing effects caused by the surrounding large scale structures (like galaxy clusters). Through this effect, the observed CMB spectra receive additional corrections coming from the matter power spectrum, which is also affected by sterile neutrinos, as described in the next Sec.~\ref{sec:cosmoeffects:LSS}. This increases the sensitivity of the CMB e.g. to the sterile neutrino mass.

Fortunately, in each given scenario, all these intricate sterile neutrino effects are consistently taken into account when fitting CMB data, thanks to appropriate modifications of the Einstein-Boltzmann solvers used to predict the theoretical spectrum of CMB anisotropies. Possible degeneracies with other cosmological parameters are also taken into account, since the quoted CMB bounds on neutrino parameters are always marginalized over all cosmological parameters. Still, CMB bounds are unavoidably derived in the framework of a given cosmological model, with a given number of free parameters (by default, the minimal \LCDM{} model with six parameters). One should keep in mind that extended models could result in looser bounds. With the precision of current CMB data, many bounds are however nearly model-independent. For example, we will see in Sec.~\ref{sec:cosmo:reconcile} that CMB bounds on the density parameter $\Neff$ of free-streaming particles are very robust.

\subsubsection{Large Scale Structure}
\label{sec:cosmoeffects:LSS}

Several LSS observables allow us to reconstruct in a more or less direct way a fundamental quantity, the two-point correlation function of density fluctuations in the recent universe, expressed in Fourier space. This quantity is called the matter power spectrum. It can be inferred from lensing corrections to the CMB temperature spectra, from a direct extraction of the CMB lensing spectrum from CMB maps, from surveys of galaxy redshifts and positions, from surveys of the deformation of galaxy images by weak lensing, from the analysis of the Lyman-alpha forest in quasar spectra, and from the monitoring of the 21~cm Hydrogen emission line across the sky. Sometimes, instead of the full matter power spectrum, people try to measure some partial information with higher precision: for instance, the scale of Baryon Acoustic Oscillations (BAOs), which appear as an oscillatory modulation feature in the matter power spectrum. All these quantities can be affected by the properties of relic active and sterile neutrinos~\cite{Lesgourgues:2018ncw}. 

The contribution of additional relics to the background density of relativistic species in the early universe produces an overall distortion of the smooth part of the matter power spectrum, as well as a shift in the scale and amplitude of BAOs. Thus LSS data is sensitive to the contribution of sterile neutrinos to $\Neff$, but that sensitivity is usually smaller than with CMB observations. More interesting is the distinct effect of active and sterile neutrino masses on the matter power spectrum. When massive decoupled particles like neutrinos propagate with high velocities in the recent universe, they cluster on large scales but not on small scales, simply because their speed exceeds the escape velocity from small-scale gravitational potential wells. As a consequence, the balance between gravity forces and Hubble friction is broken on small scales, and the other non-relativistic matter components (dark matter and baryons) cluster at a reduced rate. Thus the free-streaming of massive neutrinos suppresses the matter power spectrum on small scales. 

It is not straightforward to infer constraints on sterile neutrino masses from this effect, because it depends jointly on the mass and phase-space distribution function of all massive neutrino states. For instance, the effect of the mass of sterile neutrinos with a small number density would be weighted down compared to that of active neutrinos. The effect would further decrease for species that would be self-interacting instead of free-streaming due to non-standard neutrino interactions. 

Thus each new model would in principle deserve a dedicated fit of all its neutrino-related parameters to CMB+LSS data. However, the literature often refers to a reference model, that provides a very good approximation to many others. In this model one assumes three standard active neutrinos contributing to $\Neff$ by 3.045, with a total active neutrino mass $\sum m_{\nu,\mathrm{active}}$ that can be fixed or floated. On top of these, one introduces a fourth free-streaming neutrino $\nu_4$ with a Dodelson-Widrow phase space distribution of pre-factor $\Delta \Neff \equiv \Delta N_4 = \chi$ and a mass $m_4$. After fitting this model to the data, one could quote bounds on $(\Delta \Neff, \, m_4)$, but they would be rather specific to the Dodelson-Widrow scenario. To extend the range of validity of the results, it is more useful to report bounds on parameters that are directly controlling the effects on cosmological observables, and that can be computed in any other model.

The best suited pair of parameters matching this goal are: {\it (i)} the sterile neutrino density in the early universe, in the relativistic regime, compared to that of other relativistic relics, and  {\it (ii)} the sterile neutrino density today, in the non-relativistic regime. In first approximation, the cosmological effects described in this section and in the previous one are governed by these two quantitites.

For the first quantity, we already have a convenient parameter: $\Delta \Neff=\Delta N_4$. For the second quantity, we could quote the current energy density of $\nu_4$, $\rho_{\nu_4}(t_0)$, or equivalently the dimensionless density parameter $\omega_{\nu_4} h^2$. But following the same logic as for $\Delta \Neff$, people usually express the sterile neutrino density today in terms of that of standard active neutrinos. This is achieved by defining $\meff$, the effective mass such that standard active neutrinos with a total mass $\sum m_{\nu,\mathrm{active}} = \meff$ would have today a total density equal to $\rho_{\nu_4}(t_0)$. A fully thermalized population $\nu_4$ would have $\meff = m_4$, while a Dodelson-Widrow population would have $\meff = \Delta \Neff \,\, m_4$. For other models, it is always possible to compute the density in the relativistic and non-relativistic regime and infer $(\Delta \Neff, \, \meff)$. In Sec.~\ref{sec:cosmo:mass} we will report bounds on these effective parameters, remembering that they apply exactly to the case of thermalized or Dodelson-Widrow neutrinos, and approximately to many other models. However, for very specific cases, like models with self-interacting neutrinos, this parametrization is inaccurate. In such situations we will report results from dedicated analyses in Sec.~\ref{sec:cosmo:reconcile}.

%
%

\section{Experimental hints for eV-mass sterile neutrinos}
\label{sec:hints}

\subsection{Reactor antineutrino anomaly}
\label{sub:RAA}

The reactor antineutrino anomaly (RAA)~\cite{Mention:2011rk} constitutes an observed rate deficit in reactor experiments, which could be explained by oscillations into a sterile neutrino state on the eV mass scale. Over the past decades, reactor neutrino experiments provided important contributions for the understanding and determination of neutrino oscillation parameters in the three-flavor framework. The KamLAND experiment improved our knowledge on the ``solar'' mixing parameters $\theta_{12}$ and especially $\Delta$m$_{21}^2$~\cite{Eguchi:2002dm} for which it provides the most precise determination. The smallest of the three known neutrino mixing angles, $\theta_{13}$, was confirmed to be non-zero and determined by the $\sim$1~km baseline experiments Double Chooz~\cite{Abe:2011fz}, Daya Bay~\cite{An:2012eh}, and RENO~\cite{Ahn:2012nd}. At the same time these experiments are sensitive to the effective mass-squared difference $|\Delta m^2_{31}|$, partly with a precision comparable to that of accelerator-based experiments~\cite{Adey:2018zwh}. 

\begin{figure}[htbp]
\centering
\includegraphics[width=1\textwidth]{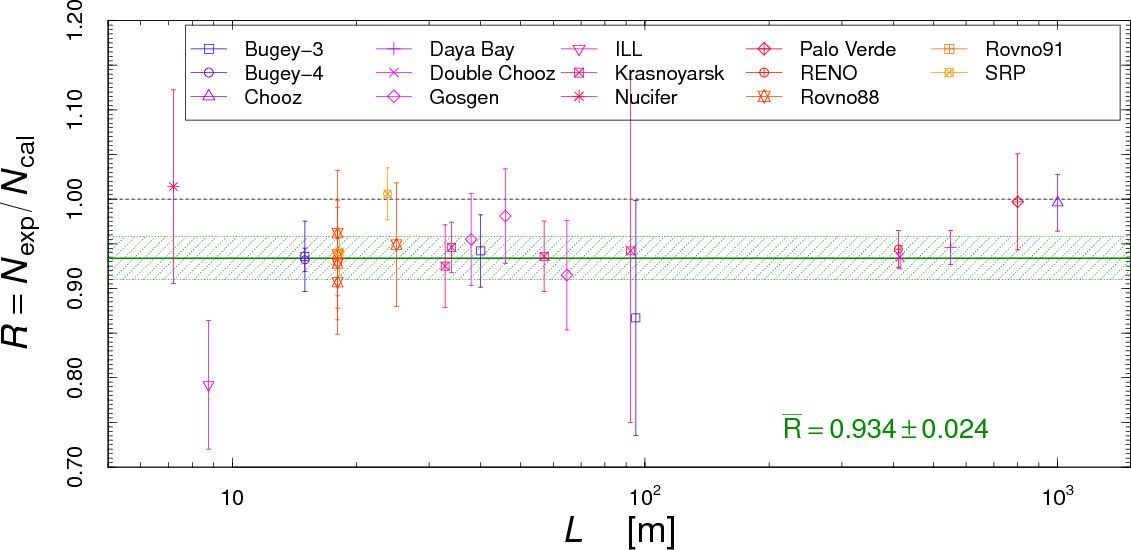}
\caption{Ratios $R$ of reactor data over predicted flux~\cite{Mueller:2011nm, Huber:2011wv} as function of the reactor-detector distance $L$~\cite{Gariazzo:2017fdh}.}
\label{fig:rateanomaly}
\end{figure}

The RAA arose in the context of the $\theta_{13}$ experiments, once the neutrino flux predictions at nuclear reactors were re-evaluated~\cite{Mueller:2011nm, Huber:2011wv}. These new calculations revealed an increase of the flux prediction of few percent as compared to the Schreckenbach et al.~predictions~\cite{Schreckenbach:1985ep, Hahn:1989zr}, which provided the reference spectra and thus rate predictions until then. Whereas experimental data were in good agreement with earlier predictions, an electron antineutrino rate deficit of more than 6\% is observed for neutrino experiments operated $6-100$~meters from the reactor when compared to the updated predictions. The rate ratios between experimental data and expected rates are shown in Fig.~\ref{fig:rateanomaly} for several experiments. The significance of this deficit knwon as the RAA is about $2.8\sigma$. 
The four main isotopes contributing to the energy production in a nuclear reactor are $^{235}$U, $^{238}$U, $^{239}$Pu, and $^{241}$Pu. The antineutrino spectra for $^{235}$U and the two Pu isotopes are obtained by a conversion method based on the beta spectra~\cite{VonFeilitzsch:1982jw} measured in the early 1980ies during exposure of Pu and U target foils with thermal neutrons at the Institut Laue-Langevin (ILL), France. In this approach, the measured electron spectra are described by a sum of virtual $\beta$-decay branches. The conversion to the antineutrino spectrum is then accomplished by the subtraction of the electron energy from the endpoint energy in each branch. By construction, this procedure reproduces the experimental electron spectrum. The $^{238}$U contribution is typically obtained either from the Mueller et al.~computation~\cite{Mueller:2011nm} or from a recent measurement from Haag et al.~\cite{Haag:2013raa} using fast neutrons at the FRM-II in Garching, Germany.

The upward shift of the normalization in the new calculations due to the harder neutrino spectra was the main contribution to the difference to previous model predictions. However, there are additional effects which enhanced the overall rate normalization increase in the simulation models. Among those are the addition of non-equilibrium effects in the calculations and a shift of the measured neutron lifetime within the last 30 years. The neutron lifetime is an input parameter in the cross-section calculations for the neutrino interaction, the inverse beta decay, and therefore has an impact on the predicted signal rate as well.  

To be independent of the measured ILL beta spectra, a complementary way to determine the expected antineutrino spectrum is to perform a computation that is just using data-base information. These ``ab initio'' summation methods allow for detailed studies of the spectrum and its contributions from the individual fission products. However, the uncertainties on such summation spectra are known to be sizeable, since they lack of experimental data, which have to be replaced by theoretical assumptions. For some important fission fragments, data bases do not provide reliable fission yields and branching ratios.

While the RAA triggered the search for oscillations involving sterile neutrinos as an exciting scenario, alternative explanations based on nuclear physics are discussed as well. The converted antineutrino spectrum depends on some assumptions. For example, $\sim$30\% of the aggregate spectra are from forbidden transitions, for which the corresponding corrections to the spectral shapes are rather uncertain~\cite{Hayes:2013wra}. Moreover, there is a critical dependence on $Z$ for the virtual branches and how the corrections from nuclear finite size and weak magnetism are implemented~\cite{Hayes:2016qnu}.  

In case $\bar{\nu}_e$ disappearance due to short-baseline oscillations provided the explanation of the RAA, the observed rate suppression should depend only on the neutrino energy but be independent of the emitting fission isotopes. This assumption was tested in Daya Bay by studying the neutrino rate for different time bins during reactor fuel evolution~\cite{An:2017osx}. A similar analysis was also done for RENO data~\cite{RENO:2018pwo}. Hence the observed neutrino deficit can be investigated for different fission fractions for the Pu and U fuel isotopes. The experiments observe a fuel-dependent variation of the inverse beta decay yield with respect to theoretical predictions. This finding disfavors the oscillation hypothesis as sole source of the RAA as well as a common mis-modelling of all fission isotopes by almost 3$\sigma$. However, a combined analysis together with global rate data shows a preference for oscillations with respect to individual isotope-dependent suppression of the mean cross section per fission~\cite{Giunti:2017yid,Giunti:2019qlt}. Hybrid models with neutrino oscillations and a prediction bias for specific fission isotopes in addition are also possible.  

\begin{figure}[htbp]
\centering
\includegraphics[width=0.8\textwidth]{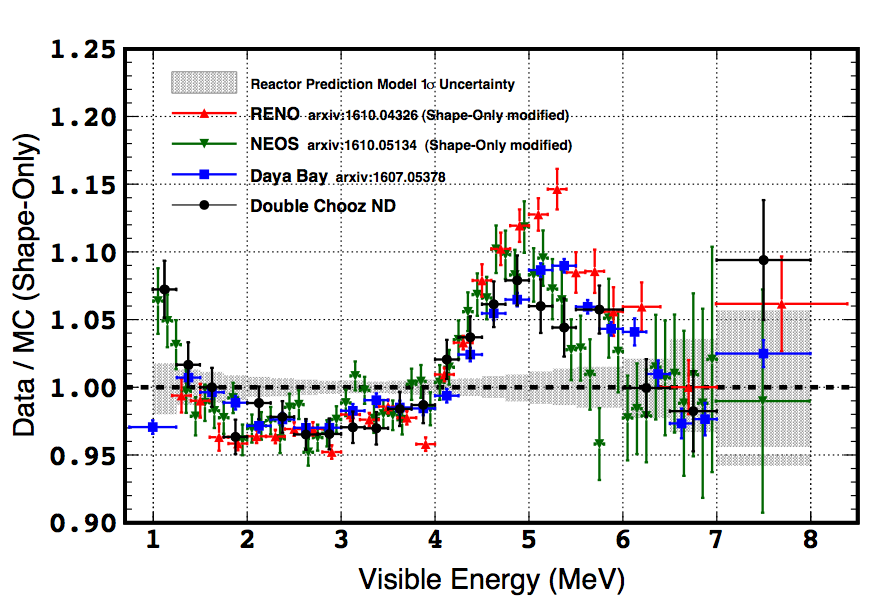}
\caption{The normalized data-to-prediction spectral ratio is shown for several experiments~\cite{DoubleChooz:2019qbj}.}
\label{fig:specdist}
\end{figure}

Double Chooz, Daya Bay, and RENO observed as well distortions in the neutrino spectrum known as the reactor ``shape anomaly''~\cite{Abe:2014bwa, An:2015nua, RENO:2015ksa}. The main feature of this distortions is an excess in the measured reactor neutrino spectrum as compared to the predicted shape around 5~MeV. A similar pattern was also observed in the NEOS data~\cite{Ko:2016owz}. However, the spectral shape measured in the Bugey~3 experiment~\cite{Achkar:1996rd} seems inconsistent with the aforementioned experiments and exhibits a flat data-to-model ratio spectrum, thus in good agreement to the prediction. For comparison, the normalized data-to-prediction spectral ratios are shown in Fig.~\ref{fig:specdist}~\cite{DoubleChooz:2019qbj} for several experiments.  

Sterile neutrinos would not explain the shape anomaly, which is more likely attributed to nuclear and reactor physics. It was suggested that a residual non-linearity in the energy response of the similarly designed detectors (see Sec.~\ref{sec:medbaseline}) could also explain the observed spectral features~\cite{Mention:2017dyq}. The new generation of experiments might be able to identify if this shape distortion is common to all fission isotopes or caused by only part of them. Some experiments are operated at highly enriched uranium (HEU) reactors, in which mainly $^{235}$U fissions contribute. Other experiments use commercial power reactors with low-enriched uranium (LEU) fuel assemblies in which several U and Pu isotopes contribute to the neutrino flux. If the neutrino spectra of HEU and LEU reactors are compared, it might show if the 5~MeV excess is solely due to the $^{235}$U contributions or similar for all isotopes~\cite{Buck:2015clx}.  

\subsection{Gallium anomaly}
\label{sub:GA}
A rate deficit of detected neutrinos as compared to the prediction is also observed in the calibration of radiochemical experiments using radioactive sources, known as the gallium (Ga) anomaly. The solar neutrino experiments GALLEX~\cite{Hampel:1998xg} and SAGE~\cite{Abdurashitov:1999zd} tested the performance of their detectors using intense neutrino sources from the decays of $^{51}$Cr and $^{37}$Ar. GALLEX and the subsequent GNO experiment~\cite{Altmann:2005ix} measured low-energy solar electron-neutrinos at the Gran Sasso Underground Laboratory (LNGS) in Italy from 1991 until 2003. As detection reaction the inverse beta decay on $^{71}$Ga producing $^{71}$Ge was used. In the GALLEX/GNO detector 30.3\,t of gallium in the form of a concentrated GaCl$_3$-HCl solution were exposed to the neutrinos. The neutrino-induced $^{71}$Ge as well as some inactive germanium (Ge) carrier atoms, which were added to the solution at the beginning of a run, were extracted from the tank in form of volatile GeCl$_4$ by a nitrogen gas stream. The nitrogen was then passed through a gas scrubber where the GeCl$_4$ was absorbed in water. The GeCl$_4$ was finally converted into GeH$_4$ and introduced into miniaturized proportional counters. There the number of $^{71}$Ge atoms was counted by the detection of its electron capture reaction with a half-life of about 11 days.

The neutrino reaction channel and detection principle in the SAGE experiment was similar to the one in GALLEX. However, in SAGE the $^{71}$Ge was extracted from metallic Ga. The solar phase of the experiment started in 1990 and took data for almost 20 years with a target mass of about 60\,t. The Ge in the SAGE experiment was extracted from the Ga metal into an aqueous solution by an oxidation reaction. By vacuum evaporation the volume of the aqueous solution was reduced by a factor of 8. The Ge was swept from this solution as volatile GeCl$_4$ by a gas flow and trapped in 1~liter of water. By a solvent extraction the Ge was concentrated into a volume of 100~ml. Finally, GeH$_4$ gas was synthesized and moved into a proportional counter in which the decays of $^{71}$Ge were counted.

In GALLEX, two intense $^{51}$Cr neutrino sources were produced by neutron capture on $^{50}$Cr. Isotopically enriched chromium (Cr) was irradiated in the core of the Siloe reactor in Grenoble for this purpose. The energies of the emitted neutrinos are about 750~keV (90\%) and 430~keV (10\%). Several different methods were used to accurately determine the source activities and a rather conservative approach was applied to estimate the systematic uncertainties. The best estimates for the source activities were ($63.4^{+1.1}_{-1.6}$)~PBq and ($69.1^{+3.3}_{-2.1}$)~PBq~\cite{Hampel:1997fc}.

In the SAGE experiment two different sources were used for calibration: a $^{51}$Cr source as in GALLEX and in addition an intense source of $^{37}$Ar. In the first calibration in 1995 a $(19.11\pm0.22)$\,PBq source of $^{51}$Cr was placed at the center of a 13.1\,t target of liquid Ga~\cite{Abdurashitov:1998ne}. The source was produced by irradiating a sample of enriched $^{50}$Cr with neutrons from the high-flux N-350 fast breeder nuclear reactor in Aktau, Kazakhstan. The activity was determined by calorimetry. Cross-checks from direct measurements of characteristic $\gamma$-lines with Ge-counters and transport calculations of the reactor neutrons showed consistent results. Almost ten years later, in 2004, the second calibration was performed with a $^{37}$Ar source, which was produced in the $(n,~\alpha)$ reaction on $^{40}$Ca~\cite{Abdurashitov:2005tb}. Calcium oxide was irradiated in the fast neutron breeder reactor at Zarechny, Russia. The $^{37}$Ar in the target was first dissolved in acid. After collection from the solution it was purified, sealed, and set up next to 13\,t of Ga. The initial activity of this source providing neutrinos with energies of 811~keV was estimated to be $(15.13\pm0.07)$\,PBq. The number is obtained from the average of different activity determination including calorimetric, counting, and mass/volume measurements.
\begin{figure}[htbp]
\centering
\includegraphics[width=0.4\textwidth]{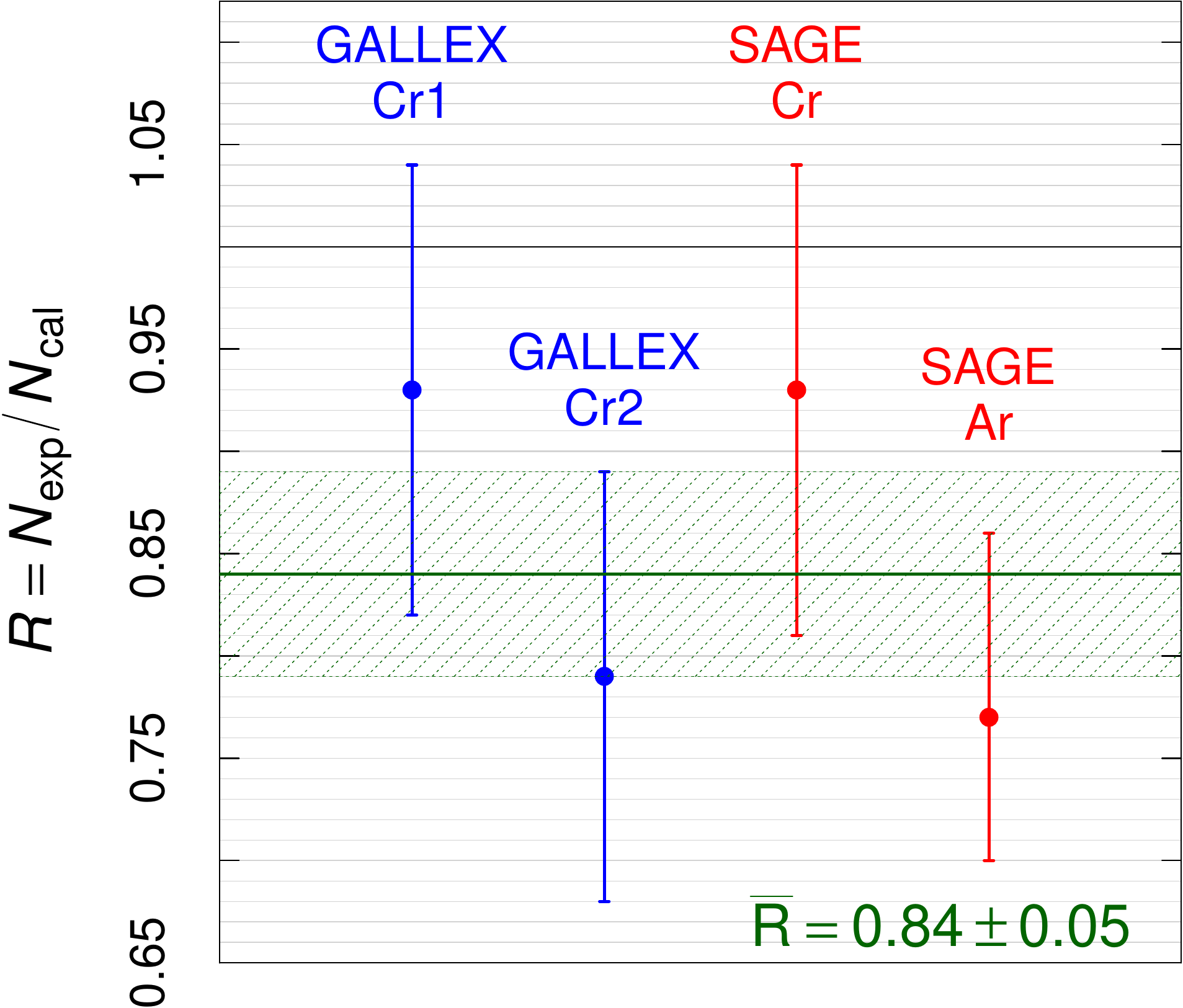}
\caption{Ratio of measured to predicted neutrino-induced signal rate in the gallium experiments GALLEX and SAGE.}
\label{fig:gallium}
\end{figure}

If the calibration runs of GALLEX and SAGE are averaged and compared to the predicted neutrino signal of the sources, a deficit of about 15\% is observed with a significance of 3$\sigma$, see Fig.~\ref{fig:gallium}~\cite{Gariazzo:2015rra}. This mismatch is known as the gallium anomaly. For each of the calibrations, the experimental uncertainty on the measured production rate is statistically dominated. Since the time of the first three of the four calibration runs, the deficit as well as its significance were continuously increasing. In the late 1990ies, the combined result of two runs in GALLEX as well as the first Cr calibration in SAGE were each compatible with the prediction within less than 1$\sigma$. With the Ar calibration in SAGE the combined deficit already reached the 2$\sigma$ level. The tension increased after a re-analysis of the GALLEX data~\cite{Kaether:2010ag} and updated estimates on the contribution of transitions to excited states of $^{71}$Ge~\cite{Frekers:2011zz}.    

The ratio of the observed-to-predicted signal in the Ga calibration runs contains two non-trivial factors to the systematic uncertainty: the neutrino interaction cross section and the chemical recovery yield of the produced isotopes. The neutrino interaction cross section on $^{71}$Ga has contributions from excited states which were originally estimated from $(p,~n)$ reactions and found to be about 5\% with a large uncertainty~\cite{Bahcall:1997eg}. More recent measurements resulted in an increasing and more precise value for these contributions of $(7.2\pm2.0)$\,\%~\cite{Frekers:2011zz}, amplifying the significance of the anomaly to the current 3$\sigma$ level. If these excited states would not be populated as strongly by the weak interaction mechanism~\cite{Hata:1995cw}, the anomaly would reduce again to less than 2$\sigma$. At least until the RAA was reported, a smaller than expected contribution of these excited states was regarded as the most likely cause for the deficit, in particular by SAGE and GALLEX experts~\cite{Kaether:2010ag, Abdurashitov:2009tn}. The high relevance of the knowledge on the cross sections is underlined by the most recent theoretical estimates. These new calculations reduce the predicted values for the cross sections by 2.5--3\% weakening the significance of the Ga anomaly from 3.0 to 2.3$\sigma$~\cite{Kostensalo:2019vmv}.

In GALLEX, the extraction efficiency was validated by the addition of $^{71}$As in the detector decaying into of $^{71}$Ge~\cite{Hampel:1997fc}. In this test the extraction yield was found to be very close to 100\%. There were also independent checks of the efficiencies in SAGE~\cite{Abdurashitov:2002nt}. Overall no indications of experimental effects lowering the results of the source experiments could be found. 

\subsection{Short-baseline appearance searches}
\label{sec:lsndminiboone}

\begin{figure}[htbp]
\centering
\includegraphics[width=0.85\textwidth]{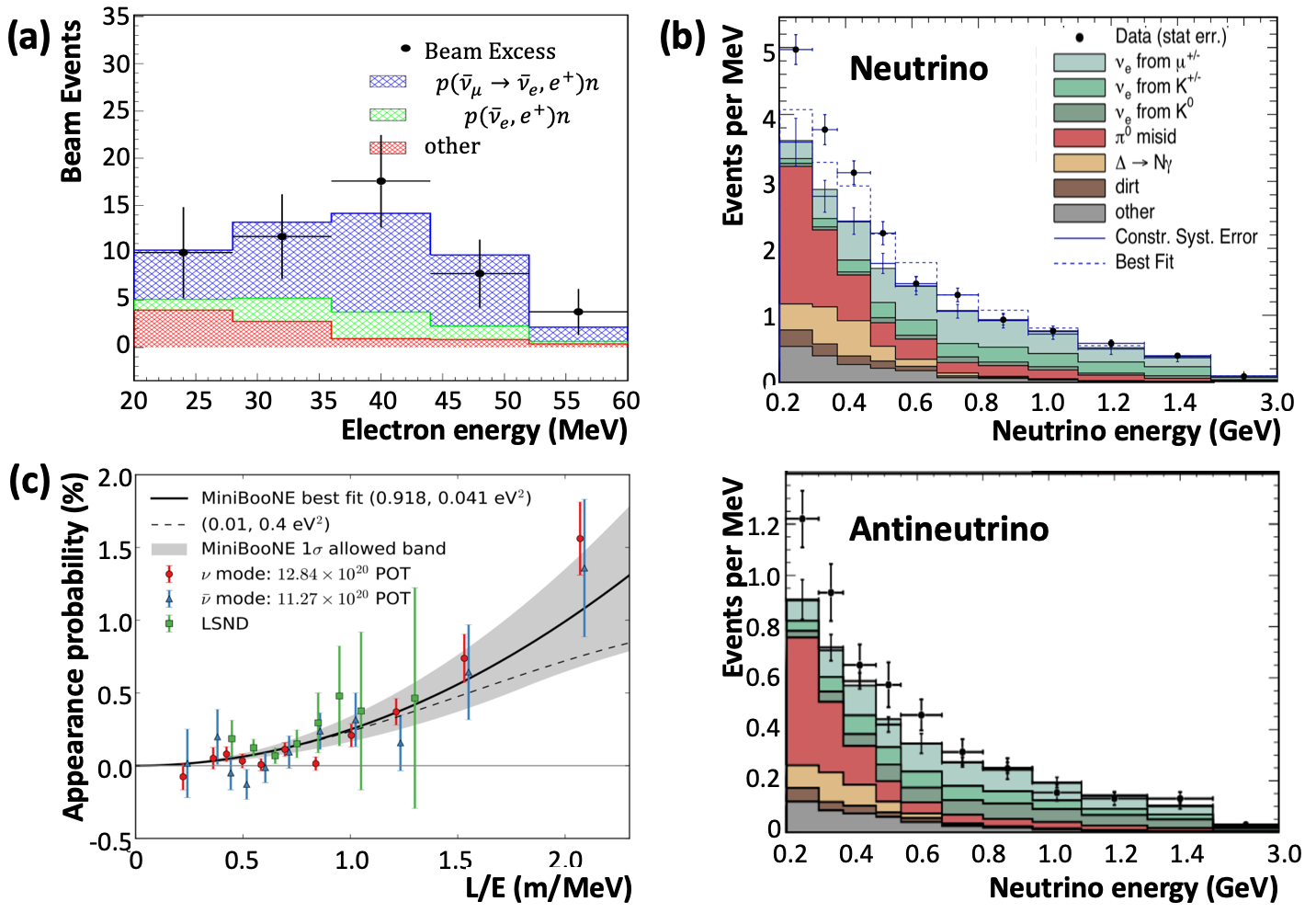}
\caption{Results for $\protect\nua{\mu}\to\protect\nua{e}$-appearance 
oscillations in the very short baseline oscillations LSND and MiniBooNE. Panel (a) shows the excess of $\bar\nu_e$ events as observed by the LSND experiment \cite{Aguilar:2001ty}, while (b) illustrates the results of MiniBooNE neutrino and antineutrino modes. A relatively small but significant $\nu_e$-like appaearance signal is observed on top of large beam-induced backgrounds \cite{Aguilar-Arevalo:2018gpe}. Panel (c) demonstrates that both results can be described by oscillations featuring a consistent $L/E$ pattern \cite{Aguilar-Arevalo:2018gpe}.}
\label{fig:lsndminiboone}
\end{figure}

\subsubsection{LSND anomaly\label{sec:lsnd}}

The oldest of the anomalies related to sterile neutrinos was found in the Liquid Scintillator Neutrino Detector (LSND) experiment: the result can be interpreted as the appearance of $\bar\nu_e$'s in a very pure $\bar\nu_\mu$ beam \cite{Aguilar:2001ty}.

LSND was realized as a beam dump experiment at the Los Alamos Meson Physics Facility (LAMPF). Protons of 800\,MeV were aimed at a fixed target to generate low-energy charged pions that were stopped within the target material before decaying. The overwhelming majority of pions (and subsequent muons) decaying were bearing positive charge, producing relatively low-energy $\nu_e$'s, $\nu_\mu$'s, and $\bar\nu_\mu$'s but crucially no $\bar\nu_e$'s. The corresponding decay of $\pi^-$'s is strongly suppressed by capture on the target atoms. The residual $\bar\nu_e$ background flux is only on the level of $\sim8\times 10^{-4}$ of the $\bar\nu_\mu$ flux.

The LSND detector was erected 30\un{m} downstream from the beam dump. The cylindrical detector held 167\un{t} of low light-yield liquid scintillator and was equipped with 1220 8''-PMTs. The diluted scintillator allowed for a simultaneous read-out of scintillation and Cherenkov light, the latter an essential asset for particle discrimination. The energy resolution at 50\un{MeV} was 7\,\%, relatively poor for a scintillation detector. LSND detected $\bar\nu_e$'s via the inverse beta decay (Eq.~\ref{eqn:IBD}), utilizing the delayed coincidence of neutron captures to achieve a substantial suppression of background.

LSND observed an unexpected excess of $\bar\nu_e$-like events that can be interpreted as $\bar\nu_\mu\to\bar\nu_e$ appearance oscillations, but at a high $\Delta m^2$ that is incompatible with our present 3-active-flavor oscillation picture. As illustrated by Fig.~\ref{fig:lsndminiboone}(a), the $\bar\nu_e$ event number and spectrum detected significantly surpass the background expectation. Background events mostly arise from the small beam contamination of $\bar\nu_e$'s in the beam and mis-identified muons from $\bar\nu_\mu+p \to n+\mu^+$ events. The number of excess events observed, $87.9\pm22.4\pm 6.0$, is $3.8\sigma$ over the background-only expectation \cite{Aguilar:2001ty,Aguilar-Arevalo:2018gpe}.

It should be noted that the KARMEN experiment operating in a similar setup but featuring overall lower statistics did not observe a corresponding excess. While at a shorter baseline of only 17\un{m}, KARMEN data does exclude a major part of the allowed parameter space when interpreting the LSND result as $\bar\nu_\mu\to\bar\nu_e$ appearance oscillations \cite{Armbruster:2002mp}. However, a combined analysis of both experimental results leaves a considerable range of allowed oscillation parameter space (Fig.~\ref{fig:LSND} \cite{Aguilar-Arevalo:2018gpe}.

\subsubsection{MiniBooNE anomaly\label{sec:miniboone}}

The Mini Booster Neutrino Experiment (MiniBooNE) was conceived to a great part to test the surprising result of the LSND experiment. In order not to be haunted by the same systematic uncertainties, both the neutrino source and the experimental conditions were changed compared to the earlier experiment. However, the $L/E$ ratio, that determines the sensitivity to the oscillation frequency and thus to $\Delta m^2$, was kept constant.

MiniBooNE \cite{Aguilar-Arevalo:2013pmq} used a conventional muon-neutrino beam setup: protons with an energy of 8\un{GeV} from the Fermilab Booster inpinged on a fixed beryllium target, producing charged $\pi$'s and $K$'s. The mesons are focused by a 60\un{cm}-long magnetic collimator and decay in flight in an evacuated decay tunnel 50\un{m} long. Polarity of the mesons can be selected by the magnetic field of the collimator, allowing operation in neutrino and antineutrino modes to search for $\nu_\mu\to\nu_e$ or $\bar\nu_\mu\to\bar\nu_e$ appearance oscillations, respectively. The resulting broad energy spectrum reaches a maximum energy of 1250\un{MeV}.

The spherical MiniBooNE detector tank is located at a distance of 541\un{m}. With a diameter of 12.2\un{m}, it holds 800\un{t} of mineral oil. The Cherenkov and scintillations photons from the charged particles in the final states of $\nu$ interactions are registered by 1280 8''-PMTs. The main information for the neutrino oscillation search is derived from charged-current quasi-elastic (CCQE) interactions. The neutrino energy is reconstructed from the visible energy of the outgoing muon or electron track and its angle relative to the beam direction \cite{Aguilar-Arevalo:2013pmq}.

As in LSND, the signature of $\nu_e/\bar\nu_e$ appearance oscillations is an excess in the number of $\nu_e/\bar\nu_e$-induced CCQE events over background. Resulting event spectra for neutrino and antineutrino modes are illustrated in Fig.~\ref{fig:lsndminiboone}(b). The vast majority of events are from beam-related backgrounds \cite{Aguilar-Arevalo:2013pmq}: in both modes, the decay of charged muons and kaons as well as K-longs in the decay tunnel leads to a sizable intrinsic contamination of the beam with electron-flavor neutrinos. Due to decay kinematics, this neutrino background is evenly distributed over the entire energy range. Contrariwise, the second-largest background of misidentified $\pi^0$'s from neutrino neutral current interactions is concentrated towards lower energies. Further backgrounds result from misidentified muons and $\nu_\mu$-induced events.  

Both in neutrino and antineutrino mode, an excess of electron-like events was discovered in the low-energy region most beset by background. Fig.~\ref{fig:lsndminiboone}(b) shows that only the lowest 4-5 energy bins are concerned that correspond to an initial muon neutrino energy of less than 500\un{MeV}. This is somewhat low (but compatible) with the $L/E$ ratio preferred from the LSND result that would the oscillation signal have appear around 500\un{MeV}, while the maximum excess in the MiniBooNE data is found in the energy region between the analysis threshold of 200\un{MeV} and 350\un{MeV}. In a sequence of results reported on the neutrino and antineutrino mode, the significance of the result steadily increased \cite{AguilarArevalo:2007it,AguilarArevalo:2010wv,Aguilar-Arevalo:2013pmq,Aguilar-Arevalo:2018gpe}. In 2018, the collaboration presented a new result including data taken with a new beam target. The excess was corroborated, placing the significance of the anomaly at $4.7\sigma$ \cite{Aguilar-Arevalo:2018gpe}. 

When expressed in terms of an oscillation signature, the putative oscillation signal translates to a slightly lower value of $L/E$ than favored by the LSND result. However, as illustrated by Fig.~\ref{fig:lsndminiboone}(c), the two results are marginally consistent \cite{Aguilar-Arevalo:2018gpe}. The overlap of the preferred oscillation parameter regions for both experiments are displayed in Fig.~\ref{fig:MB}. Best agreement is found for two islands: either a small oscillation amplitude $\sin^2(2\theta)<10^{-2}$ at high $\Delta m^2_{41}\approx 5\,{\rm eV}^2$ or considerably larger amplitudes for lower values of $\Delta m^2_{41}<1\,{\rm eV}^2$. The significance of the combined LSND and MiniBooNE anomalies has been determined to $6.0\sigma$ \cite{Aguilar-Arevalo:2018gpe}.

The difficulty of explaining the MiniBooNE low-energy excess of electron-like events
with neutrino oscillations compatible with the constraints of other experiments
has prompted several studies of alternative explanations.
One possibility is that the excess is not due to $\nu_{e}$-induced electrons,
but to photons that could have been produced by the decay of $\pi^{0}$'s generated
by neutral-current $\nu_{\mu}$ interactions in the detector
\cite{Hill:2009ek,Hill:2010zy,Harvey:2007rd,Zhang:2012xn},
and cannot be distinguished from electrons in the MiniBooNE detector.
This possibility will be investigated in the Liquid Argon Time Projection Chamber (LArTPC)
detectors of the SBN program (see Section~\ref{sub:sblacc}),
that can distinguish electrons from photons.
Another possibility is that the
MiniBooNE low-energy excess is generated by non-standard particles produced in the MiniBooNE target
or in the MiniBooNE detector.
The first case has been excluded in Ref.~\cite{Jordan:2018qiy}
by noting that non-standard particles produced in the MiniBooNE target
must decay in the MiniBooNE detector producing a visible electron-like signal
that is rather strongly peaked in the forward direction and
cannot fit the angular distribution of the MiniBooNE electron-like excess~\cite{Aguilar-Arevalo:2018gpe}.
The remaining possibility is the production in the MiniBooNE detector of a non-standard heavy particle
that decays producing an electron-like signal.
The existing hypotheses involve a heavy neutrino $\nu_{4}$
with a mass $m_{4}$ larger than about 40 MeV,
that is mostly sterile and
is produced in the detector by neutral-current $\nu_{\mu}$ interactions
through a small mixing $U_{\mu4}$.
The possibility of a radiative decay of the heavy $\nu_{4}$
\cite{Gninenko:2009ks,Gninenko:2010pr,Masip:2012ke}
is excluded by the angular distribution of the MiniBooNE electron-like excess~\cite{Radionov:2013mca}.
Recent studies~\cite{Bertuzzo:2018itn,Ballett:2018ynz,Arguelles:2018mtc,Coloma:2019qqj}
considered the possibility of a heavy $\nu_{4}$
that is charged under a new $U(1)'$ gauge group.
The corresponding gauge boson $Z'$ can have kinetic mixing with the standard hypercharge gauge boson
that lead to electromagnetic interactions of the $Z'$,
which can produce a collimated $e^{+}e^{-}$ pair that is misidentified as a single electron-like event
in the MiniBooNE detector.
If the $Z'$ is lighter than the $\nu_{4}$,
it can be produced by $\nu_{4}$ decay into a lighter neutrino
and it can subsequently decay into a $e^{+}e^{-}$ pair.
this case was proposed in Ref.~\cite{Bertuzzo:2018itn} and shown to be compatible
with the energy and angular distributions of the MiniBooNE electron-like events.
The parameter space of the model has been recently restricted
in Ref.~\cite{Arguelles:2018mtc} to a region around
$|U_{\mu4}|^2 \approx 10^{-8}$
and
$m_{4}$ between about 40 and 250 MeV
(for $ m_{Z'} = 30 \, \text{MeV} $)
by considering MINER$\nu$A~\cite{Park:2015eqa} and CHARM-II~\cite{Vilain:1994qy}
neutrino-electron scattering measurements.
On the other hand,
the authors of Ref.~\cite{Ballett:2018ynz} considered a $Z'$ that is heavier than the $\nu_{4}$
and mediates the decay of the $\nu_{4}$ into a lighter neutrino and a $e^{+}e^{-}$ pair.
This model can fit the energy and angular distributions of the MiniBooNE electron-like events
for $m_{4}$ between about 100 and 200 MeV,
$m_{Z'}$ larger than about 1 GeV,
and mixings
$|U_{\mu4}|^2 \approx 10^{-6}$ and $|U_{\tau4}|^2 \approx 10^{-3}$.
Such a large value of $|U_{\tau4}|^2$ can be probed in
atmospheric neutrino experiments,
where it generates an excess of neutral-current events in the up-going sample
\cite{Coloma:2019qqj}.

\subsection{Preferred parameter regions of the anomalies}

\begin{figure}[!t]
\begin{minipage}[b]{0.56\linewidth}
\subfigure[The reactor anomaly \protect\cite{Mention:2011rk}.]{\label{fig:RAA}
\includegraphics*[width=\linewidth]{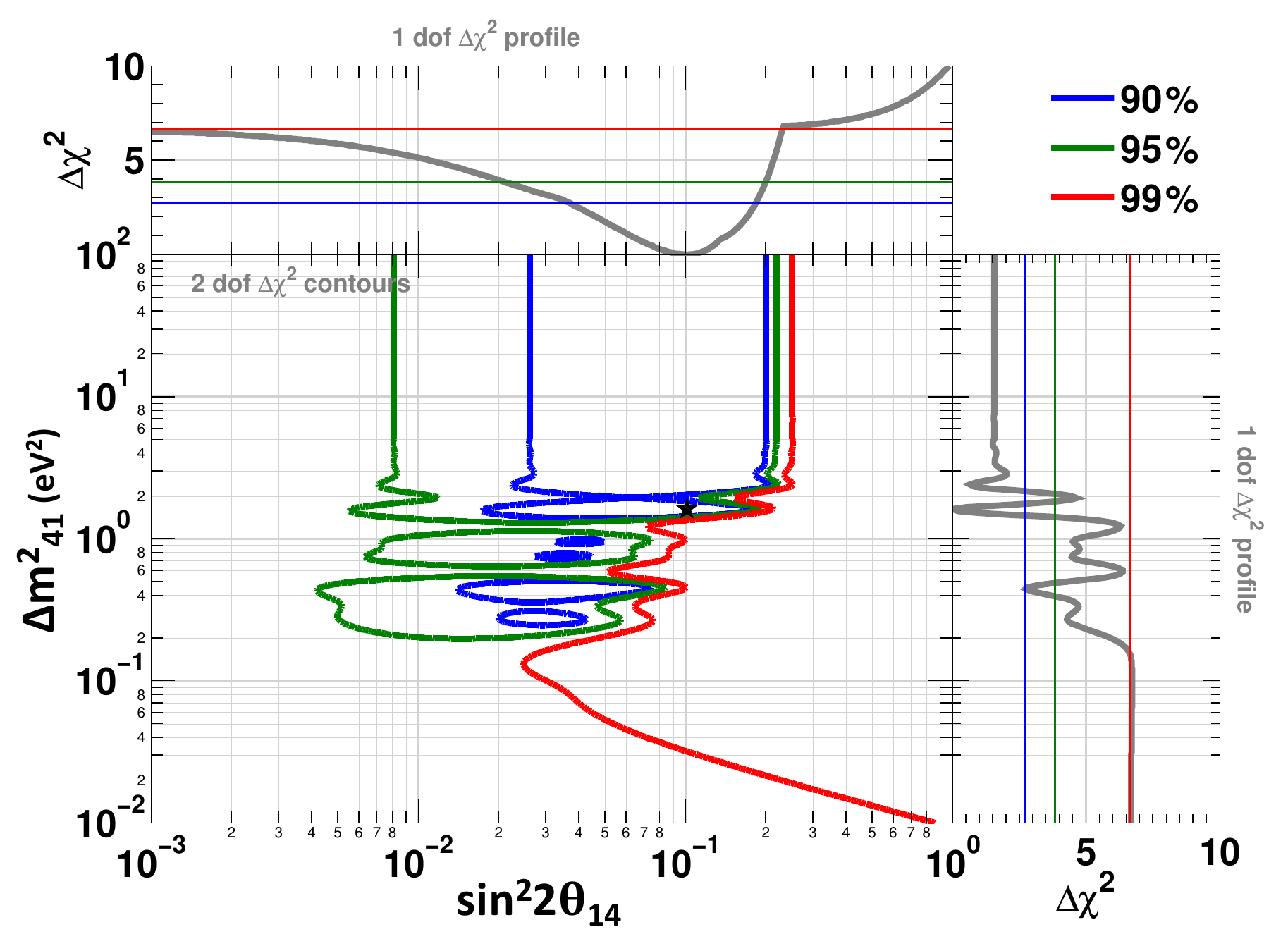}
}
\end{minipage}
\hfill
\begin{minipage}[b]{0.4\linewidth}
\subfigure[The gallium anomaly \protect\cite{Giunti:2010zu}.]{\label{fig:GA}
\includegraphics*[width=\linewidth]{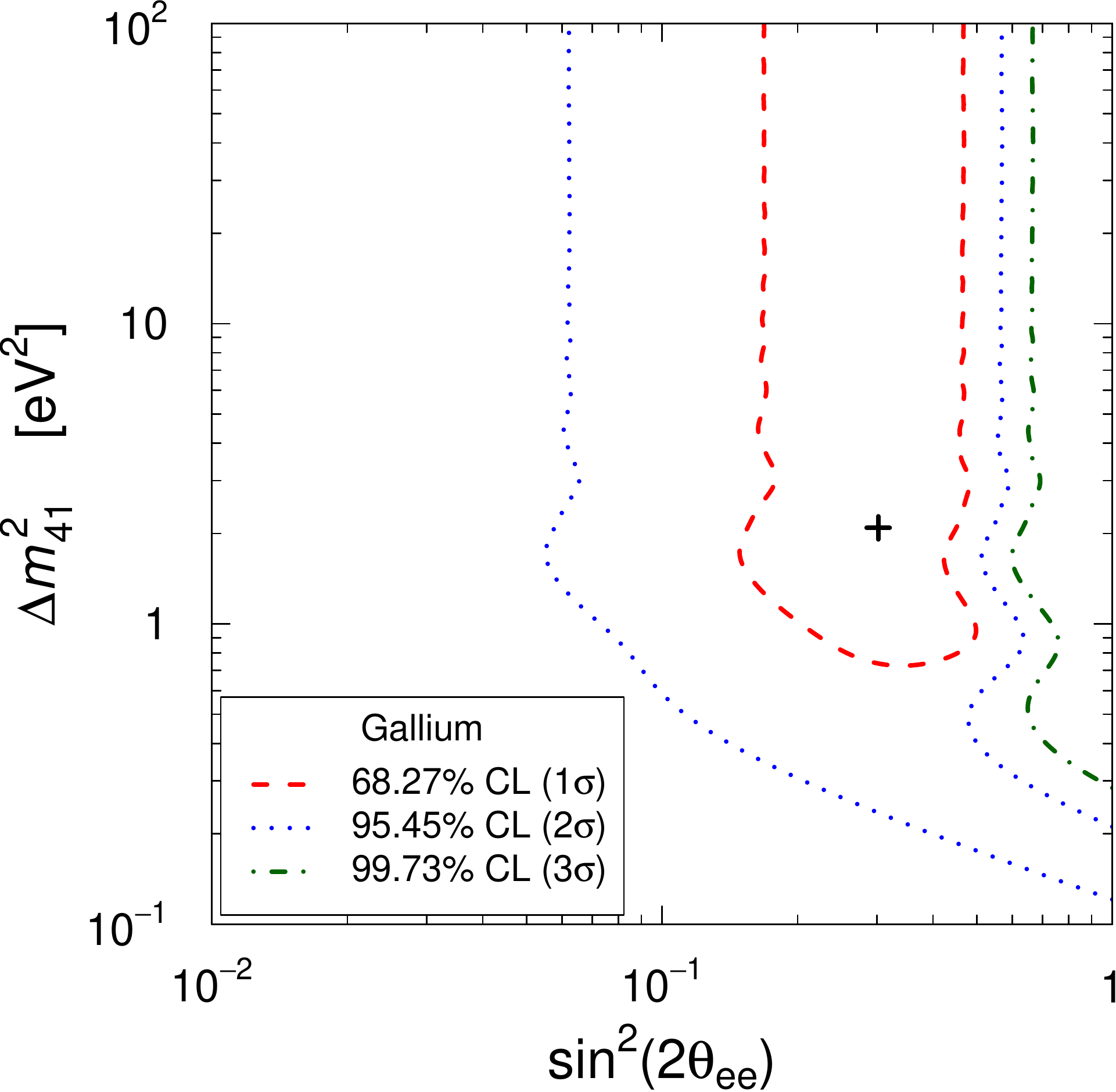}
}
\end{minipage}
\\
\begin{minipage}[b]{0.49\linewidth}
\subfigure[The LSND anomaly \protect\cite{Aguilar:2001ty}.]{\label{fig:LSND}
\includegraphics*[width=\linewidth]{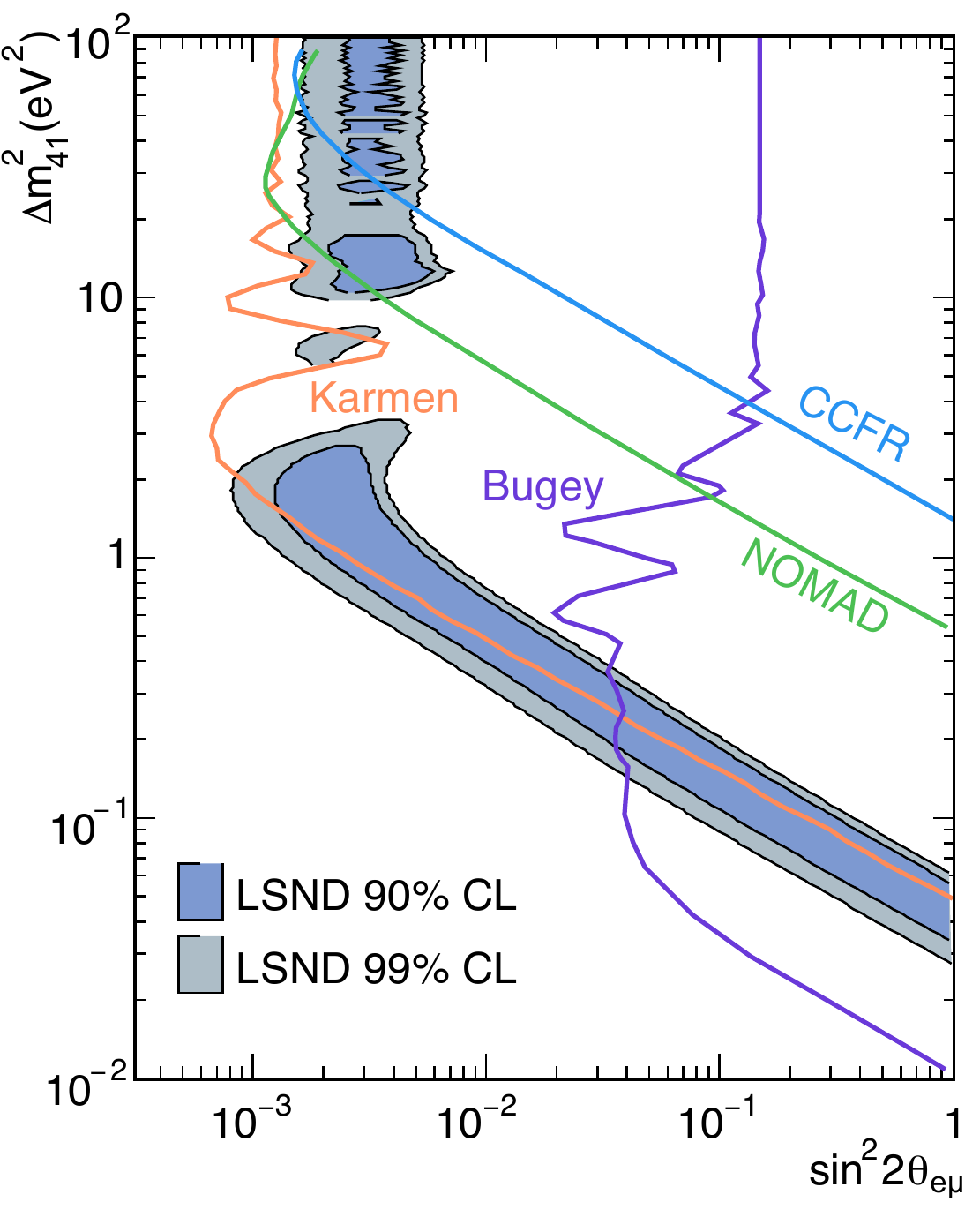}
}
\end{minipage}
\hfill
\begin{minipage}[b]{0.48\linewidth}
\subfigure[The MiniBooNE anomaly \protect\cite{Aguilar-Arevalo:2018gpe}.]{\label{fig:MB}
\includegraphics*[width=\linewidth]{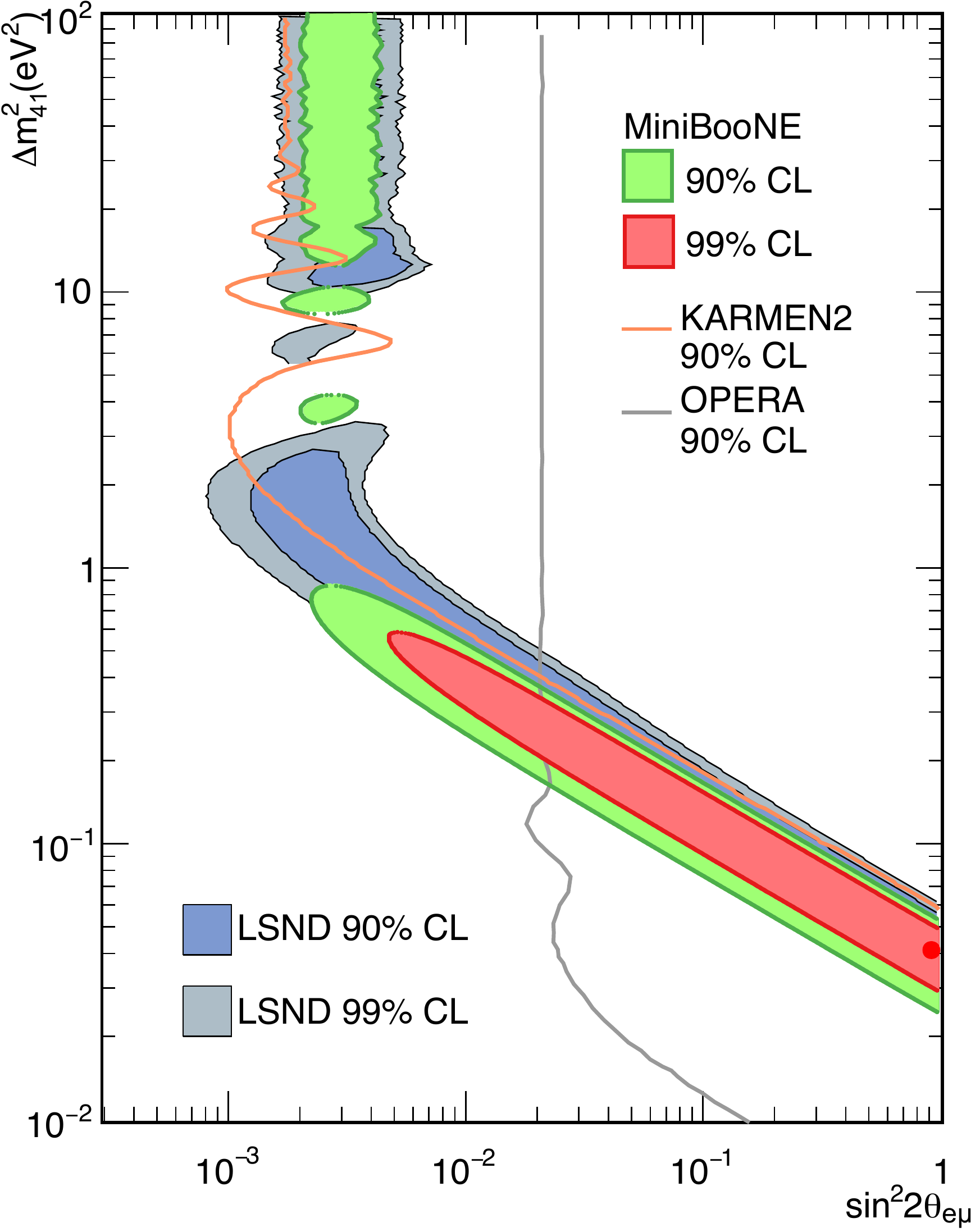}
}
\end{minipage}
\caption{ \label{fig:anomalies}
Best-fit regions of the anomalies.
\subref{fig:RAA}~The regions allowed by the reactor anomaly \protect\cite{Mention:2011rk}.
The blue and green lines enclose the regions allowed at 90\% and 95\% CL.
The red line excludes the region on the right at 99\% CL.
\subref{fig:GA}~The lines enclose the regions allowed by the gallium anomaly \protect\cite{Giunti:2010zu}.
\subref{fig:LSND}~The shaded regions are allowed by the LSND anomaly \protect\cite{Aguilar:2001ty}.
\subref{fig:MB}~The lines with the colors indicated in the legend enclose
the regions allowed at different CL's by the MiniBooNE anomaly \protect\cite{Aguilar-Arevalo:2018gpe}.
}
\end{figure}

 If the anomalies detailed above are interpreted in terms of active-to-sterile neutrino oscillations, the measured rate deviations and spectral distortions of the underlying neutrino data can be interpreted to obtain the preferred regions of the oscillation parameters shown in the four plots in Fig.~\ref{fig:anomalies}\cite{Mention:2011rk,Giunti:2010zu,Aguilar:2001ty,Aguilar-Arevalo:2018gpe}.
\medskip\\
Panels (a) and (b) display the allowed parameter space of the reactor and gallium anomalies, respectively. Their interpretation as sterile neutrino oscillations relies on electron (anti-)neutrino disappearance. Therefore, they are sensitive to the mixing amplitude  $\sin^2(2\theta_{14})$, corresponding to the mixing matrix element $U_{e4}$, as well as the corresponding mass square splitting $\Delta m^2_{41}$. While the gallium anomaly prefers slightly larger values for both mixing parameters, there is a broad region of overlap with the reactor anomaly.
\medskip\\
On the other hand, panels (c) and (d) indicate the preferred parameter regions for the LSND and MiniBooNE $\nua{\mu}\to\nua{e}$ appearance searches. The agreement between the results is well recognizable in the overlay of panel (d). While the effective mixing angle in this case corresponds to the product of mixing matrix elements $|U_{e4}|^2|U_{\mu4}|^2$, the observed $\Delta m^2$ can be almost directly related to that of the $\nu_e$ disappearance anomalies. In fact, all four results are very well described under the assumption of a (3+1) oscillation framework including one additional sterile neutrino. Compared to the mass eigenstates associated to the active flavors, the preferred $\Delta m^2_{41}$ is large and of order $1\,{\rm eV}^{2}$. 
\medskip\\
At the time of publication of the reactor neutrino anomaly in 2011, the explanation of the anomalies in the (3+1) framework appeared very compelling. The consistent picture resulting from fully compatible $\nu_e\to\nu_e$ and $\nu_\mu\to\nu_e$ mixing parameters was somewhat disturbed by the non-observation of the disappearance of muon neutrinos. As detailed in Sec.~\ref{sec:status}, an amplitude of comparable level to $\nu_e\to\nu_e$ disappearance would be required to maintain a large enough conversion in $\nu_\mu\to\nu_e$ appearance channel \cite{Dentler:2018sju}. Moreover, even before the release of the PLANCK data \cite{Aghanim:2018eyx} the cosmological limits on the sum of the neutrino masses disfavored a new mass eigenstate on the eV-scale. On the other hand, the existing limits on $N_{\rm eff}$ were easily compatible with (and at times seemed to favour) a fourth relativistic neutrino state (Sec.~\ref{sec:cosmology}) \cite{Komatsu:2010fb}.

This slightly ambiguous situation provided the starting point for a broad, worldwide experimental program that aimed to verify or disprove the explanation of the anomalies in terms of eV sterile neutrino oscillations. Since the observation of the characteristic $L/E$-dependence would provide a bullet-proof confirmation of the oscillation hypothesis, most experiments concentrated at very short baselines from their respective neutrino sources. The results and limits of this present generation of experiments as well as their final sensitivities form the content of Sec.~\ref{sec:oscillations}.

%
%

\section{Oscillation experiments}
\label{sec:oscillations}

\subsection{Reactor experiments}
\label{sub:reactorexp}

Experiments search for sterile neutrinos at nuclear reactors by checking for oscillation effects in the energy spectrum, in the neutrino rate at different baselines or both. Nuclear reactors are intense, continuous, and pure sources of electron antineutrinos. These particles are produced in the $\beta$-decay of the neutron-rich fission fragments in the reactor core. An isotropic flux of more than $10^{20}$ neutrinos per GW of thermal power in every second is provided. For a good neutrino flux prediction, a precise knowledge on the thermal power and the time-dependent fractional fission rates is needed. However, the main uncertainty in the flux models is coming from the predicted neutrino energy spectra of the different fission isotopes (see Sec.~\ref{sub:RAA}).     

One can distinguish between two types of nuclear reactors, research
reactors and power reactors. The latter have the advantage that they provide typically an about two orders of magnitudes higher flux. Spectral studies require high statistics samples, therefore an intense flux is mandatory given the tiny cross sections for neutrino interactions. On the other hand, it is often simpler to get access at short baselines for research reactors. Moreover, the fuel in research reactors are often highly enriched in $^{235}$U. Therefore, contributions from the Pu isotopes and $^{238}$U to the total neutrino flux are negligible. Moreover, there are normally more frequent and longer periods with the reactor turned off for research reactors. During those, valuable measurements can be performed to obtain a good understanding of the background sources and rates. Since the oscillation length required to explain the RAA is on the meter scale and baselines are very short, a compact reactor core is advantageous. 

The standard reaction used to detect the electron antineutrinos is the inverse beta decay (IBD) on protons, typically in organic liquid scintillator (LS) detectors:

\begin{equation}
    \bar{\nu}_e + H^+ \rightarrow e^+ + n.
    \label{eqn:IBD}
\end{equation}

In this reaction, a coincidence signal of a prompt positron and a delayed neutron event is produced. Since the neutron is heavier than target proton, the IBD interaction has a kinematic threshold of 1.8\,MeV. The prompt positron deposits its kinetic energy in the neutrino detector and finally annihilates with an electron producing two 511\,keV gammas. The ``visible energy'' $E_{\rm prompt}$ is directly correlated with the incident antineutrino energy $E_{\bar{\nu}_e}$: 

\begin{equation}
E_{\rm prompt} \sim E_{\bar{\nu}_e}- 0.784\un{MeV},
\label{Eq:Epro}
\end{equation}

\noindent in which the offset results mostly from the difference between the 1.8\,MeV, absorbed from $E_{\bar{\nu}_e}$ in order to make the IBD kinematically possible, and the energy released during the positron annihilation. 

The neutron produced in the IBD reaction thermalizes within several \un{\mu s} and is then mainly captured on hydrogen for the case of an unloaded LS. In this delayed event, a 2.2\,MeV gamma is emitted after a mean capture time of about 200\un{\mu s}. To improve the probability for neutron captures and for better background discrimination, the LS is sometimes doped with gadolinium (Gd). In this case the gamma energy upon neutron capture is increased to about 8\,MeV. Moreover, due to the very high cross section for thermal neutron capture for $^{155}$Gd and $^{157}$Gd, the coincidence time is significantly reduced. For typical Gd-concentrations at the order of 0.1\,wt.\%, the mean capture time is shortened to less than 30~$\mu$s. For improved event localization $^6$Li can be used instead of Gd. The $^6$Li isotope decays into a triton and an alpha particle after neutron capture. To test the sterile neutrino hypothesis as origin of the RAA, the general requirements on the detector are low background environment, high energy resolution, precise energy scale knowledge, and a baseline in the 10\,m range or even below. Segmentation and modularity also help to improve the sensitivity. Table~\ref{tab:VSBE} summarises several of the most important parameters for the different experiments discussed below. 

\begin{table}
\begin{center}
\begin{tabular}{l|cccccc}  
Experiment &  P$_{th}$ [MW] & L [m] & Depth [mwe] & M [t] & Technique & S/B \\ \hline
NEOS & 2700 & 25 & 20 & 1 & Gd-LS & 22\\ 
DANSS & 3100 & 10--12 & 50 & 0.9 & Gd-PS & $\sim$20\\
Neutrino-4 & 100 & 6--11 & 5--10 & 1.5 & Gd-LS & $<$1\\
\textsc{Stereo} & 57 & 9--11 & 10 & 1.7 & Gd-LS & 0.9\\
SoLid & 80 & 6--9 & 10 & 1.6 & $^6$Li-PS & 0.3\\ 
PROSPECT & 85 & 7--9 & $<$1 & 4 & $^6$Li-LS & $>$1\\ \hline
\end{tabular}
\caption{Comparison of sterile neutrino experiments at reactors including the thermal power of the reactor (P$_{th}$), baseline (L), overburden, target mass (M), detection technique, and signal-to-background ratio (S/B). All of the experiments in the table except NEOS are using segmented detectors.}
\label{tab:VSBE}
\end{center}
\end{table}

\subsubsection{NEOS}
The NEOS detector~\cite{Ko:2016owz} was installed close to the reactor unit 5 of the Hanbit Nuclear Power Complex in Yeonggwang, Korea, the same reactor complex being used for the RENO experiment. It is a commercial LEU reactor with a thermal power of 2.8\,GW. The active core size is 3.1\,m in diameter and 3.8\,m in height. The detector has an unsegmented target volume consisting of about 1\,m$^3$ of 0.5\% Gd-loaded organic liquid scintillator (Gd-LS). It is located at a baseline of 24\,m and 10\,m below ground level. In addition with the building structure directly above, the minimum overburden corresponds to 20\,m water equivalent (m\,w.e.). The signal to background ratio at this detector site is 22. The systematic error on the energy scale is estimated from the difference of calibration data and simulations. It was reported to be 0.5\% only~\cite{Ko:2016owz}.

\begin{figure}[htbp]
\centering
\includegraphics[width=0.4\textwidth]{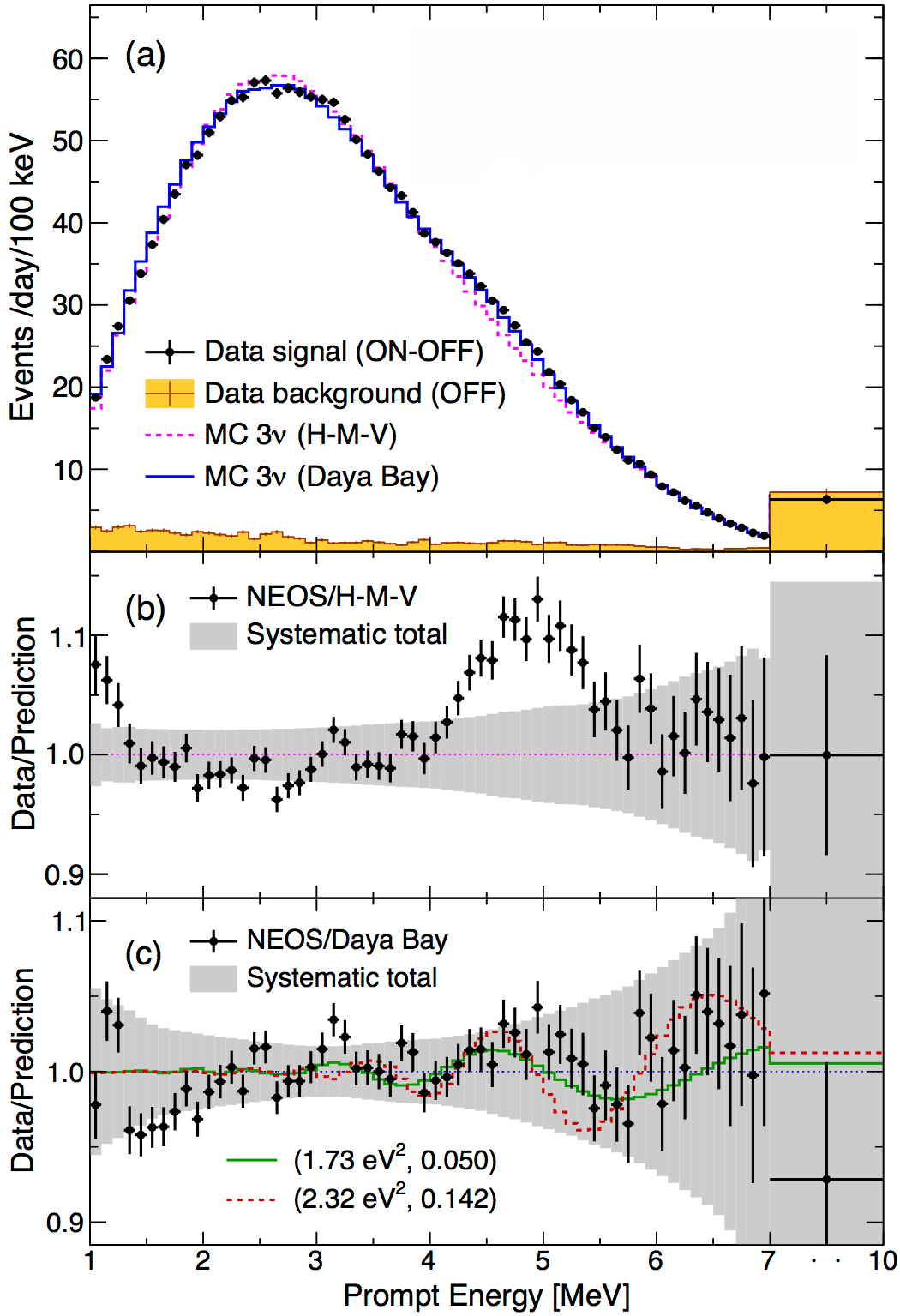}
\caption{(a) The NEOS IBD prompt energy spectrum and comparison to Daya Bay. (b) Ratio of the NEOS spectrum to the Huber/Mueller flux prediction assuming no sterile neutrinos. The predicted spectrum is normalized to the data excluding the 5~MeV excess region. (c) Ratio of the data to the expected Daya Bay spectrum. The solid green line shows the best fit of the data including a 4th neutrino state. The dashed red line corresponds to the RAA best fit parameters~\cite{Ko:2016owz}.}
\label{fig:NEOS}
\end{figure}

In the years 2015 and 2016, the experiment took data for 8~months with 6~months of reactor ON. The rather high statistics of almost 2000 antineutrinos per day in the detector allowed NEOS to confirm the spectral distortion around 5~MeV with high significance. Concerning the search for sterile neutrinos, the measured prompt energy spectrum is compared with the Daya Bay unfolded spectrum~\cite{An:2016srz} as shown in Fig.~\ref{fig:NEOS}. From an analysis of this spectral ratio, the NEOS collaboration reported an exclusion of the parameter space below 0.1 for sin$^2(2\theta_{14})$ in a $\Delta$m$^2_{41}$ region ranging from 0.2\,eV$^2$ to 2.3\,eV$^2$ with a confidence level of 90\%~\cite{Ko:2016owz}. This exclusion area already disfavours some of the best fit points in global analyses. The minimum $\chi^2$ value in a (3+1) hypothesis was found for the pair (sin$^2(2\theta_{14})$, $\Delta$m$^2_{41})= (0.05, 1.73$~eV$^2$). Since the spectra in NEOS and Daya Bay were obtained at different reactors, some spectral modelling is required. The anomalous behavior of the measured shape and its unknown origin might therefore have an impact on the analysis. Overall no strong evidence for (3+1) neutrino oscillations were observed in this experiment. After a longer break the NEOS collaboration restarted data taking using the same detector at the same site. The plan is to cover one full burnup cycle with this new phase-II data. 

\subsubsection{DANSS}
The DANSS experiment~\cite{Alekseev:2016llm} is operated at the 3.1~GW LEU reactor of the Kalinin nuclear power plant in Russia. The high power of the industrial reactor in combination with a rather short baseline that can be varied between 10.7--12.7\,m provides a high statistics neutrino signal of almost 5000 IBD events per day. Three other reactors at the site are at distances of few hundred meters. They contribute with less than 1\% to the measured rate. The reactor core is 3.7\,m high at a diameter of 3.2\,m, leading to some smearing of a hypothetical oscillation pattern.

After a prototype phase with the DANSSino detector, the full scale experiment started data taking in 2016. The highly segmented detector including 1\,m$^3$ of plastic scintillator is installed below the reactor core. The whole setup can be moved in vertical direction. The positions are changed typically 3 times per week to study the antineutrino spectrum at distances of 10.7~m (top), 11.7\,m (middle) and 12.7\,m (bottom). By the comparison of the IBD positron energy spectra that have been taken at these positions, it is possible to exclude a wide range of the sterile neutrino parameters. The DANSS approach minimises the dependence on the predicted shape and normalization of the neutrino spectrum as well as on detector effects.

The site has an overburden corresponding to a 50\,m\,w.e.~shielding which reduces the cosmic muon flux by a factor of 6 and limits the cosmic background to $\sim$5\% of the neutrino signal. The DANSS detector consists of 2500 plastic scintillator strips (1~x~4~x~100~cm$^3$), which are co-extruded with a white layer for light containment. This polystyrene-based coating does not only serve as light reflector, but contains 6\% of Gd oxide (0.35\,wt.\% pure Gd) to capture the IBD neutrons. Light is collected using a combination of Silicon photomultipliers and 'classical' photomultiplier tubes (PMTs) providing a total response of 38 photoelectrons per MeV. This implies some limitations as regards the energy resolution (about 34\% at 1~MeV) as compared to other experiments.   

\begin{figure}[htbp]
\centering
\includegraphics[width=0.6\textwidth]{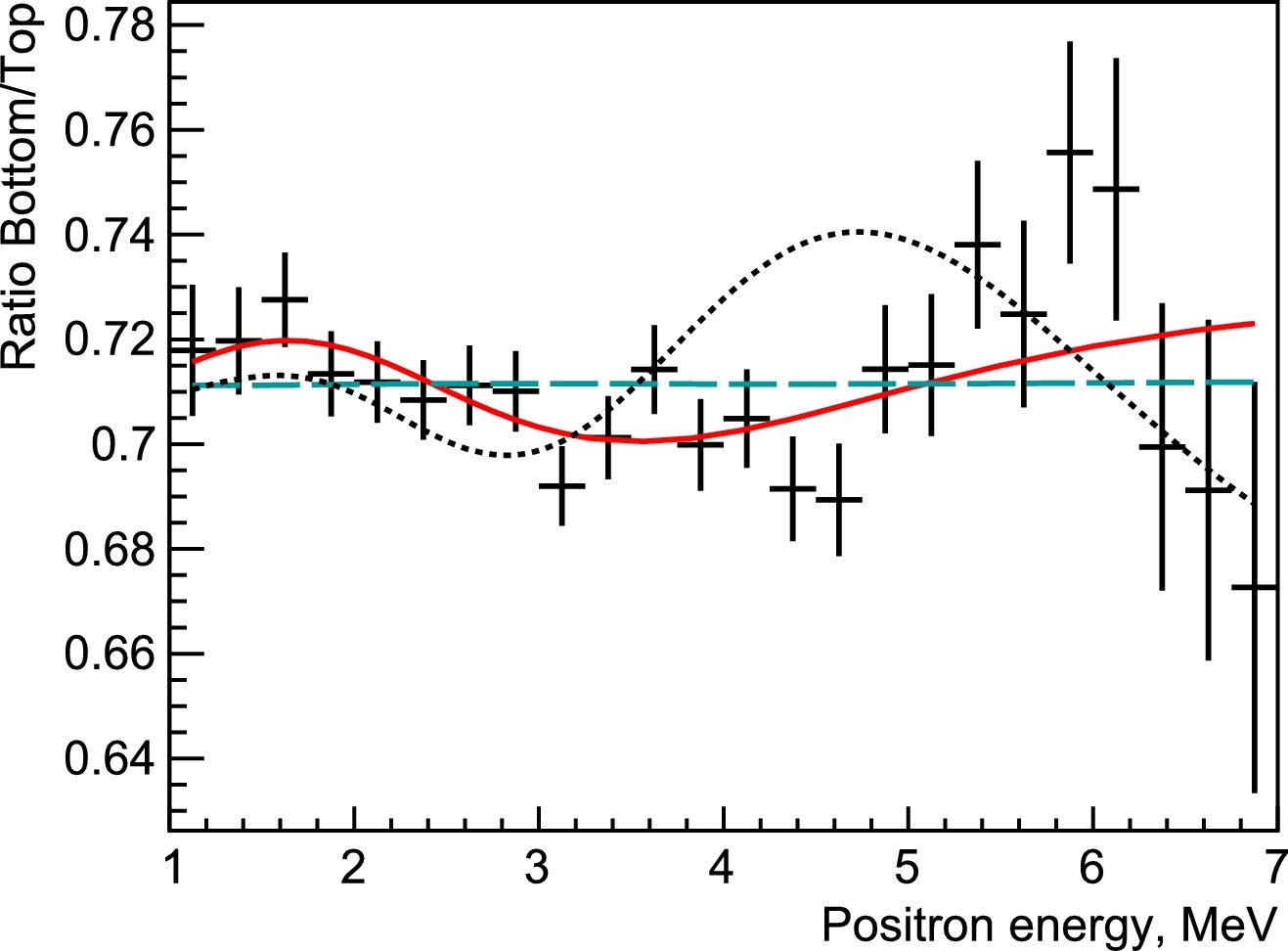}
\caption{Ratio of positron energy spectra in DANSS measured at the bottom and top detector positions (statistical errors only)~\cite{Alekseev:2018efk}. The flat dashed curve is the prediction for the three active neutrino case, the red solid curve corresponds to the best fit in a (3+1) neutrino scenario and the black dotted curve is the RAA expectation.}
\label{fig:DANSS}
\end{figure}

The recently published oscillation analysis in DANSS is based on almost a million antineutrino events~\cite{Alekseev:2018efk}. With this data, the  parameter space for a mass splitting from $0.5-2.5$\,eV$^2$ is excluded for mixing angles with sin$^2(2\theta_{14})$ above $0.01-0.1$ at 95\%~CL This exclusion area includes the best fit point of the RAA. The result is driven by the ratio of positron energy spectra at the bottom and top detector positions as shown in Fig.~\ref{fig:DANSS}. The middle position adds only little to the sensitivity. The data points in Fig.~\ref{fig:DANSS} can be fitted with a straight line (3$\nu$ scenario) resulting in a $\chi^2$ of 35 for 24 dof. The best fit for a (3+1) neutrino case would give a parameter combination with a rather low value of sin$^2(2\theta_{14})=0.05$ and a mass splitting of $\Delta$m$^2_{41}=1.4$~eV$^2$ ($\chi^2=22$). In an analysis update including more events and a full set of systematic uncertainties~\cite{DANSS:eps}, the $\Delta\chi^2$ between the best fit point and the 3$\nu$ case diminished notably. The significance of a neutrino oscillation signal around the new best fit parameters is less than 2$\sigma$. DANSS has the highest neutrino interaction rate of all the reactor experiments listed here and a very good signal to background ratio. Therefore, the experiment has very high potential in terms of sensitivity towards small mixing angles, given the collaboration can control the systematic uncertainties. 

Concerning the shape anomaly including the observed 5\,MeV excess in other experiments, the DANSS spectrum so far exhibits no significant distortions as compared to the model predictions. However, for a quantitative spectral analysis further studies on the systematic uncertainties and improvements in the Monte Carlo simulations are required.

\subsubsection{\textsc{Stereo}}
The \textsc{Stereo} experiment at the Institut Laue-Langevin (ILL) in Grenoble, France, is measuring neutrinos 10\,m away from a compact 58\,MW HEU research reactor. The neutrino target volume of the detector~\cite{Allemandou:2018vwb} is segmented in six identical cells filled with a 0.2\% Gd-LS. The more than 1800~liters of neutrino target are surrounded by another 2100~liters of Gd-free LS to detect escaping gammas. The produced scintillation light is collected by a set of 48 photomultiplier tubes (PMTs) of 8~inch diameter which are separated from the LS by an acrylic buffer and mineral oil. In \textsc{Stereo}, neutrino oscillation involving an eV sterile state would manifest itself in relative distortions of the neutrino energy spectra in the cells at different baselines.

Due to the nearby reactor and the surrounding neutron beam lines, the \textsc{Stereo} environment has a rather high background level of neutrons and gammas. For that reason a heavy shielding made of B$_4$C, lead and borated polyethylene surrounds the detector. In addition, a water-Cherenkov veto on top of the detector tags cosmic muons at the shallow depth of this site. The \textsc{Stereo} experiment started data taking in November 2016 and detects 400~neutrinos per day in reactor ON phases with a signal-to-background ratio of about 0.9.

\begin{figure}[htbp]
\centering
\includegraphics[width=0.6\textwidth]{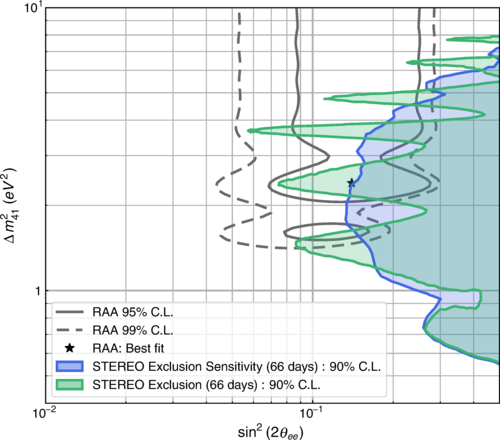}
\caption{\textsc{Stereo} sensitivity and exclusion contour of the oscillation parameters~\cite{Almazan:2018wln}.}
\label{fig:Stereo}
\end{figure}

The exclusion contour shown in Fig.~\ref{fig:Stereo}~\cite{Almazan:2018wln} was obtained from a raster scan method. The 2D oscillation parameter space is divided into slices for different $\Delta$m$^2_{41}$ bins. For each slice, the $\chi^2$ is computed as a function of sin$^2(2\theta_{14})$. The $\Delta\chi^2$ values are obtained from the minimum value of each slice. The 90\%~CL~exclusion contour in Fig.~\ref{fig:Stereo} corresponds to the parameter space where the $\Delta\chi^2$ is higher than the value giving a one sided p value of 0.1 in the probability density function obtained from pseudo-experiments for each bin. The oscillations in the contour line are due to statistical fluctuations. Before the data fit, the corresponding sensitivity contour was computed in a similar way. 

The final sensitivity of the experiment is expected to improve significantly with the envisaged total statistics of about 300~days at nominal reactor power and a similar reactor OFF period. With the current 66~days of analysed \textsc{Stereo} reactor ON data, the original RAA best fit can already be excluded at 97.5\%~CL~\cite{Almazan:2018wln}.

\subsubsection{PROSPECT}
The antineutrino source in PROSPECT~\cite{Ashenfelter:2018zdm} is the 85\,MW High Flux Isotope Reactor (HFIR) at the Oak Ridge National Laboratory in the US. As in \textsc{Stereo}, the cylindrical HEU reactor core (diameter = 0.435\,m, height = 0.508\,m) provides a pure $^{235}$U spectrum (fission fraction $>$99\%). This spectrum is measured with a fixed detector position in a baseline range of 7--9\,m using a 4000~liters (3000~liters fiducial volume) segmented LS detector loaded with 0.1\% of $^6$Li. The rectangular target volume consists of 154 optically isolated segments ($14.5\times14.5\times117.6$\,cm$^3$) including double-ended 5~inch PMT readout. 

Similarly to the other short baseline experiments, one of the key challenges of PROSPECT with an overburden of less than 1\,m\,w.e.~is the background reduction. The experiment achieves efficient background suppression by using vertex information and a powerful pulse shape discrimination (PSD) technique. There is a good separation of electron/gamma-like events and the ones from the heavy recoils in the delayed events. This way, a signal to background ratio of better than 1:1 could be accomplished. After background subtraction the measured IBD rate is 771~events/day.

\begin{figure}[htbp]
\centering
\includegraphics[width=0.6\textwidth]{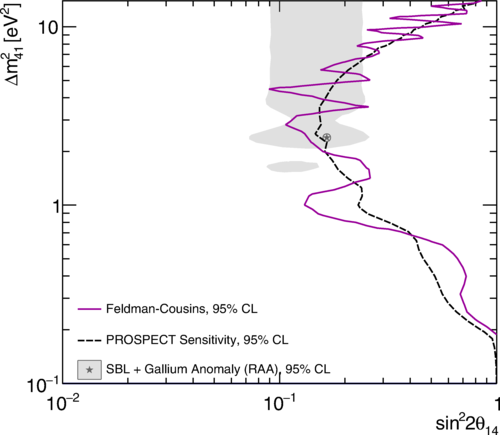}
\caption{PROSPECT sensitivity and neutrino oscillation exclusion contour with 33 live-days of reactor-ON data~\cite{Ashenfelter:2018iov}.}
\label{fig:Prospect}
\end{figure}

Also here, the existence of sterile neutrino oscillations is tested by the comparison of the measured prompt energy spectra at different baselines to a common reference. First results with 33~days of reactor ON and 28~days of reactor OFF data are in good agreement with a no-oscillation hypothesis. A significant part of the sterile neutrino oscillation parameter space can be constraint at 95\% CL~as shown in Fig.~\ref{fig:Prospect}. The best fit of the RAA is excluded at 2.2$\sigma$ CL~\cite{Ashenfelter:2018iov}. The sensitivity so far is statistically limited. In the oscillation analysis of PROSPECT, a global minimum is found for the parameter combination sin$^2(2\theta_{14})=0.35$ and $\Delta$m$^2_{41}=0.5$~eV$^2$. With the compact reactor core and the rather short baseline PROSPECT has one of the best prospects concerning the sensitivity to the parameter region at higher mass splittings.

Moreover, the PROSPECT collaboration recently published first results on the measured $^{235}$U antineutrino energy spectrum~\cite{Ashenfelter:2018jrx}. With 3426~MW-days corresponding to~40.2 days of reactor-ON, the experiment detected $31678\pm304$~(stat.) neutrino-induced IBDs. The spectrum is broadly in reasonable agreement with the Huber prediction, in particular in the 5~MeV region where most LEU reactor experiments observed an event excess. However, due to the limited statistics the spectrum is within uncertainties also consistent with the shape of other experiments like Daya Bay. 

\subsubsection{Neutrino-4}
The Neutrino-4 experiment~\cite{Serebrov2019} is running since 2016 at the compact ($35\times 42 \times 42\un{cm^3}$ core size) 100\,MW SM-3 HEU reactor in Dimitrovgrad, Russia. The neutrinos interact via the IBD reaction inside a 0.1\% Gd-LS detector with a total volume of 1.8\,m$^3$ divided in $10\times5$ sections. The first and the last of the 10 detector rows are used as active shielding reducing the fiducial volume to 1.42\,m$^3$. The detector is mounted on a moveable platform on rails providing a baseline range from 6--12\,m. It is surrounded by 60\,t of passive shielding including layers of steel, lead, and borated polyethylene. Nevertheless, due to the lack of PSD capabilities, the S/B ratio in Neutrino-4 is about 0.5 only.

\begin{figure}[htbp]
\centering
\includegraphics[width=0.7\textwidth]{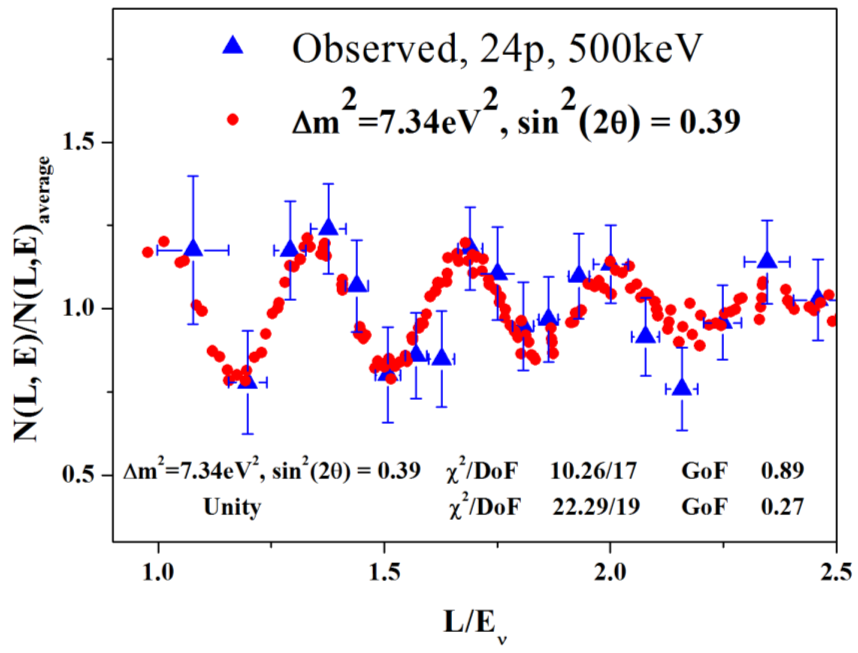}
\caption{The plot shows the L/E dependence for the Neutrino-4 data points (blue triangles) compared to the expected oscillation signal for the best fit values (red dots)~\cite{Serebrov2019}.}
\label{fig:Neutrino4}
\end{figure}

The analysis is performed by comparing the spectra recorded at the various distances of each detector section to an averaged spectrum. Compared to the simulation predictions, this averaged spectrum shows an excess at the lowest energy bin (1.5\,MeV) and a deficit (about 20\%) compared to the model around 3\,MeV energy. The experiment reported an oscillation signal with about 3$\sigma$ significance at sin$^2(2\theta_{14})=0.39$ and a rather large mass splitting of $\Delta m_{14}^2 =7.3$\,eV$^2$. Fig.~\ref{fig:Neutrino4} shows the corresponding L/E behavior. This analysis is based on a data set with 480~days of reactor ON and 278~days of reactor OFF with the reactor switched on and off 58 times. However, the quoted result is in tension with the limits obtained by the other measurements.

\subsubsection{SoLid}
The sterile neutrino search in the SoLid experiment~\cite{Abreu:2017bpe} is performed at baselines from 6--9\,m from the BR2 HEU reactor in Belgium at surface level (10\,m\,w.e.~overburden). The reactor core with a height of 90\,cm and 50\,cm diameter provides up to 80\,MW thermal power. The novel detector technology applied in SoLid was first tested in a 288\,kg prototype module~\cite{Abreu:2018pxg} deployed at the research reactor during reactor ON and OFF phases. The construction and commissioning of the SoLid Phase~1 detector with a sensitive mass of 1.6\,t was then completed in February 2018. It is enclosed in a shielding of water filled PE tanks at the sides and a 50\,cm PE ceiling on top. The setup is cooled to 10$^\circ$C for reduction of the dark count rate in the silicon photomultipliers. 

The experimental concept is based on precise localisation of the IBD events combined with a high neutron-gamma discrimination capability. To achieve this goal, a composite scintillator design is applied. The neutrino target is made out of 5\,cm cubes of polyvinyl toluene (PVT) scintillator, which are optically isolated by reflective Tyvek wrapping. In the phase~1 there are 10 PVT layers each consisting of 16~x~16 cubes. The energy depositions of annihilation gammas in neighbouring cubes can be used to tag the prompt positron of the IBD interaction. On two faces of each PVT cube there is a 250\un{\mu m} thick neutron sensitive layer of $^6$LiF:ZnS(Ag). After thermalization in the PVT cube, the IBD neutron can be captured on a $^6$Li nucleus in these layers within a coincidence time of about 60\un{\mu s}. Most of the produced alpha and triton energy is deposited inside the ZnS(Ag) inorganic scintillator. The time profile of the photon production in the ZnS is much slower (\un{\mu s} scale) than the one of PVT signals (ns scale). These characteristic time signatures can be used for effective background discrimination. The IBD efficiency in SoLid was found to be about 30\% with a signal to background ratio of 1:3.  

\subsubsection{Medium-baseline experiments}
\label{sec:medbaseline}

Beyond the new generation of short baseline experiments, there are also the $\theta_{13}$ experiments Double Chooz~\cite{Abe:2011fz}, Daya Bay~\cite{An:2012eh}, and RENO~\cite{Ahn:2012nd} with some sensitivity for sterile neutrino oscillations. These experiments are based on a concept with near and far detectors searching for the disappearance of electron antineutrinos. Whereas the far detectors are positioned at the km scale from the reactor, the near detectors are placed at baselines of few hundred meters. Due to the these longer baselines, the experiments are sensitive towards lower mass splitting as compared to the very short baseline experiments. 

The detector concepts are similar for the three experiments with identically designed near and far detectors consisting of concentric cylinders. The central target volumes contain several tons of liquid scintillators with a Gd-loading of about 0.1\% in an acrylic vessel. The target cylinders are surrounded by metal-free liquid scintillators in the second volume and a mineral oil-like transparent buffer liquid in the third zone. The scintillation light produced in the neutrino signal is observed by PMTs mounted on the inside of the steel walls of the buffer.     

An oscillation including a sterile neutrino might manifest itself in relative differences of the measured spectra for the different baseline configurations or in deviations from the predicted spectrum. The sterile neutrino search in Daya Bay is conducted via both analyses~\cite{An:2016luf}. The antineutrino rates and energy spectra are compared for their eight detectors distributed in three experimental halls. In each of the two near underground halls, data was taken with two detectors, whereas the far hall has four identical detectors. The Daya Bay configuration of multiple baselines allowed to explore a mass splitting range over three orders of magnitude between $2\cdot10^{-4}\leq \Delta m^2_{41} \leq 0.3$~eV$^2$. Both of the applied analyses yield consistent results and no indication for the existence of sterile neutrinos could be found~\cite{An:2016luf}. A similar study was performed in RENO with comparable findings~\cite{Yeo:2017ied}. 

\subsubsection{Outlook}
All of the short-baseline experiments described above continue to take data or even plan upgrades of their setups. Therefore, the sensitivity will further improve and should cover all of the preferred parameter space suggested by the RAA and Gallium calibration data. From the predicted final sensitivities, very high potential is expected from the DANSS and PROSPECT experiments. DANSS will profit from high statistics samples at different baselines combined with a good signal-to-noise ratio. PROSPECT has the advantage of a much higher energy resolution in their detector segments. 

All reactor neutrino experiments so far used the IBD reaction for detection. Currently several running and upcoming experiments investigate the possibility to detect the neutrinos by coherent elastic neutrino nucleus scattering. In this reaction, the cross-section is about two orders of magnitude higher than the one of the IBD. Therefore, the mass of the detector could be strongly reduced given the energy threshold is low enough. This new approach might open the possibility for a new set of reactor experiments with modular design at very short baselines. They might provide complementary insights on the reactor anomalies, in particular for the high energy part of the antineutrino spectrum.

\subsection{Radioactive source experiments}
\label{sub:sourceexp}

\begin{table}
\begin{center}
\begin{tabular}{l|l|llll}  
$\nu$ type & Isotope & Lifetime & Decay& $E_{\nu}$ [MeV] & Production\\ \hline
$\nu_e$ & $^{51}$Cr & 40 d& EC & 0.75 (90\%) &  $^{50}$Cr irradiation in reactor \\
           &  &  &  &  0.44 (10\%) & \\
$\nu_e$ & $^{37}$Ar & 35 d& EC & 0.811 &   Ca oxide irradiation in reactor \\ \hline
$\bar{\nu}_e$ & $^{144}$Ce-$^{144}$Pr & 411 d& $\beta^{-}$ & $< 2.996$ &  Extraction from reactor spent fuel \\
$\bar{\nu}_e$ & $^{106}$Ru-$^{106}$Rh & 538 d& $\beta^{-}$ & $< 3.54$ &  Extraction from reactor spent fuel \\
$\bar{\nu}_e$ & $^{90}$Sr-$^{90}$Y & 40 y & $\beta^{-}$ & $< 2.28$ &  Extraction from reactor spent fuel \\
$\bar{\nu}_e$ & $^{8}$Li & 868 ms & $\beta^{-}$  & $< 12.9 $ &  Beam-generated neutrons on $^7$Li \\ \hline
\end{tabular}
\caption{Possible isotopes that could be used as (anti)neutrino sources in the search of light sterile neutrino. The $^{51}$Cr and $^{37}$Ar neutrino sources were used in the 90ies by the Gallex and SAGE collaborations to calibrate their radiochemical detectors measuring solar neutrinos.  The first three antineutrino sources were suggested in~\cite{Cribier2011}. The IsoDAR style $^{8}$Li source have been suggested in~\cite{Basov1985,Lutostansky2011,Bungau:2012ys}.}
\label{tab:sources}
\end{center}
\end{table}

An alternative way to search for $\nu_e\to\nu_e$ disappearance due to oscillations of MeV-scale neutrino at meter-scale baseline is to apply strong artificial radioactive sources emitting $\nua{e}$ next to (or even inside) large-volume neutrino detectors. Sterile neutrinos of eV mass would cause a deficit in the measured interaction rate of electron-flavor neutrinos. In addition, in case of large neutrino detectors, a {\it smoking gun} signature would be the observation of an continuous oscillation pattern across the target volume. This approach is fully complementary to other oscillation experiments searching for sterile neutrinos. Comparing to the very short-baseline reactor experiments (Sec.~\ref{sub:reactorexp}), the advantage here is certainly the fact that the neutrinos originate from a single isotope and their energy spectrum is thus less complex. The fact that source experiments can be placed in deep underground laboratories is fundamental for the suppression of cosmogenic background, typically dominant background for shallow reactor experiments. In addition, the irreducible background (neutrons, gammas) is much smaller compared to the environment next to reactors. Furthermore, it is not possible to place large-scale detectors at meter-scale distance from reactors. The typical size of the radioactive sources is smaller with respect to the cores of nuclear reactors and thus the systematics related to the uncertainty about the origin of emitted neutrinos is smaller in source experiments. An obvious disadvantage is the limited neutrino flux compared to nuclear reactors, as well as a limited measurement time due to the short lifetime of some isotopes. Intrinsic is the complexity of the source production, considering very high activity and radio-purity levels that are required. Non-negligible is also the factor of administrative difficulties related to the permissions for transport and usage of the source.  

Table~\ref{tab:sources} summarizes the basic properties of the isotopes that have been considered in the search for light sterile neutrinos. Electron-capture decays of $^{51}$Cr and $^{37}$Ar produce sub-MeV mono-energetic $\nu_e$'s. Such sources have been actually produced and have been used in the calibration of the radiochemical experiments detecting solar neutrinos~\cite{Hampel:1997fc,Abdurashitov:1998ne,Abdurashitov:2005tb}, as described in Sec.~\ref{sub:GA}. Isotopes decaying via $\beta^{-}$ process produce $\bar{\nu}_e$ with continuous spectra. The $^{144}$Ce-$^{144}$Pr, $^{106}$Ru-$^{106}$Rh, and $^{90}$Sr-$^{90}$Y systems have been identified in~\cite{Cribier2011}. The IsoDAR (Isotopic Decay At Rest) experimental program aims at the application of the very short-lived $^{8}$Li isotope~\cite{Basov1985,Lutostansky2011,Bungau:2012ys}, continuously produced via irradiation of the $^7$Li target with neutrons produced by a proton beam impinging on $^{9}$Be target.

So far, no experimental results are available with this experimental technique. In Sec.~\ref{sub:SOX} we describe the SOX~\cite{SOX} project, that was approved to apply the source below the Borexino detector at Laboratori Nazionali del Gran Sasso (LNGS) in Italy. In spite of the fact that the project was cancelled due to the problems with the source production, a discussion of its sensitivity may be valuable for similar experiments. In Sec.~\ref{sub:BEST} we briefly describe the BEST experiment~\cite{Barinov:2016znv,Barinov:2017ymq} in Baksan in Russia: the new $^{71}$Ga radiochemical experiment plans to apply 3\,MCi $^{51}$Cr neutrino source fron July 2019. In Sec.~\ref{sub:JUNO} we review the potential with source application at JUNO 20\,kton liquid scintillator detector~\cite{JUNOYellowBook} planning to start data taking in Jiangmen in China in 2021, including IsoDAR@JUNO.

\subsubsection{SOX}
\label{sub:SOX}


The SOX (Short distance neutrino Oscillations with BoreXino) project~\cite{SOX} considered the deployment of $^{51}$Cr and $^{144}$Ce-$^{144}$Pr sources close to the Borexino detector~\cite{ALIMONTI2009568} at LNGS in Italy. The main advantage with respect to the Gallium calibration experiments (Sec.~\ref{sub:GA}), observing only the deficit of the integrated rate above the experimental threshold, is the ability to reconstruct the energy and position of each individual interaction. This means, in principle, a possibility to observe an oscillation pattern across the detector. The project was the only approved experiment of this kind and the production of the $^{144}$Ce-$^{144}$Pr source via the extraction from the spent nuclear fuel
had been contracted with FUSE PA Mayak in Russia. Due to significant delays in the production of the source with requested activity and radiopurity, the project has been closed in January 2018.

Borexino is an extremely radio-pure detector filled with 280\,t of liquid scintillator held in a thin nylon sphere with 4.25\,m radius. Borexino detects about 500 photoelectrons per MeV, resulting in the energy resolution of 5\% at 1\,MeV. The position of each event is reconstructed by the time-of-flight method with the resolution of 10\,cm at 1\,MeV. 
%
Borexino is taking data since May 2007 with the main aim to measure solar neutrinos~\cite{Agostini:2018uly} via elastic scattering off electrons. 
It has detected as well geo-neutrinos, i.e.~$\bar{\nu}_e$'s from the $^{238}$U and $^{232}$Th decay chains occurring inside the Earth~\cite{Agostini:2015cba}, via the IBD interaction (Eqn.~\ref{eqn:IBD}). The same mechanism was to be applied for detection of $\bar{\nu}_e$'s from the $^{144}$Pr source.


The life-time of $^{144}$Pr is way too short to allow the fabrication of a pure $^{144}$Pr source. The parent $^{144}$Ce nucleus has a much longer life-time (half-life of 250 days) which fits the needs. Only the $\bar{\nu}_e$ emitted from $^{144}$Pr extend above the IBD kinematic threshold. The required total activity of the source was 100-150\,kCi, corresponding to order of $10^4$ IBD events in 1.5 years of data taking, with an expected antineutrino background of about 6 geoneutrinos and 12 reactor antineutrinos.
About 4\,kg of CeO$_2$ were to be pressed into a stainless steel capsule of 170\,mm height. With about 0.7\% branching ratio, $^{144}$Pr decays to an excited state of $^{144}$Nd, accompanied by the emission of a 2.2\,MeV gamma. A 19\,cm thick tungsten-alloy shielding, attenuating this gamma by a factor of $10^{12}$ had been already produced at Xiamen Ltd. in China. The total power emitted by the source, a key ingredient in the analysis, was to be measured by the two redundant calorimeters already constructed. Their performance was proven to be better than the required 1\% precision for the decay heat emitted by the source.

The electron - and therefore also the $\bar{\nu}_e$ - spectrum of the main $^{144}$Pr decay branch follows a non-unique first forbidden decay that cannot be directly determined from the theory and the published measurements show large disagreements.
%
%
An uncertainty in the spectral shape can affect as well
%
the transformation of the calorimetrically measured source power to the activity.
%
%
Therefore, two table-top measurements of $^{144}$Pr electron energy spectrum are ongoing. 

\begin{figure}
\begin{minipage}[b]{0.45\linewidth}
{
\includegraphics*[width=\linewidth]{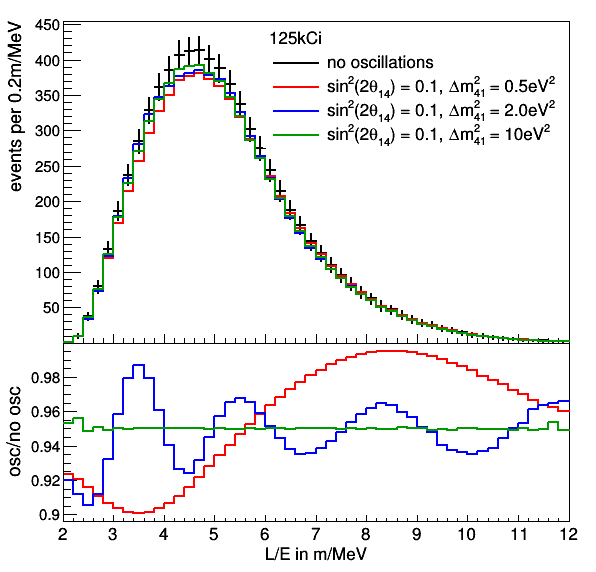}
}
\end{minipage}
%
%
\begin{minipage}[b]{0.5\linewidth}
{
\vspace{-5mm}
\includegraphics*[width=\linewidth]{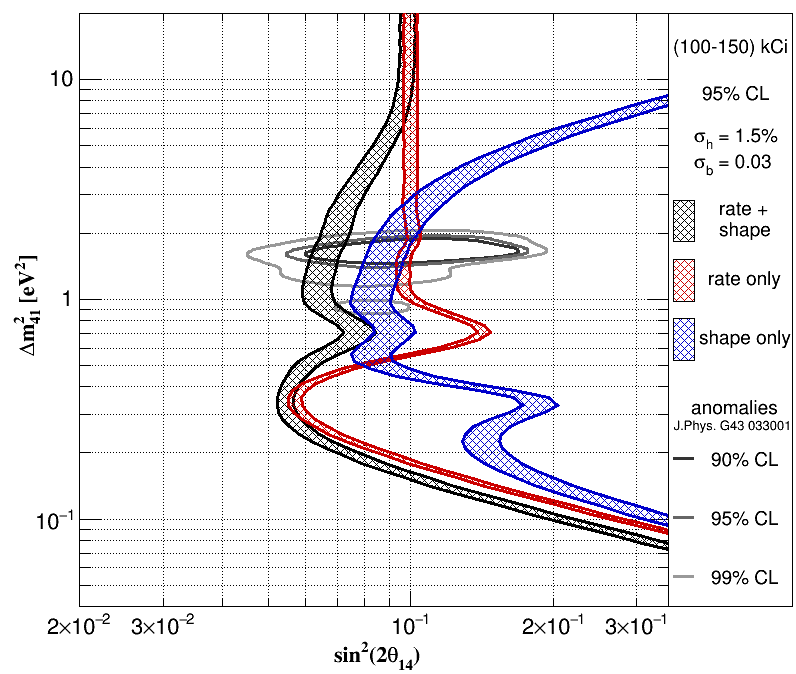}
}
\end{minipage}
\caption{SOX project: Left: expected signature as a function of baseline ($L$) over reconstructed antineutrino energy ($E = E_{\bar{\nu}_e}$) for 125 kCi $^{144}$Ce-$^{144}$Pr source underneath Borexino. Right: 95\% CL exclusion limits for a 100-150 kCi source in the $\Delta m^2_{41}$ - $\sin^2(2\theta_{14})$ plane compared to the preferred regions of a global fit to the anomalies~\cite{Gariazzo:2015rra}. From~\cite{SOX-ICHEP2016}.}
\label{fig:sox_sensitivity}
\end{figure}
 
In the analysis, almost all scintillator could serve as the fiducial volume, thus providing the baselines $L$ between 4.25 and 12.75\,m. Fig.~\ref{fig:sox_sensitivity} shows in its left part the Monte Carlo simulations~\cite{SOX-ICHEP2016} of the expected signal ($L$ over reconstructed $\bar{\nu}_e$ energy) for different oscillation parameters. The lower pad gives the ratio between the oscillation hypothesis and the no-oscillation hypothesis. For $\Delta m^2_{41} \sim$0.5 - 5\,eV$^2$, the {\it smoking-gun} oscillation wave could be observed across the detector, being referred to as "shape analysis". 

The 95\% CL exclusion limits~\cite{SOX-ICHEP2016} for a (3+1) sterile neutrino model are shown in the right part of Fig.~\ref{fig:sox_sensitivity}. A rate (in red) and a shape analysis (in blue) can be performed independent from each other. 
%
%
For oscillation lengths smaller than the spatial resolution (high $\Delta m^2_{41}$-values), only an average rate deficit can be measured. This explains the drop in the shape sensitivity, but the remaining stable sensitivity in the rate analysis. Also for oscillation lengths larger than the detector size (small $\Delta m^2_{41}$-values), the sensitivity of a shape analysis is lost, but still the rate deficit can be analyzed. For the combination of a rate and shape analysis (in black) almost the whole best-fit region~\cite{Gariazzo:2015rra} of the anomalies can be excluded. In this analysis the expected main systematic uncertainties have been taken into account.

Employment of a $^{51}$Cr source has been also considered. Since $\nu_e$ detection via elastic scattering off electrons does not provide the golden channel of delayed coincidences as IBDs and has nearly two orders of magnitude smaller cross section, the requested activity is much higher, about 2-4\,MCi, in order to overcome the backgrounds from radioactivity and solar $\nu$'s. 

\begin{figure}[htbp]
\centering
\includegraphics[width=0.45\textwidth]{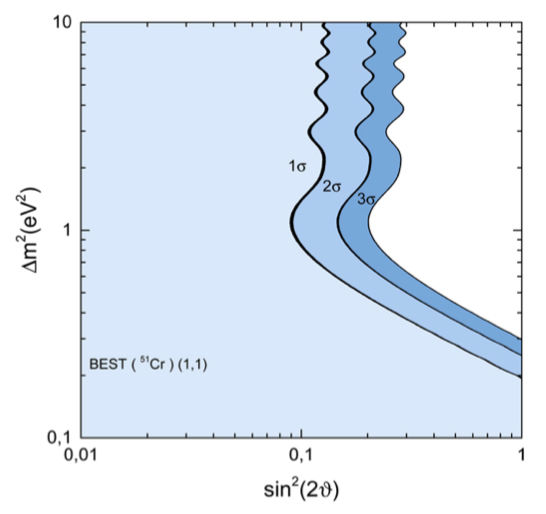}
\caption{BEST sensitivity to electron neutrino disappearance based on a single run with a $^{51}$Cr source \cite{Barinov:2017ymq}.}
\label{fig:best}
\end{figure}

\subsubsection{BEST}
\label{sub:BEST}

A new gallium experiment, BEST, has been proposed for the Baksan Neutrino Observatory  \cite{Barinov:2016znv,Barinov:2017ymq}. BEST aims to provide a definite experimental test of the gallium anomaly (Sec.~\ref{sub:GA}). Similarly to the source calibration runs of the SAGE and GALLEX experiment that are at the heart of the gallium anomaly, BEST plans to measure the rates of mono-energetic neutrinos from a $^{51}$Cr electron-capture (EC) source of 3\,MCi interacting with the isotope $^{71}$Ga in a gallium target. Differently from the former experiments, BEST foresees a measurement using not one but two concentric vessels filled with a total of 50\un{t} of metallic gallium. This allows to sample the rate at two different distances from the source, at 0.4\,m and 0.8\,m. At the time of writing, the $^{50}$Cr source material is irradiated at the SM-3 reactor in Dimitrovgrad. Measurements are due to start in July 2019 \cite{Barinov:2019vmp}.

Like for solar neutrinos, electron neutrinos from the $^{51}$Cr source will interact in the target by inducing the conversion of the isotope $^{71}$Ga to $^{71}$Ge. The germanium is chemically extracted and the $\nu_e$ flux can be determined via the measurement of the number of the $^{71}$Ge re-decays. Given the finite half-life of the $^{51}$Cr of 27.7\,d, measurements are to be limited to ten extraction cycles, each 9 days in length. In the first run, the combined event rate in both gallium vessels is expected to be $\sim$65 per day. Based on the statistics expected for the total exposure, the $\nu_e$ neutrino flux crossing both vessels can be determined with an accuracy of 4.2\,\% \cite{Barinov:2017ymq}. Fig.~\ref{fig:best} shows the corresponding exclusion contours in case of non-observation of oscillations.

While this sensitivity is relatively low compared to many of the running short-baseline reactor antineutrino experiments, BEST is nevertheless very interesting because it is sensitive to the disappearance of electron neutrinos (instead of antineutrinos). Assuming not one but two source runs, the second potentially with a $^{65}$Zn neutrino source, the authors state that they can exclude the gallium anomaly at a $5\sigma$ significance level~\cite{Barinov:2017ymq}.

\subsubsection{JUNO and IsoDAR}
\label{sub:JUNO}

\begin{figure}
\begin{minipage}{.6\textwidth}
  \centering
\includegraphics[width=\textwidth]{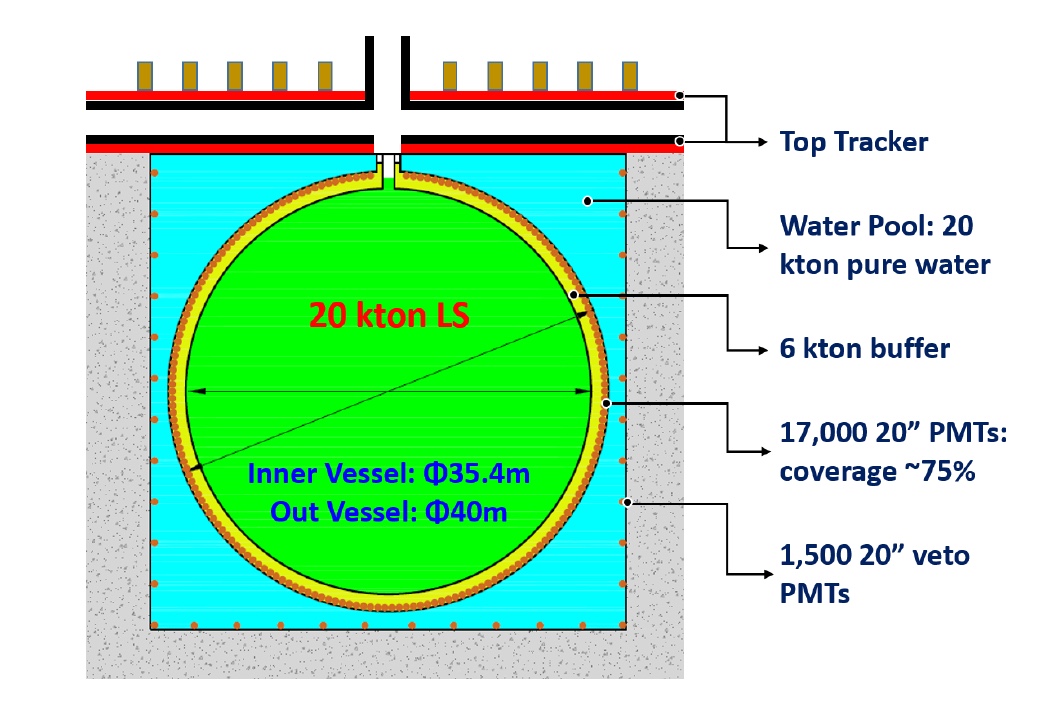}
\end{minipage}
\begin{minipage}{.38\textwidth}
  \centering
  \vspace{0.5cm}
\includegraphics[width=\textwidth]{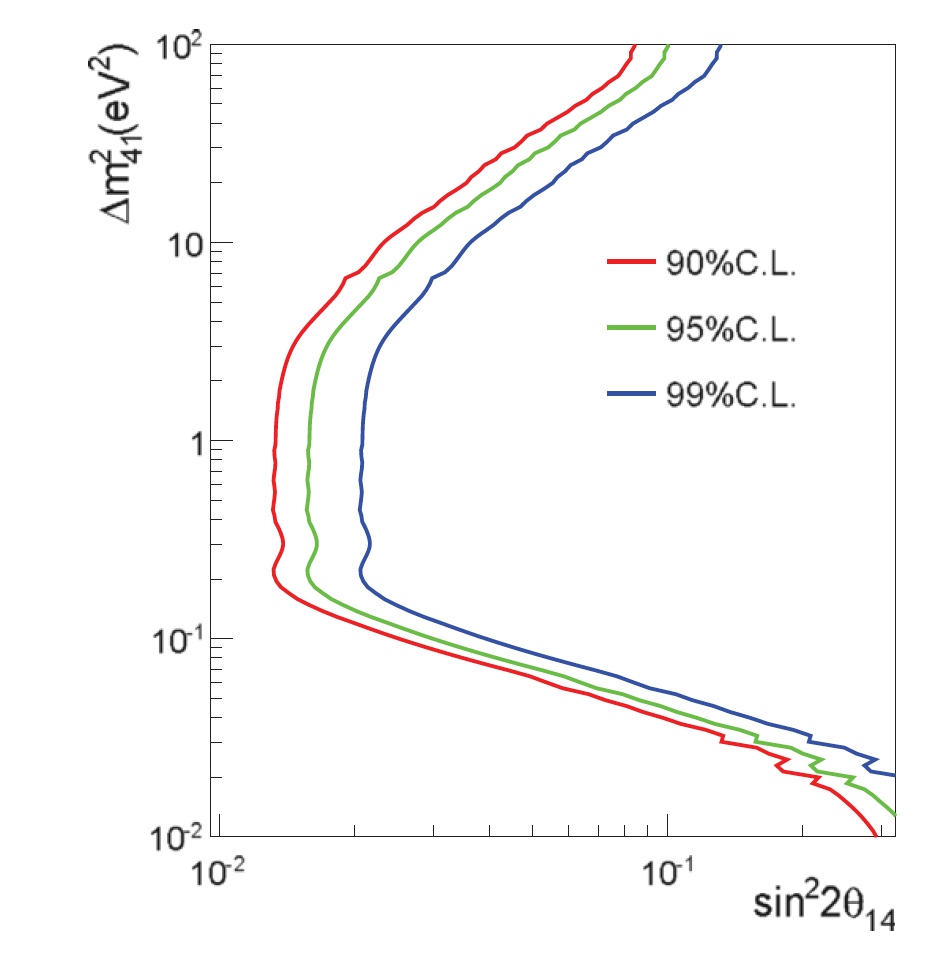}
\end{minipage}
\caption{Left: sketch of the JUNO detector~\cite{Wang_2016}. Right: sensitivity of a $\bar{\nu}_e$ disappearance search at JUNO to the oscillation parameters $\Delta m^2 _{41}$ and $\sin^2(2\theta_{14})$ assuming a 50\,kCi $^{144}$Ce-$^{144}$Pr source at the detector center with 450\,days of data taking~\cite{JUNOYellowBook}. }
\label{fig:JUNO}
\end{figure}

The Jiangmen Underground Neutrino Observatory (JUNO)~\cite{JUNOYellowBook} will be a next generation multi-purpose detector currently under construction in Jiangmen in China. It plans to start data taking in 2021. The central detector (left part of Fig.~\ref{fig:JUNO}) will contain 20\,kton of liquid scintillator. The main goal of the project is to determine the neutrino mass hierarchy within six years of run time with 3-4$\sigma$ significance by measuring reactor $\bar{\nu}_e$'s at 53\,km baseline. In order to achieve this goal, the unprecedented energy resolution of 3\% at 1 MeV is required. 

Similarly to the SOX project (Sec.~\ref{sub:SOX}), the JUNO detector could be for a sterile neutrino search employing radioactive neutrino sources. Thanks to its large dimensions, JUNO's sensitivity both to $\Delta m^2 _{41}$ and $\sin^2(2\theta_{14})$ is largely superior with respect to SOX. The right panel of Fig.~\ref{fig:JUNO} demonstrates the sensitivity of a $\bar{\nu}_e$ disappearance search with a 50\,kCi $^{144}$Ce-$^{144}$Pr source at the detector center with 450\,days of data taking~\cite{JUNOYellowBook}. The background by reactor antineutrinos is considered. A spatial resolution of 12\,cm at 1\,MeV is assumed. Other systematic effects considered are a 2\% uncertainty for detector efficiency, and 2\% and 3\% for the source and reactor normalization, respectively.

\begin{figure}[htbp]
\begin{minipage}{.55\textwidth}
  \hspace{-10mm}
\includegraphics[width=1.1\textwidth]{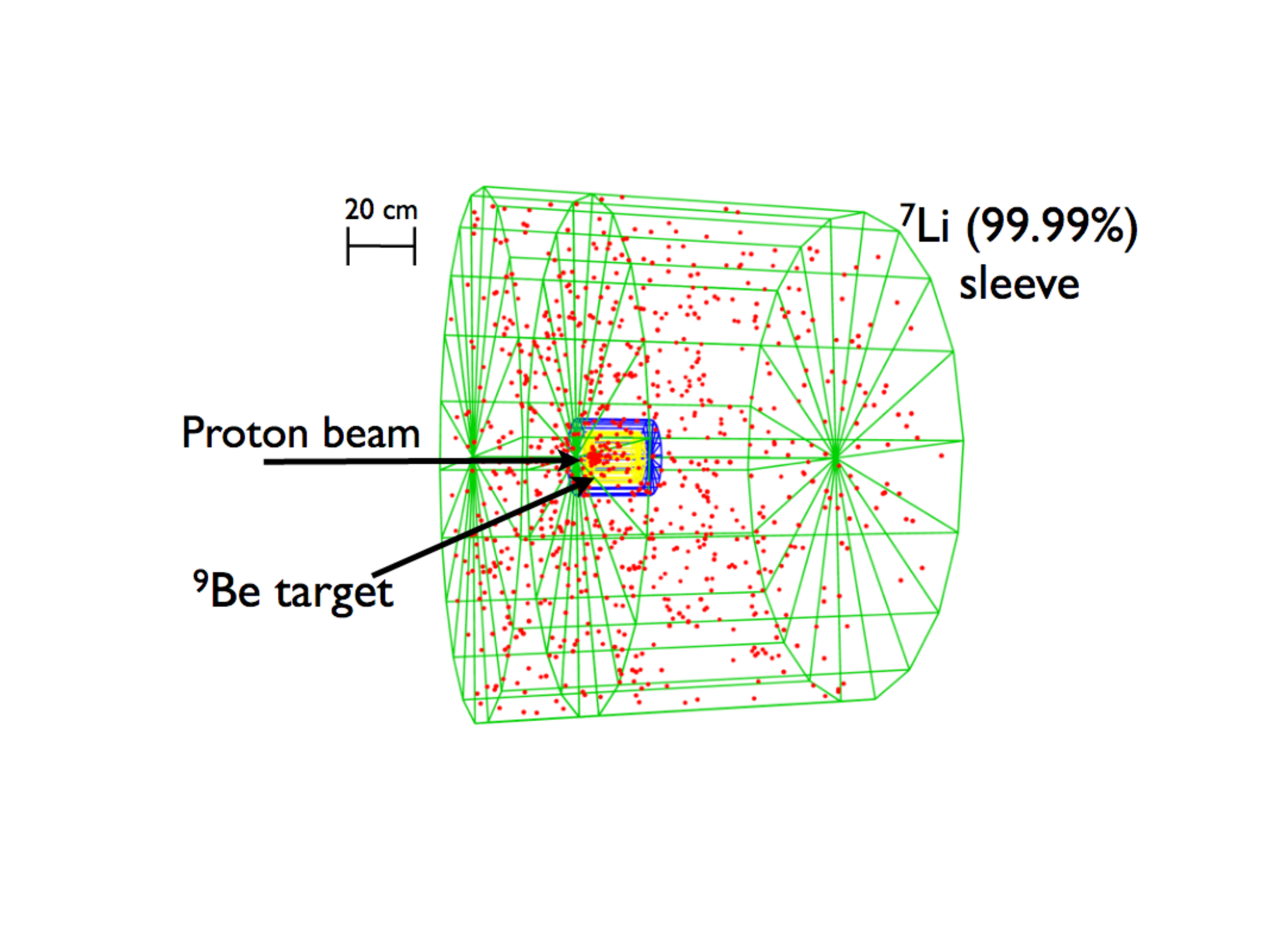}
\end{minipage}
\begin{minipage}{.56\textwidth}
\hspace{-10mm}
\includegraphics[width=\textwidth]{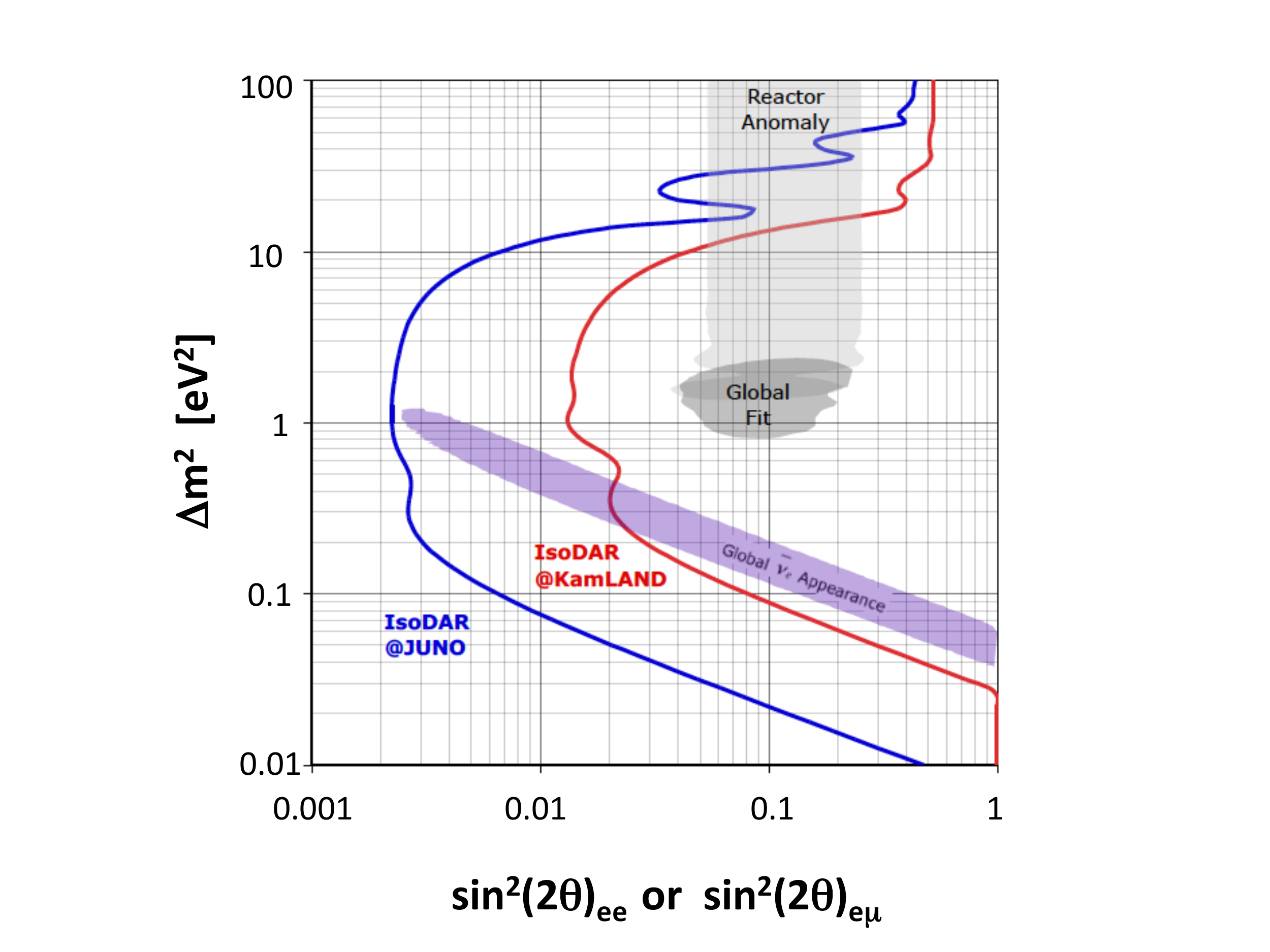}
\end{minipage}
\caption{Left: a scheme of the IsoDAR target and surrounding volumes~\cite{Bungau:2012ys}. The dots represent $^8$Li creation points, obtained with $10^5$ 60-MeV protons simulated on target. Right: the red and blue curves indicate $\Delta m^2_{41}$ and $\sin^2(2\theta_{14})$ ($\theta_{ee}=\theta_{14}$) boundaries where the null oscillation hypothesis can be excluded at 5$\sigma$ with IsoDAR@KamLAND and IsoDAR@JUNO experiments, respectively, for five years of data taking. Also, shown by the light (dark) gray areas are the 99\% allowed regions for the reactor anomaly~\cite{Mention:2011rk} (global oscillation fit~\cite{Sorel:2003hf}). The purple area corresponds to the $\Delta m^2_{41}$ and $\sin^2(2\theta_{e\mu})$ allowed region at 99\% CL from a combined fit to all $\bar{\nu}_e$ appearance data~\cite{Conrad:2012qt}. From~\cite{PhysRevD.89.057301}.}
\label{fig:IsoDar}
\end{figure}

The IsoDAR (Isotopic Decay At Rest) experimental program aims at the application of the $^{8}$Li isotope (see Table~\ref{tab:sources}) with an existing large-volume neutrino detector. Its advantage is a high end-point of the antineutrino energy spectrum (12.9\,MeV) and the mean energy of 6.4\,MeV, well above the 1.8\,MeV threshold of the IBD interaction. Due to its short lifetime, $^8$Li has to be continuously produced via irradiation of the $^7$Li target with neutrons, in turn produced by a proton beam impinging on a $^{9}$Be target. This technical complication has the advantage that the data taking can be extended to several years, impossible with "traditional" neutrino sources. This kind of source has been considered in~\cite{Basov1985,Lutostansky2011}, while Ref.~\cite{Bungau:2012ys} suggests a specific design of the IsoDAR source applied in combination with the KamLAND detector. 

The main idea of this kind of source can be described as follows: Firstly, a 60\,MeV/amu cyclotron accelerates 5\,mA of H$^2+$ ions. The 60\,MeV proton beam impinges on a cylindrical $^9$Be target (see Fig.~\ref{fig:IsoDar}, left) that is 20\,cm in diameter and 20\,cm long and produces copious neutrons. These are then moderated and multiplied by a surrounding 5\,cm thick layer of D$_2$O. Secondary neutrons enter a cylindrical sleeve of solid lithium ($^7$Li enriched to 99.99\%), 150\,cm long and 200\,cm in outer diameter, enveloping the target and D$_2$O layer.
%
Fig.~\ref{fig:IsoDar} shows this geometry as well as the Monte Carlo simulation of the production of about 1500 $\bar{\nu}_e$ from $^8$Li decay. As suggested in~\cite{PhysRevD.89.057301}, using deuterons instead of protons gives an enhanced $\bar{\nu}_e$ rate by a factor 2.7.

Fig.~\ref{fig:IsoDar} (right part) shows the  5$\sigma$ exclusion curves of the null-oscillation hypothesis for  IsoDAR@KamLAND and IsoDAR@JUNO, assuming five years of data taking~\cite{PhysRevD.89.057301}. IsoDAR@JUNO would provide a sensitivity not only covering the reactor anomaly~\cite{Mention:2011rk} but also the entire short-baseline appearance allowed region~\cite{Conrad:2012qt} at 5$\sigma$. This represents a decisive test of the LSND and MiniBooNE appearance signals within all models that are $CPT$ invariant~\cite{PhysRevD.89.057301}. 

\subsection{Atmospheric neutrino experiments}
\label{sub:atmospherics}

Atmospheric neutrinos are created in abundance by the ubiquitous flux of highly energetic charged cosmic rays that interact with the atomic nuclei in the earth atmosphere~\cite{Honda:2015fha, Honda:2011nf, Gaisser:fk, Battistoni:2002ew}. In the ensuing hadronic air shower that develops through the atmosphere neutrinos are predominantly created in the decay of light charged mesons such as pions or kaons
\begin{alignat*}{3}
  p + N \to X + \,& \pi^{\pm},K^{\pm}   \\
            & \decayto[2.5] && \mu^{\pm} + \nu_\mu \\
            & && \decayto e^{\pm} + {\nu_e} + \nu_\mu  \\
\end{alignat*}
resulting in a ratio of $\nu_\mu : \nu_e \sim 2$ up to energies of 1\un{GeV}. At
higher energies, the muon neutrino flux becomes increasingly abundant as more and more muons reach the ground instead of decaying in the atmosphere. Due to energy redestribution in the hadronic shower, the spectrum of atmospheric neutrinos is steeper than the spectrum of cosmic rays with a spectral index of $\Phi(\nu) \propto E_\nu^{-3.7}$. 

Due to the rather large (and potentially maximal) mixing angle $\sin^2(2\theta_{23}) \sim 1$~\cite{Esteban:2018azc}, the atmospheric mass splitting $\Delta m^2_{32} \simeq \Delta m^2_{31} \simeq 2.5\cdot10^{-3}\un{eV^2}$ dominates the oscillations of atmospheric muon neutrinos. At GeV energies, the oscillation length is many hundreds or thousands of kilometers. Oscillations are either studied by long-baseline accelerator neutrino beams (see Sec.~\ref{sub:lblacc}) or based on the atmospheric neutrino flux itself, where they are not observed for neutrinos from interactions in the atmosphere above the detector, but only for neutrinos that travel a significant distance through the earth. For distances of the order of the earth radius the variations in neutrino production height become negligible, which has the advantage that the zenith angle under which the neutrino is observed becomes a direct measure of the neutrino propagation distance. A good understanding of the atmospheric neutrino flux is thus required in these two key observables: the energy and the zenith angle distribution.

While the flux of cosmic rays is homogeneous in its arrival direction at the $10^{-3}$ level~(\cite{Erlykin:2019iuz} and references therein) over the energy range of interest, neutrinos from zenith angles closer to the horizon stem from air showers with a larger slant depth. Hence in these showers the mesons and muons have more time to decay, enhancing both the neutrino flux as well as the muon-to-electron neutrino abundance towards the horizon. While at energies above $\sim10\un{GeV}$, this effect is symmetric around the horizon, for lower energies this symmetry is broken. This is due to the geomagnetic field: low energy cosmic-ray particles are bent significantly by the geomagnetic field, and only cosmic-ray particles above certain rigidity (i.e. energy-to-charge ratio) can enter into the atmosphere. Together with the magnetic field, this depends on the position in the earth and on the direction of the cosmic-ray particles and hence neutrinos. Furthermore the flux is affected by the density and density variation of the atmosphere which changes the fraction of mesons that interact instead of decaying. Detailed models are therefore required to determine the neutrino flux at each specific location~\cite{Honda:2015fha}. The sensitivity of atmospheric neutrino experiments such as Super-Kamiokande and IceCube benefits from the fact that the zenith- and energy-dependence of the systematic effects are different from the pattern induced by neutrino oscillations and their results are among the most constraining to the so-called {\it atmospheric} mixing angle $\theta_{23}$ and the atmospheric mass splitting $\Delta m^2_{32}$~\cite{Aartsen:2017nmd, Abe:2017aap}.

\twofig[tbp!]
    {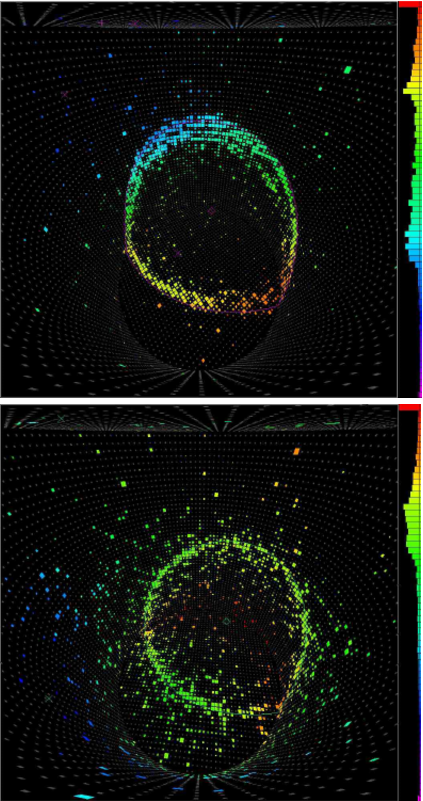}
    {Two simulated events displayed for the Super-Kamiokande detector. Due to the spread of shower particles, the Cherenkov ring of the electron event (bottom) is more fuzzy than for the muon case (top).\label{fig:sk-events}}
    {atmospheric/IceCube-detector}
    {Graphical representation of the IceCube detector at the geographic South Pole. While the majority of modules is deployed with a horizontal spacing of 125\un{m} and a vertical spacing of 16.6\un{m}, the eight more densly packed strings in the center that form IceCube-DeepCore extension provide a volume with reduced energy threshold.\label{fig:icecube} }

\subsubsection{Super-Kamiokande}
The Super-Kamiokande detector~\cite{Fukuda:2002uc} located in the Kamioka mine in Japan consists of a cylindrical stainless steel tank filled with 50\un{kton} pure water. More than 11,000 photomultiplier tubes with the largest available diameter of 20 inch monitor this water volume for neutrino interactions. Charged particles with velocities above the Cherenkov threshold are generated in both charged and neutral current interactions of neutrinos. The emitted Cherenkov light is observed under the Cherenkov angle $\theta_c = \arcsin\frac{1}{n} \sim 42\un{^\circ}$ in water, leading to a characteristic ring-like pattern. Muons generated in charged current interactions of $\nu_\mu$'s can be well separated from the electromagnetic and hadronic showers generated by charged current interactions of $\nu_e$ and $\nu_\tau$ and neutral current interactions: at the energies discussed here, the muons travel in a straight line, so that the light emitted by the muon and its secondaries forms a sharper ring-like structure than for showers in which multiple scattering of the particles creates a more fuzzy ring~(Fig.~\ref{fig:sk-events}). The rock overburden with a height of 1,000\un{m} (corresponding to 2,700\un{m} water equivalent) provides excellent shielding against atmospheric muons and reduces their flux to $6\times 10^{-8} \un{cm^{-2}} \un{s^{-1}} \un{sr^{-1}}$. To discriminate the atmospheric and radioactive backgrounds (mostly radon), a fiducial volume is defined 2\un{m} inside from the physical walls\footnote{Newer analyses~\cite{Jiang:2019xwn} expand this volume using a new likelihood selection technique which has not been employed in the search for sterile neutrinos yet.}. For the atmospheric neutrino analysis, events are classified in three categories: fully contained, partially contained and up-going muon-like. The {\it sub-GeV fully contained} events with energies from few hundred MeV to about 10 GeV dominate the sample with a rate of $\sim 25\un{\mu{}Hz}$ per 22.5\un{kton} for muon-like and electron-like events each. While the muon sample reaches a purity of 95\% for $\nu_\mu$ charged current events, the purity of the electron sample is reduced to 89\% due to contamination with neutral current events~\cite{Wendell:2010md}. A different event categorization is used for the {\it multi-GeV fully contained} sample which contributes another $20\un{\mu{}Hz}$. At a few \%, the energy resolution in these fully contained samples is rather good, but the directional resolution is limited by the kinematic angle between neutrino and outgoing lepton and becomes as large as 100\un{^\circ} for the lowest energies. In contrast, the {\it partially contained} events ($\sim 7\un{\mu{}Hz}$) and particularly the {\it upgoing events} ($\sim 16\un{\mu{} Hz}$) are almost exclusively $\nu_\mu$ charged current induced. While in partially contained events with the interaction vertex inside the fiducial volume the energy resolution is degraded due to outgoing muons leaving the detector, the directional resolution is improved due to the longer muon track. Finally, for upgoing muon events with a vertex outside the detector, only a lower limit can be set by the energy deposited in the detector. On the other hand, due to the much larger energies of 10-1,000\un{GeV}, the directional resolution improves to few degrees. In combination with the propagation baselines which vary from few to 12,800\un{km} for earth-crossing neutrinos, these data samples allow to probe a wide range of $L/E$ values ranging from 1 -- $10^4$\un{km/GeV}, thus covering well the first oscillation maximum of the atmospheric mixing at $L/E \simeq 500 \un{km/GeV}$.

\subsubsection{IceCube}
The IceCube detector~\cite{Aartsen:2016nxy} employs the natural glacial ice sheet at the geographic South Pole as a neutrino detection medium. Holes are drilled to a depth of 2,500\un{m} in which a total of 5,160 Digital Optical Modules (DOMs), each containing a 10 inch photomultiplier and readout electronics, are deployed~(Fig.~\ref{fig:icecube}) in the range between 1450--2450\un{m}. The bulk of the detector consists of 86 strings arranged in a triangular grid with a horizontal spacing of 125\un{m} and a vertical spacing between DOMs of 16.6\un{m}. This sparse instrumentation results in an energy threshold of $\sim100\un{GeV}$, but provides an instrumented volume of $1\un{km}^3 \simeq 10^6\un{kton}$. However, this threshold is to high to observe oscillations even for atmospheric neutrinos crossing the full diameter of the earth, with a first $\nu_\mu$ dissappearence minimum at 25\un{GeV}. In the central part of the detector the eight additional strings of the subarray IceCube-DeepCore~\cite{Collaboration:2011ym} reduce the horizontal spacing to 40--50\un{m}. At the same time, the 480 DOMs on these strings are equipped with photomulitpliers featuring a 35\% higher quantum efficiency and are spaced at a vertical distance of 6\un{m}. While at trigger level, the resulting effective volume is still $\sim10\un{Mt}$, the very sparse sampling of the emitted Cherenkov light, scattering of the photons in the ice and the huge abundance of atmospheric muon background results in a rather low purity at this level. A series of harsh event selection criteria are required to reject the muon background to below 1\% and select well reconstructed events~\cite{Aartsen:2014yll}. For the same reason, no direct flavor identification is possible and events are categorized only in {\it track-like} when the muon track from a $\nu_\mu$ charged current interaction can be identified and {\it cascade like} for all $\nu_e$, $\nu_\tau$ and neutral current interactions. For track-like events, the neutrino direction can be estimated from the direction of the muon which is well aligned with the neutrino at the energies discussed here. The median neutrino zenith resolution is 12$^\circ$ at 10\un{GeV} and 6$^\circ$ at 40\un{GeV}. On the other hand, only a rather poor estimate of the direction is possible for electron neutrinos, so that they are neglected in the analysis~\footnote{More recent anlyses~\cite{Aartsen:2017nmd} using a more sophisticated likelihood approach for the event reconstruction include cascade-like events.}. At the final selection level, $\nu_\mu$ charged current events are identified at a rate of $100\un{\mu{}Hz}$ with a $\nu_e$ contribution of $18\un{\mu{}Hz}$ and a $\nu_\tau$ fraction of $5\un{\mu{}Hz}$ estimated from simulation in the range from $6-56\un{GeV}$. The energy resolution, which is mainly based on the length of the muon track emerging from the shower at the neutrino interaction vertex, is about 25\% over most of the energy range. While the narrower energy range compared to Super-Kamiokande only allows to probe $L/E$ in the range from 10--2,000\un{km/GeV}, the first muon neutrino dissappearence maximum is still well covered in this range.

\setlength{\figwidth}{1.0\textwidth}
\onefig[htb]
{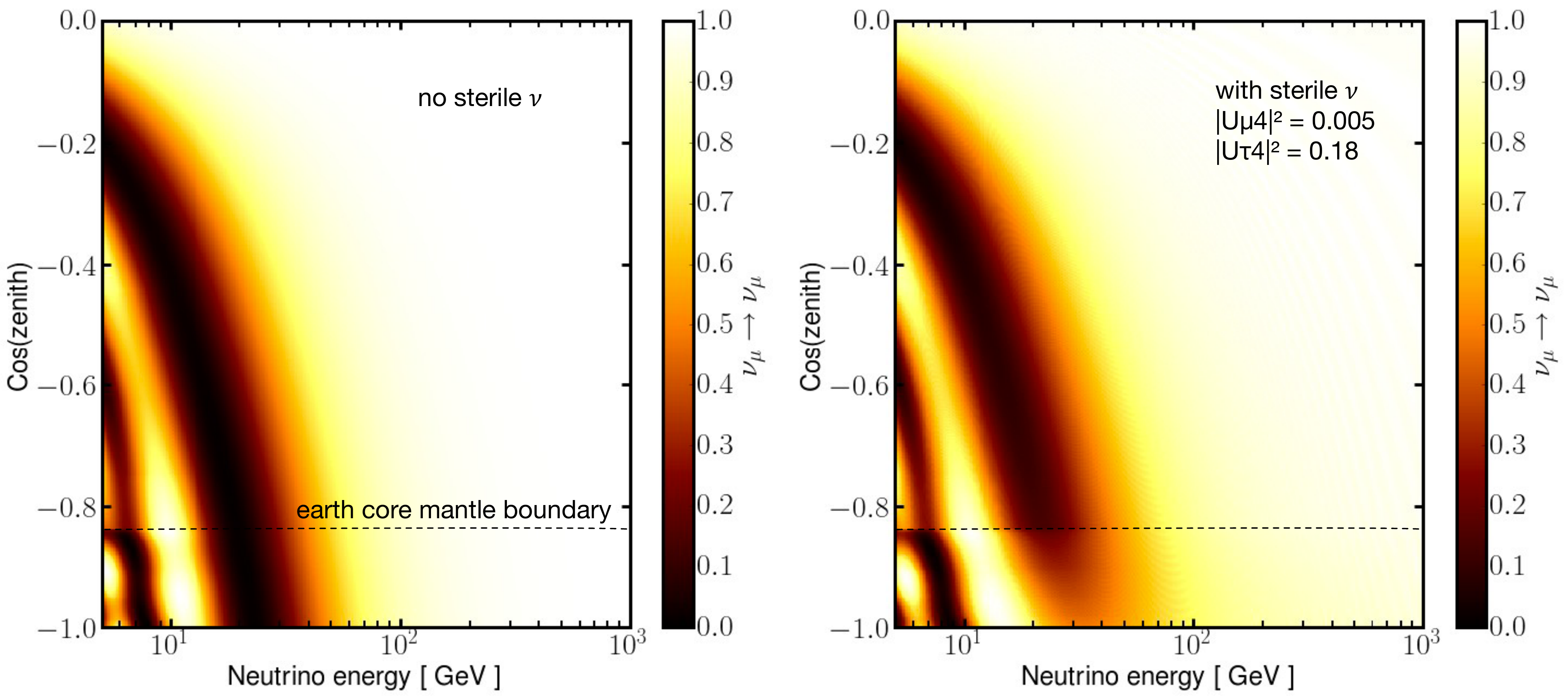}
{Atmospheric $\nu_\mu$ survival probability without sterile neutrinos (left) and with a sterile neutrino (right). Standard $\nu_e$ charged current matter effects in the earth core create a distortion around a few GeV in the most upward-going zenith angles $\cos(\theta_z) < -0.8$. In the presence of a sterile neutrino with coupling to $\nu_\mu$ ($\left|U_{\mu 4}\right|^2 > 0$) and $\nu_\tau$  ($\left|U_{\tau 4}\right|^2 > 0$), the amount of $\nu_\mu$ disappearance reduces for the most upward-going bins in the tens of GeV region. There is also a small amount of extra disappearance on top of the standard oscillations introduced by the nonzero $\left|U_{\mu 4}\right|$ which is most visible in the slight darkening of the right part of the plot corresponding to the overall $\nu_\mu \to \nu_\mu$ disappearance effect.\label{fig:atmomudiss}}
\setlength{\figwidth}{0.9\textwidth}

\subsubsection{Low-energy oscillation searches~\label{sec:IClowensterile}}
The addition of a sterile neutrino state modifies the neutrino oscillations in two ways that are relevant for the atmospheric neutrino analyses presented here~\cite{Abe:2014gda,Aartsen:2017bap}. 

The first is the {\it disappearance} signature where active neutrinos oscillate into s sterile state (Sec.~\ref{sec:signature-dissappear}). For $\Delta m_s^2 \sim 1\un{eV^2}$, the first oscillation minimum is at a value of $L/E \simeq 1 \un{km/GeV}$, outside the range covered with high statistics by Super-Kamiokande and IceCube-DeepCore. The signature of this vacuum oscillations-like disappearance is hence only observable in the unresolved regime in a change of the overall flux normalization. Hence $\sin^2\left(\Delta m_s^2 L/E\right) = 0.5$ is used in good approximation for $\Delta m^2 > 0.1\un{eV}^2$ for Super-Kamiokande. The IceCube-DeepCore analysis uses a fixed value of $\Delta m_s^2 = 1\un{eV}^2$ but demonstrates that the result only depends weakly on this choice in the range from $0.1\un{eV^2}$ to $10\un{eV^2}$. In both cases no constraint on $\Delta m_s^2$ is possible though. 

The second effect is caused by the different effective matter potential experienced by sterile and active neutrinos when crossing the Earth. In their propagation, the active neutrinos {\it feel} a potential due to neutral current interactions with electrons and nuclei with an additional potential for electron neutrinos due to charged current interactions with the electrons of the Earth's matter~(Fig.~\ref{fig:neutrinoforward}). In contrast, sterile neutrinos have no interactions with the earth matter. This modifies not only the energy but also the amplitude of the oscillation minimum with a strength proportional to the amount of matter along the neutrino trajectory. The effect is therefore most pronounced for neutrinos crossing the Earth’s core, i.e. zenith angles $\cos\theta_z < -0.8$. Note that even in the absence of sterile neutrinos, a distortion of the oscillation pattern in the Earth will be observed due to the charged-current interaction potential only present for electron neutrinos. However, sterile neutrinos modify the pattern and extend it to higher energies (Fig.~\ref{fig:atmomudiss}).

\setlength{\figwidth}{0.7\textwidth}
\onefig[htbp]{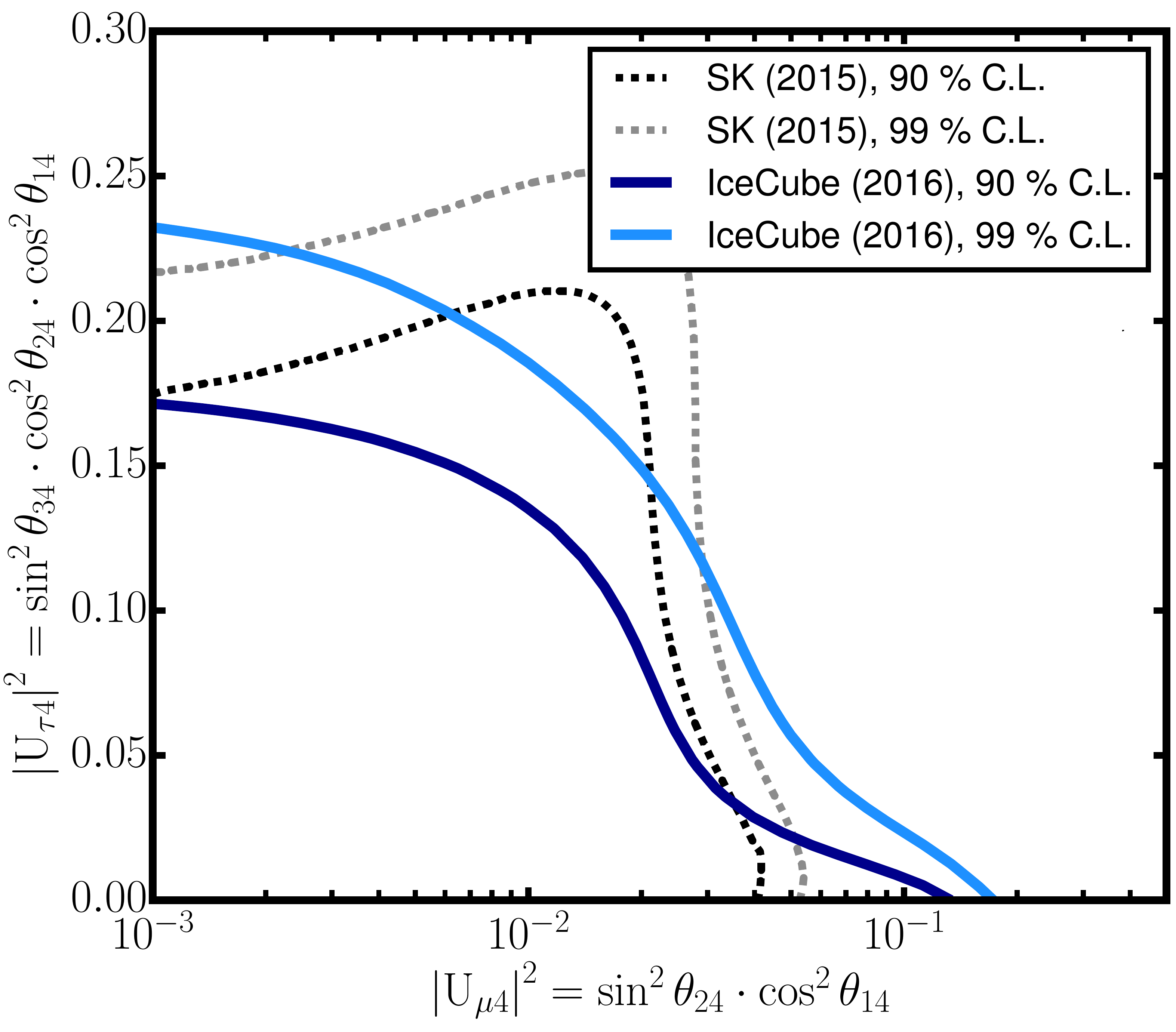}{Constraints from Super-Kamiokande~\cite{Abe:2014gda} and IceCube-DeepCore~\cite{Aartsen:2017bap} on the mixing of a fourth mass state to muon and tau neutrinos assuming $\left|U_{e4}\right|^2 = 0$ and normal hierarchy ($\Delta m^2_{32} > 0$).\label{fig:atmolowensterile}}
\setlength{\figwidth}{0.9\textwidth}
In both analyses, $\left|U_{e4}\right|^2=0$ is chosen such that the new sterile state does not couple to electrons. This is mostly motivated by the reduced sensitivity to electron neutrino oscillations because of the smaller value of $\sin^2\left(2\theta_{13}\right) < \sin^2\left(2\theta_{23}\right)$, but generally somewhat improves the constraints obtained for the remaining two parameters of the model $\left|U_{\mu 4}\right|^2$ and  $\left|U_{\tau 4}\right|^2$. Figure~\ref{fig:atmolowensterile} shows the resulting constraints on these parameters. We note that in~\cite{Abe:2014gda} the further {\it no-$\nu_e$ approximation} is made by setting the mixing parameters of the active neutrinos to the electron $\sin^2(2\theta_{12}) = \sin^2(2\theta_{13}) = 0$. This reduces the oscillations to an effective two-flavor system, allowing to solve the oscillation probabilities analytically even in the presence of a matter potential. This also eliminates the normal $\nu_e$ matter effects in the absence of a sterile neutrino~(compare~Fig. 2(b) in~\cite{Abe:2014gda}). Because $\nu_e$ are present at a few percent level, matching the overall normalization to the observed data thus introduces a systematic bias towards lower measured $\left|U_{\mu 4}\right|^2$. In contrast, in~\cite{Aartsen:2017bap} the neutrino oscillation probabilities are determined by numerically solving the Schrödinger equation using the GLoBES software~\cite{Huber:2004ka, Huber:2007ji} and a 12-layer approximation of the earth density model~\cite{Dziewonski:1981xy}. This more realistic treatment results in a modified shape of the exclusion contour, in particular at $\left|U_{\tau 4}\right|^2 = 0$. We also note that choosing the inverted hierarchy ($\Delta m^2_{32} < 0$) reduces the sensitivity for the analysis in~\cite{Aartsen:2017bap} as it moves the matter effect for the active flavors from neutrinos to antineutrinos. While IceCube can not distinguish between neutrinos and antineutrinos and their atmospheric flux is roughly equivalent, the cross-section for neutrinos is larger by about a factor of two, thus enhancing the event rate and sensitivity to the matter effects.

Constraints from the water-Cherenkov neutrino detector ANTARES located in the Mediteranean sea are in preparation~\cite{Albert:2018mnz} and expected to yield comparable constraints. In order to significantly improve on these results though, improved energy resolution and statistics in the few-GeV energy region is required. Both of these may be provided by the next generation of neutrino telescopes that are planned or even under construction in the Mediteranean (ORCA~\cite{Adrian-Martinez:2016fdl}), underground (Hyper-Kamikande~\cite{Abe:2018uyc}) and at the South Pole (PINGU~\cite{Aartsen:2014oha}).

\subsubsection{High energy Earth-Core resonance searches}
Another option to search for a fourth mass eigenstate using atmospheric neutrinos is by looking for the signature of $\nu_\mu$ disappearance. Due to uncertainties in the hadronic interaction models~\cite{Gaisser:fk,Fedynitch:2012fs}, the overall normalization of this flux is rather poorly constrained for such a search. It is therefore best carried out in the {\it oscillation} regime with a first oscillation minimum at a few TeV for $\Delta m_s \sim 1\un{eV}^2$~\footnote{Using the same approach, constrained have also been obtainted for so-called (1+3)-models where $\Delta m_s^2 < 0$, i.e. the fourth eigenstate is lighter than the active mass eigenstates. Note that unless new physics are involved, these models are strongly constrained by the bounds on $\Sigma m_i$ from cosmology.}. 

While the uncertainty on the shape of the flux from both hadronic interactions and primary cosmic rays systematically limits the sensitvity of this approach, the earth-core resonant effect discussed in Sec.~\ref{sec:signature-dissappear}, which makes the disappearance nearly maximal for Earth crossing neutrinos, significantly boosts the sensitivity of this search. Fig.~\ref{fig:atmohedissappear} shows the fractional disappearance of the neutrino flux in the absence and presence of a sterile neutrino.

\twofigone[htbp]
{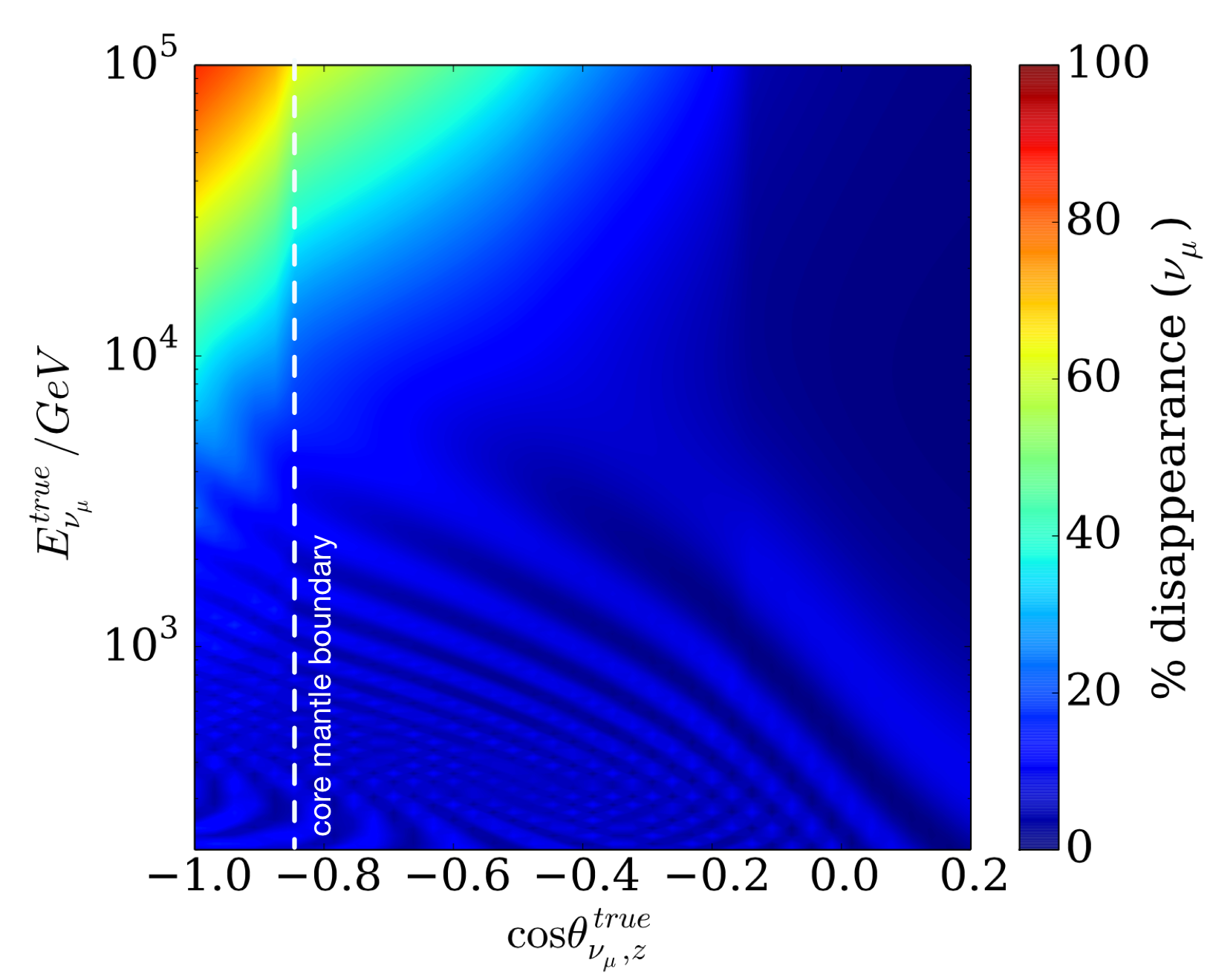}
{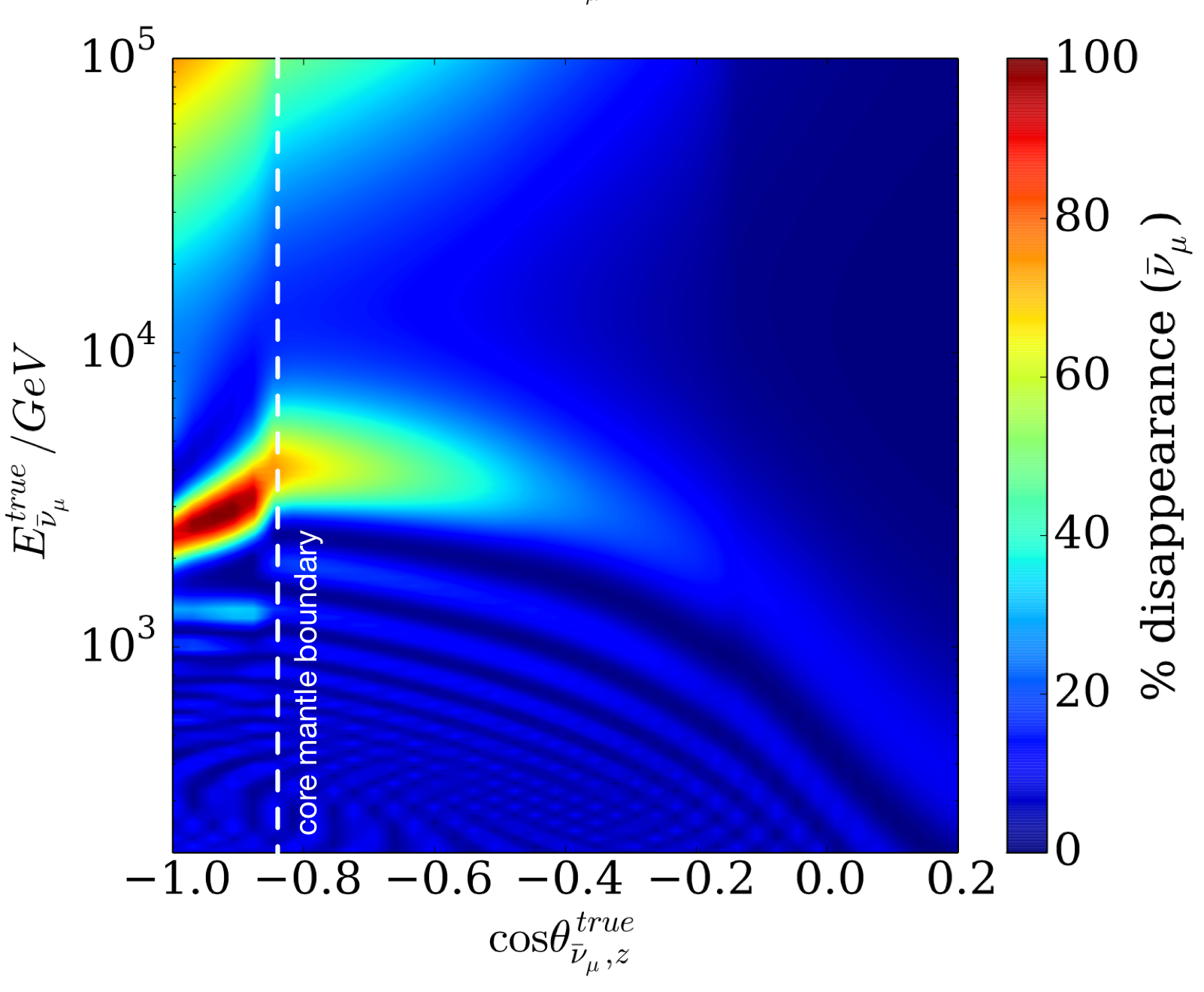}{Disappearance fraction of atmospheric muon neutrino flux. Left for three active neutrinos, right in the presence of an additional mass eigenstate at $\Delta m_s^2=1\un{eV}^2$. Despite $\sin^2\left(2\theta_{24}\right) = 0.1$, the earth core resonance effect makes the dissapearance at the first oscillation maximum at $E \sim 3\un{TeV}$ nearly maximal for $\cos(\theta_z)<-0.8$. The dissappearence of upgoing neutrinos at $E > 10^4\un{GeV}$ is due to neutrino absorption in the earth.\label{fig:atmohedissappear}}

With its large volume the IceCube neutrino telescope is very well suited to perform this search. In the study presented in~\cite{TheIceCube:2016oqi}, a sample of atmospheric neutrinos passing an event selection developed for a search for diffuse astrophysical muon neutrinos~\cite{Aartsen:2015rwa} is used. At energies ranging approxmitely from 320\un{GeV} to 20\un{TeV}, the muons emerging from charged current $\nu_\mu$ interactions travel many hundred meters in the glacial ice. As these muons are essentially co-linear with the neutrino direction due to the kinematic boost the directional resolution reaches below 1\un{^\circ} at the highest energies. This not only yields and excellent path length estimate for the oscillation, but also makes the sample essentially background free by restriciting the search to upward going tracks, eliminating the abundant flux of downward going atmospheric muons. Because the events are not contained within the detector, the energy has to be estimated from stochastic losses~\cite{Aartsen:2013vja} which become more frequent with higher energies. The resulting muon energy resolution is approximately $\sigma(\log_{10} E_\mu/\un{GeV}) \sim 0.5$. 

Care has to be taken to consider all systematic effects that affect the shape of the observed flux spectrum. In addition to detector effects such as the efficiency of the DOMs, the modelling of scattering and absorption in the ice and particularly in the drill holes, uncertainties in the atmospheric flux model also have to be taken into account. Using the MCEq cascade equation technique~\cite{Fedynitch:2015zma} variations in the primary spectral index, $\bar{\nu}/\nu$ ratio and $\pi / K$ production ratio, hadronic interactions and atmospheric models all can be propagated to uncertaintes in the flux at affordable computational effort. Figure~\ref{fig:atmohelimits} shows the result from an analysis~\cite{TheIceCube:2016oqi} using 20,145 reconstructed up-going muon events collected in one year of operation of the full IceCube detector. Strong exlusion limits are obtained in the range of $\Delta m_s^2 = 0.02 - 0.3 \un{eV}^2$ corresponding to the energy range with the highest statistics in the sample. As due to the nearly maximal mixing of $\nu_\mu$ and $\nu_\tau$, oscillations of $\nu_\tau \to \nu_s$ lead to additional disappearance, the conservative choice of $\sin^2\left(2\theta_{34}\right) = 0$ was assumed in this study.

\twofig[htbp]
{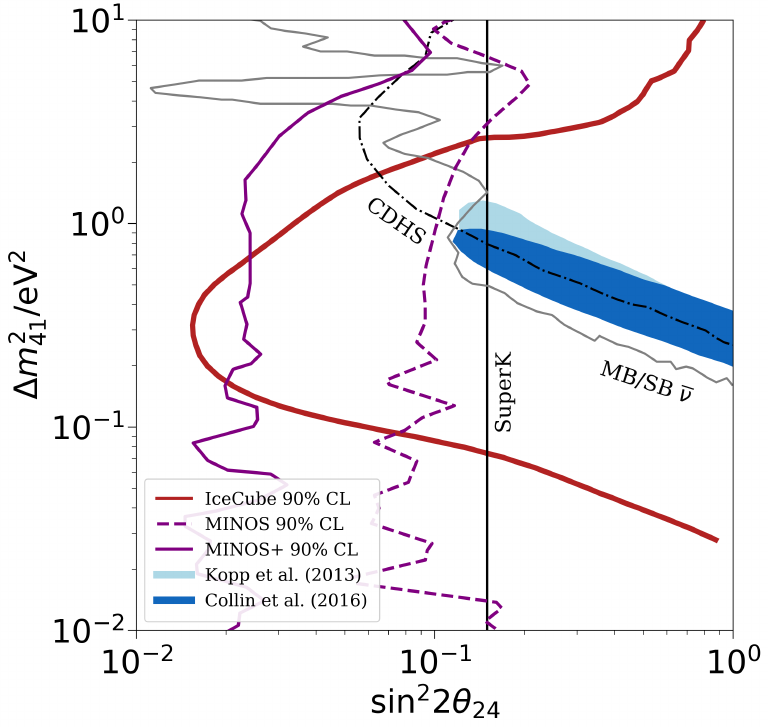}
{90\% CL limit from the IceCube high-energy sterile neutrino search of one year of data compared with allowed
regions from appearance experiments (blue) and constraints from accelerator long-baseline disappearance searches (purple)~\cite{Jones:2019nix}.\label{fig:atmohelimits}}
{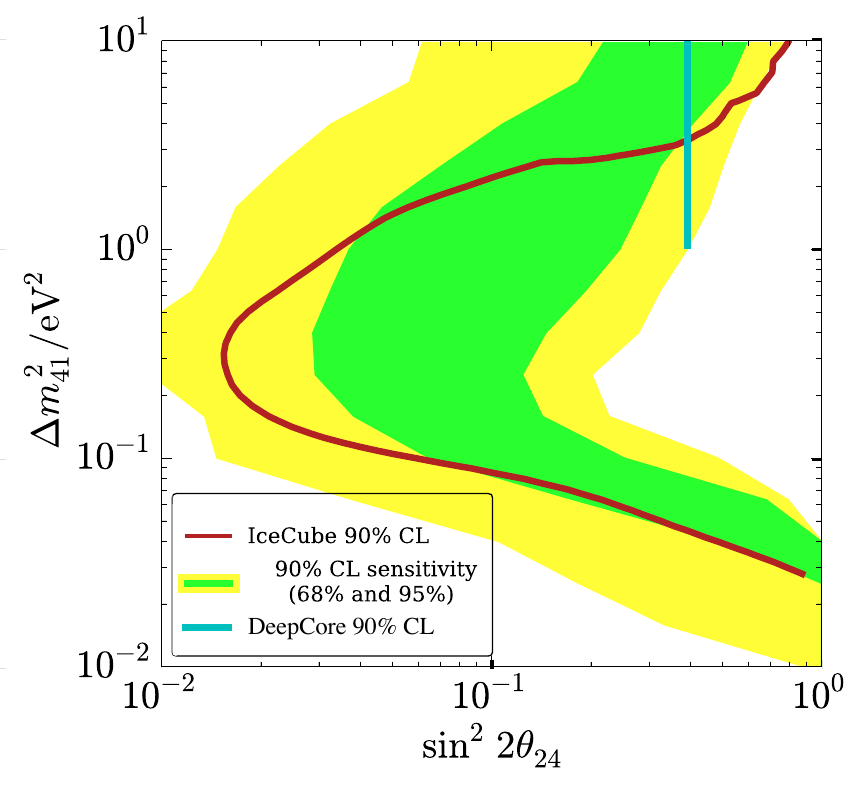}
{IceCube high-energy sterile neutrino limit compared to the sensitivity of the analysis and constraints on $\left|U_{24}\right|^2$ obtained in the
IceCube-DeepCore analysis (adapted from~\cite{TheIceCube:2016oqi}).\label{fig:limitcompare}}

While proper treatment of the systematic errors is challenging, this analysis is still statistics limited. Figure~\ref{fig:limitcompare} shows the obtained limit compared to the band in which 68\% and 95\% of the pseudo-experiments fall. The large width of this sensitivity band and the stronger than average constraints obtained over some range of $\Delta m_s^2$ is mostly the result of volatility to statistical fluctuations in the chosen statistical approach. The low-energy search described in~Sec.~\ref{sec:IClowensterile} works in the unresolved regime and is thus independant of $\Delta m_s^2$.  While more data will be collected with IceCube, stronger constraints in particular at higher values of $\Delta m_s^2$ are expected mostly from the next generation of underwater- and underice neutrino telescopes KM3Net~\cite{Adrian-Martinez:2016fdl} and IceCube-Gen2~\cite{Aartsen:2014njl}. 

\subsection{Accelerator experiments}

Accelerator neutrino experiments use a beam of GeV muon or muon antineutrinos produced by high-energy protons from an accelerator complex as a very directed source of neutrinos primarily for oscillation experiments. These experiments are also sensitive to sterile neutrinos in two different baseline regimes. On so-called short baselines of a few 100\,m, the experiments are looking for active-to-active neutrino appearance $\nu_{\mu} \rightarrow \nu_e$ (see Sec.~\ref{sec:signature-appear}) and also active-to-sterile disappearance $\nu_{\mu} \rightarrow \nu_{\mu}$ (see Sec.~\ref{sec:signature-dissappear}). 
The Short Baseline Neutrino (SBN) Program which is going to operate at these baselines is described in Sec.~\ref{sub:sblacc}. SBN is directly following up on the LSND and MiniBooNE results discussed in Sec.~\ref{sec:lsnd} and \ref{sec:miniboone}.

Long-baseline neutrino oscillation experiments are primarily designed for precision measurements of the three-flavor neutrino mixing parameters. Such experiments operate on baselines of a several 100\,km and are sensitive to active-to-sterile disappearance of muon neutrinos produced in the neutrino beam (see Sec.~\ref{sec:signature-dissappear}). The long-baseline experiments which have performed searches for sterile neutrinos are discussed in Sec.~\ref{sub:lblacc}.

\subsubsection{Short Baseline Neutrino Program at Fermilab\label{sub:sblacc}}

The Short Baseline Neutrino Program (SBN) \cite{Antonello:2015lea, Machado:2019oxb} is a three-detector setup aiming to search for light sterile neutrinos causing $\nu_e$ appearance and $\nu_{\mu}$ disappearance in a muon neutrino beam, and in particular to probe the result of the MiniBooNE experiment (see Sec.~\ref{sec:miniboone}). Thus, SBN has intentional commonalities with MiniBooNE such that it is located in the same neutrino beam line at Fermilab. The three-detector experiment is designed to be able to confirm or exclude the allowed parameter space in $\Delta m^2$ and $\sin^2 2\theta_{\mu e}$ by the MiniBooNE and LSND results with $5\sigma$ confidence (see Fig.~\ref{fig:sbnsens}).

All detectors are located in the Booster Neutrino Beamline (BNB)\cite{AguilarArevalo:2008yp}. The neutrino beam is created by accelerated protons hitting a beryllium target and producing neutrinos through the decay of secondary particles in the beamline. Fermilab’s Booster synchrotron delivers protons with 8\,GeV kinetic energy to the target at 5\,Hz. The beryllium target is embedded within a pulsed electromagnet (the “horn”) that produces a toroidal magnetic field to focus positive secondary particles and defocus negative secondary particles emerging from proton-beryllium interactions to obtain a $\nu_{\mu}$-enhanced beam. The polarity of the horn current can be changed to produce an antineutrino-enhanced beam as was done in the MiniBooNE experiment. The secondary particles decay in a 50\,m long decay tunnel followed by an absorber, and a beam of neutrinos with an average energy of 800\,MeV remains. In neutrino-mode, the flux is dominated by $\nu_{\mu}$'s ($\sim$93.6\%), with a ~5.9\% wrong-sign contamination of $\bar \nu_{\mu}$'s. The contamination of intrinsic $\nu_e$'s and $\bar \nu_e$'s, which is an important background in the appearance search, is at the level of 0.5\% at energies below 1.5\,GeV\cite{Antonello:2015lea}.
 
\begin{figure}[htbp]
  \centering
\includegraphics[width=\textwidth]{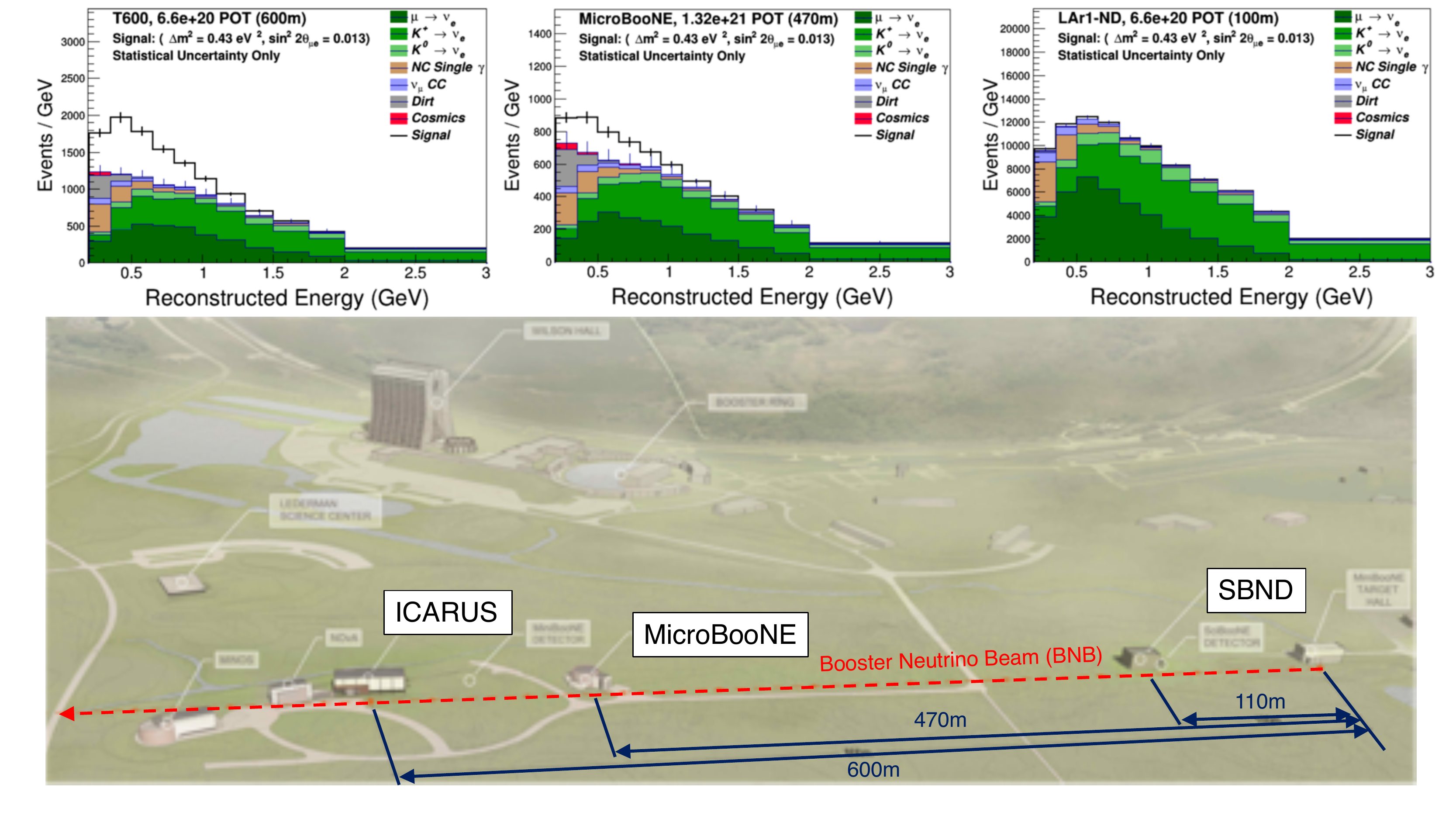}
\caption{The graphic shows the positioning of the SBN experiments SBND, MicroBooNE, and ICARUS in the BNB. The beam direction follows from right to left in the above picture. The plots show a stacked histogram of simulated signal and backgrounds for an idealized version of each detector and illustrate the expected event distribution assuming an oscillation signal of $\Delta m^2 = 0.43$~eV$^2$ and $\sin^2 2\theta_{\mu e} = 0.013$. The expected signal from a sterile neutrino (white) increases with the baseline and is largest in the ICARUS T600 detector. The background of electromagnetic signatures from $\gamma$s produced in $\nu_{\mu}$ interactions (brown) is greatly reduced compared to the MiniBooNE experiment by the LArTPC technology. The main background are electromagnetic showers from interactions of $\nu_e$ and $\bar \nu_e$ intrinsic to the beam (green). Note, that the plots are taken from the SBN design proposal\cite{Antonello:2015lea} and serve illustrative purposes. They will be superseeded once fully developed calibration and event reconstruction is available for all three detectors.}
\label{fig:sbnschematic}
\end{figure}

The three detectors are positioned at 110\,m, 470\,m, and 600\,m from the proton target, respectively (see Fig.~\ref{fig:sbnschematic}). With a baseline of 470\,m, the MicroBooNE experiment\cite{Acciarri:2016smi} is located closely to the MiniBooNE experiment, which operated at a baseline of 541\,m. The far detector ICARUS, which previously operated in the CNGS neutrino beam\cite{Amerio:2004ze, Antonello:2015zea} and was later refurbished and moved to Fermilab in 2017, is five times more massive than MicroBooNE and will collect higher statistics. At 110\,m from the proton target, the near detector SBND will observe an unoscillated neutrino spectrum, and provide the necessary flux and cross section constraints for the SBN program. 

The near detector will in particular enable a $\nu_{\mu}$ disappearance search, which is of interest since while there are hints on sterile neutrinos from $\nu_e$ appearance experiments, only exclusion limits have been reported in $\nu_{\mu}$ disappearance searches (see Sec.~\ref{sub:global}). The analysis of both channels within the same experiment will therefore provide very important insights to this tension. The SBN program will be able to increase the sensitivity to $\nu_{\mu}$ disappearance by an order of magnitude compared to the MiniBooNE + SciBooNE measurement (see Fig.~\ref{fig:sbnsens} and \ref{fig:DHKMMMS-2018-JHEP-1808-010-f5}).

\begin{figure}[htbp]
\begin{minipage}{.6\textwidth}
  \centering
\includegraphics[width=\textwidth]{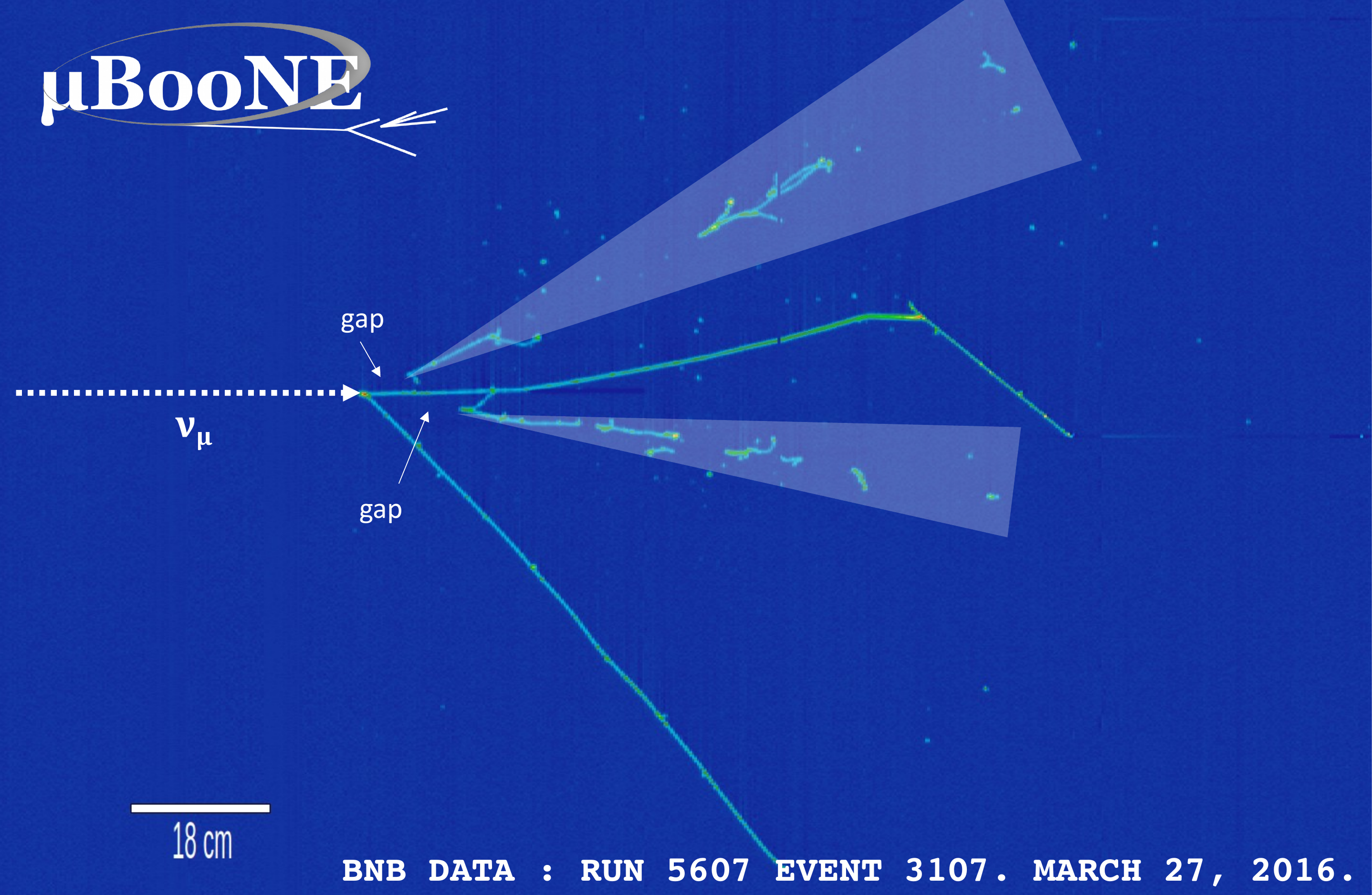}
\end{minipage}
\begin{minipage}{.4\textwidth}
  \centering
\includegraphics[width=\textwidth]{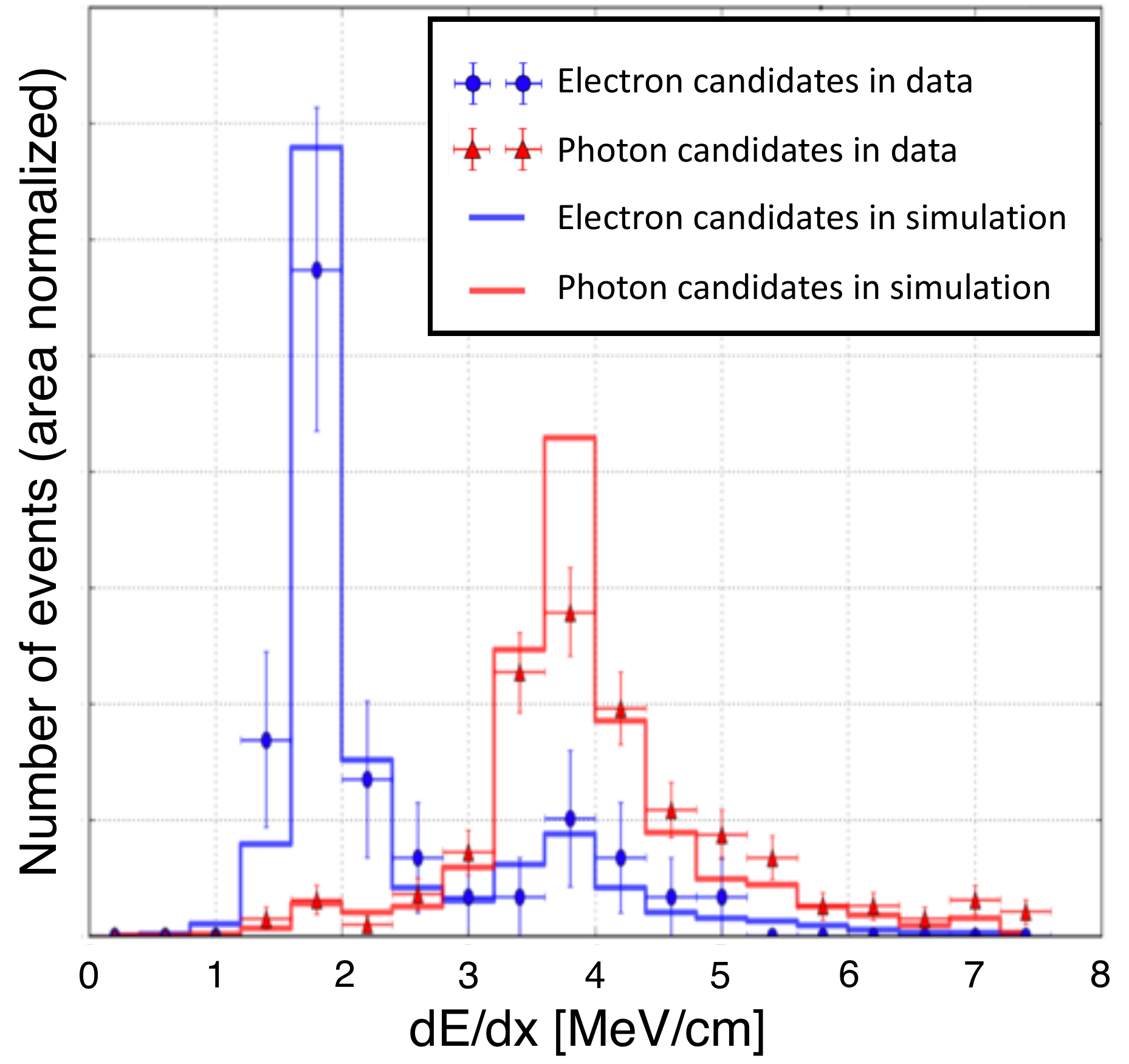}
\end{minipage}
\caption{Left: this LArTPC event display from MicroBooNE shows the interaction of a $\nu_{\mu}$ producing a $\pi^0$\cite{Adams:2018sgn}. The distinct signature of the $\pi^0$ is the gap between the start point of the showers (indicated by the cones) from the interaction vertex caused by the neutral and unseen $\gamma$s produced in the decay of the $\pi^0 \rightarrow \gamma \gamma$. The color represents a measure of the deposited charge. Right: demonstration of $e / \gamma$ separation in LArTPCs by the ArgoNeuT experiment\cite{Acciarri:2016sli}. ArgoNeuT was a smaller-size experiment advancing the LArTPC technology. Electrons and $\gamma$s are categorized by an observed gap between the interaction vertex and the start of the electromagnetic shower in data and simulation. For both sub-samples, the energy loss distribution is shown. Photon showers are characterized by twice as much energy loss than an electron induced shower due to the shower starting with pair production ($\gamma \rightarrow e^+ e^-$). }
\label{fig:dedx}
\end{figure}

\begin{figure}[htbp]
  \centering
\includegraphics[width=\textwidth]{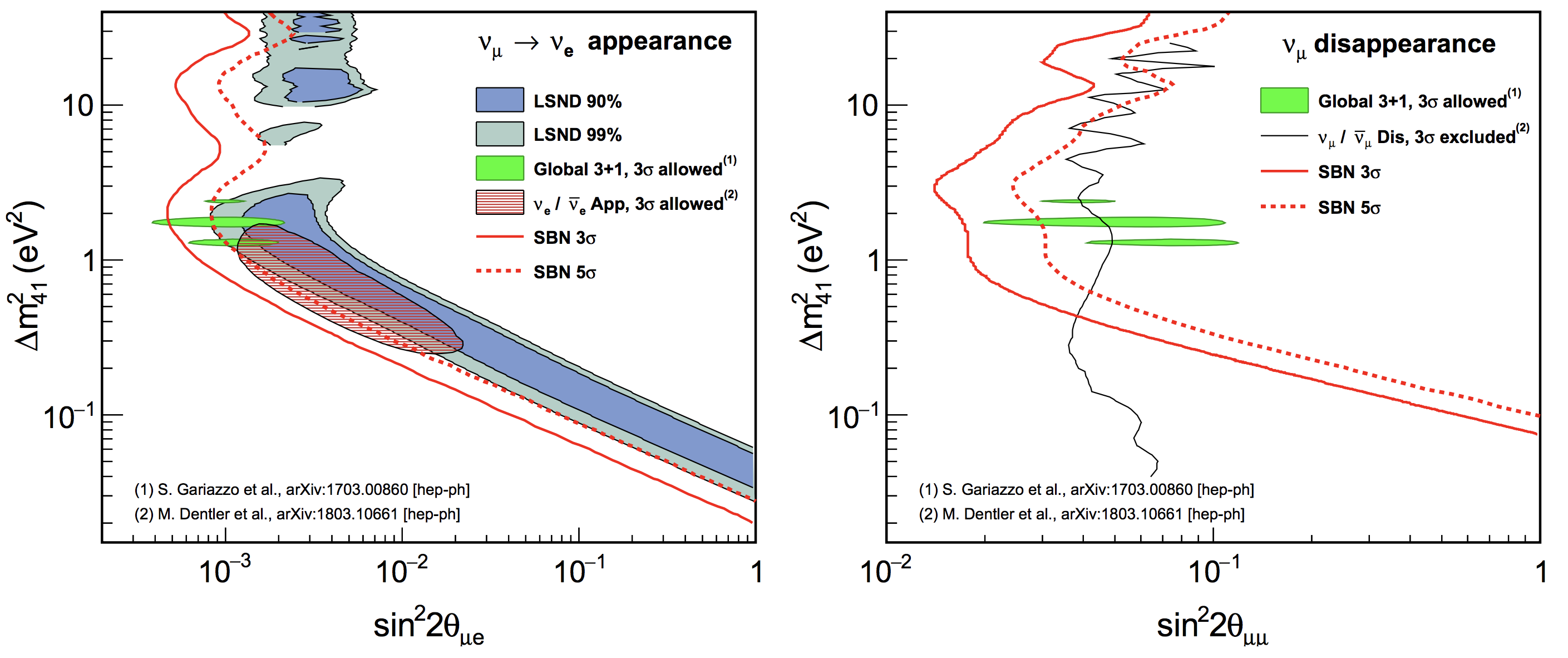}
\caption{Sensitivity plots for the combined $\nu_e$ appearance (left) and $\nu_{\mu}$ disappearance (right) analyses of all three detectors in the SBN program (SBND, MicroBooNE, ICARUS)\cite{Machado:2019oxb}. The solid red line corresponds to a $3\sigma$, the dashed red line to a $5\sigma$ sensitivity to a light sterile neutrino. The LSND preferred regions at 90\% confidence level (blue shaded) and 99\% confidence level (gray shaded) are shown for comparison. The $3\sigma$ global best fit regions are shown in green\cite{Gariazzo:2017fdh}.}
\label{fig:sbnsens}
\end{figure}

All three detectors are liquid argon Time Projection Chambers (LArTPC). The liquid argon serves as target for the neutrino interactions as well as detection medium. Secondary charged particles produced in the neutrino-argon interaction ionize the argon atoms, and ionization electrons are drifted to one side of the detector by an electric field, where they leave a projected image of the interaction as they deposit charge on the anode plane wires. The benefit of this technique is its superior imaging capability providing mm-resolution views of the interaction that allow to track even short secondary particles such as protons emerging from the nucleus\cite{Adams:2018gbi, Adams:2018dra, Acciarri:2017hat}. In this particular application of a $\nu_e$ appearance search, the advantage is that the discrimination of signatures produced by a single electron from a $\nu_e$ interaction against the signature of photons produced for example in the decay of secondary $\pi^0$s is superior to the Cherenkov technique used in the MiniBooNE experiment (see Fig.~\ref{fig:dedx}).

The challenges for the $\nu_e$ appearance analysis are backgrounds from various sources: a large background are beam-intrinsic $\nu_e$'s and single photon backgrounds from $\nu_{\mu}$ interactions. The near detector will help to constrain these with its flux and cross section measurements. Cosmic rays are another background since the SBN experiments are operating on the surface and the LArTPC technology requires a rather long readout window of order ms. All detectors use external cosmic tagging systems to reject cosmic events.

MicroBooNE started neutrino beam data taking in October 2015, ICARUS and SBND are expected to start physics running in 2020 and 2021, respectively. During the years MicroBooNE has been operating without the near and far detector, the experiment is studying the spectrum of electromagnetic events – the analogous measurement to the MiniBooNE result but with LArTPC detection technology providing separation power between electron and photon induced event signatures. MicroBooNE by itself will be able to confirm or exclude an electromagnetic event excess as seen in MiniBooNE, and determine if the excess is indeed caused by electrons, which is imperative to linking the observation to the hypothesis of sterile neutrinos. With SBND and ICARUS, the SBN Program will perform $\nu_e$ appearance and $\nu_{\mu}$ disappearance searches and reach its full sensitivity covering the favored parameter regions of the LSND and MiniBooNE results (see Fig.~\ref{fig:sbnsens}).

\subsubsection{Constraints from long-baseline experiments\label{sub:lblacc}}

Long-baseline neutrino accelerator experiments are built with the purpose to take precision measurements of three flavor neutrino oscillations. However, these experiments also provide sensitivity to sterile neutrinos by studying the disappearance of muon neutrinos from a neutrino beam similar to experiments studying $\nu_{\mu}$ disappearance with atmospheric neutrinos (see Sec. \ref{sub:atmospherics}). No anomalies have been observed in $\nu_{\mu}$ disappearance measurements, and the MINOS/MINOS+ as well as the NO$\nu$A experiments have been able to set exclusion limits in the phase space of $\Delta m^2_{41}$ and $\sin^2  \left( \theta_{24} \right)$.

MINOS/MINOS+ and NO$\nu$A both operate in the Fermilab NuMI beam\cite{Aliaga:2016oaz}, which produces few-GeV muon neutrinos. The energy spectrum of the beam is configurable. MINOS is a steel-scintillator sampling calorimeter made out of alternating planes of magnetized steel and plastic scintillators\cite{Michael:2008bc}. It consists of a near and far detector of identical technology. While the far detector weighs 5.4\,kt and is located underground in Soudan mine in Northern Minnesota at a baseline of 735\,km, the near detector is with 980\,t much smaller and located on Fermilab site. Both detectors are situated on-axis in the NuMI beam. From 2005 to 2012, MINOS operated in a 1-6 GeV muon neutrino beam studying three flavor oscillations. The upgrade MINOS+ operated from 2013 to 2016 in a broader and higher energy beam (4-10\,GeV), which improves the sensitivity to sterile neutrinos wrt MINOS. 

The NO$\nu$A experiment is located in the same neutrino beam but 14\,mrad off-axis\cite{Ayres:2007tu}. NO$\nu$A consists of a near and far detector, both made of cells filled with liquid-scintillator. The far detector is located in Minnesota at 810\,km baseline and has a total mass of 14\,kt. Again, the NO$\nu$A near detector is located on Fermilab site. Due to its operation in the off-axis beam, the neutrino flux at the NO$\nu$A far detector peaks at 2\,GeV and has a more narrow energy spectrum that the on-axis flux. NO$\nu$A's main purpose is to measure $\nu_e$ appearance in the three flavor oscillation paradigm. 

The MINOS sampling calorimeter is best suited to measure muons from $\nu_{\mu}$ charged-current interactions, but can also detect products of neutral-current interactions. In the charged-current $\nu_{\mu}$ disappearance channel, a hypothetical oscillation
due to $\Delta m^2_{41}$ would be
observed on top of the three-flavor oscillation. The observed neutral current rate is unaffected by three-flavor oscillations but sensitive to sterile oscillations. The effect of sterile oscillation strongly depends on the mass of the sterile neutrino. For light sterile neutrinos with $\Delta m^2_{41} <~ 1$\,eV$^2$, the effects are seen in the far detector. For larger masses, the oscillations become more rapid and start to be seen in the near detector. MINOS/MINOS+ published several iterations of the sterile neutrino analysis with MINOS and MINOS+ neutrino and antineutrino beam data\cite{Adamson:2017uda}. Exclusion limits are shown in Fig.~\ref{fig:numdis-DHKMMMS} and are discussed in the context of global $\nu_{\mu}$ disappearance fits in Sec.~\ref{sub:global}.
The NO$\nu$A sterile neutrino analysis utilizes the neutral current channel with improved event reconstruction and identification, and background rejection compared to MINOS. The results are consistent with three-flavor oscillations and no hints on sterile neutrinos are found\cite{Adamson:2017uda}. Future analyses with additional data will increase the sensitivity of the analysis.

\subsubsection{Implications for future long-baseline experiments}
\label{sub:lblfut}

The possible existence of light sterile neutrinos have potential implications for long-baseline
experiments~\cite{deGouvea:2014aoa,Klop:2014ima,Berryman:2015nua,Gandhi:2015xza,Palazzo:2015gja,Agarwalla:2016mrc,Agarwalla:2016xxa,Agarwalla:2016xlg,Dutta:2016glq,Capozzi:2016vac,Coloma:2017ptb,Choubey:2017cba,Choubey:2017ppj,Agarwalla:2018zko,Gupta:2018qsv},
especially concerning the search of CP violation.
The key observation is that although long-baseline experiments
are not sensitive to the value of $\Delta m^2_{41}$
because the corresponding oscillations are averaged at long distances,
the averaged terms in the oscillation probabilities depend on the
mixing angles that connect the active flavor neutrinos to $\nu_{4}$
and on the related CP-violation phases.

CP violation is not observable in short-baseline experiments,
because the effective oscillation probabilities in Eq.~(\ref{eqn:probSBL})
depend on the absolute values of the fourth column of the mixing matrix.
This is due to the fact that in short-baseline experiments oscillations are
driven by only one squared-mass difference, $\Delta m^2_{41}$,
since the source-detector distance is too short
for the emergence of the oscillations driven by the other two squared-mass differences
$\Delta m^2_{31}$ (the atmospheric squared-mass difference)
and
$\Delta m^2_{21}$ (the solar squared-mass difference).
Hence, the effective short-baseline oscillation probabilities in Eq.~(\ref{eqn:probSBL})
have a two-flavor form without CP violation effects.
Since CP violation is a property of mixing only for more than two flavors,
it can be observed only when the oscillations driven by more than one $\Delta m^2$
are operative and appear in their interference.
This is what can occur in long-baseline experiments,
which are sensitive to standard three-flavor CP violation
through the interference of the oscillations driven by
$\Delta m^2_{31}$ and $\Delta m^2_{21}$,
and can be sensitive to the new CP-violation phases in four-flavor mixing through the interference
of the oscillations driven by
$\Delta m^2_{31}$ and $\Delta m^2_{41}$,
even if the oscillations due to $\Delta m^2_{41}$ are averaged.

The exact oscillation probabilities in long-baseline experiments are very complicated
and can be calculated numerically,
taking into account also the matter effects for neutrino beams propagating
below the Earth surface.
However,
one can understand the physics involved by considering the following
approximation of the long-baseline probability of $\nu_{\mu}\to\nu_{e}$ transitions,
that is the main observable sensitive to CP violation in long-baseline experiments.
Since the global fits of short-baseline neutrino oscillation data
(see Section~\ref{sub:global})
indicate that
$|U_{e4}| \sim |U_{\mu4}| \sim |U_{e3}| \sim \varepsilon \sim 0.15$
and
$
\Delta{m}^2_{21}/|\Delta{m}^2_{31}|
\simeq 0.03 \sim \varepsilon^2
$,
we can calculate the leading terms of the vacuum probability
$P_{\nu_{\mu}\to\nu_{e}}^{\text{LBL}}$
in an expansion over $\varepsilon$.
At order $\varepsilon^3$, we obtain
\begin{align}
P_{\nu_{\mu}\to\nu_{e}}^{\text{LBL}}
\simeq
\null & \null
4
\,
\sin^2\theta_{13} \sin^2 \theta_{23} \sin^{2}\Delta_{31}
\nonumber
\\
\null & \null
+
2
\sin\theta_{13}
\sin 2\theta_{12}
\sin 2\theta_{23}
(\alpha\Delta_{31})
\sin\Delta_{31}
\cos(
\Delta_{32}
+
\delta_{13}
)
\nonumber
\\
\null & \null
+
4
\,
\sin\theta_{13}
\sin\theta_{14}
\sin\theta_{24}
\sin \theta_{23}
\sin\Delta_{31}
\sin(
\Delta_{31}
+
\delta_{13}
-
\delta_{14}
)
,
\label{probLBL}
\end{align}
where
$
\alpha
\equiv
\Delta{m}^2_{21}/|\Delta{m}^2_{31}|
$
and
$\Delta_{kj} \equiv \Delta{m}^2_{kj} L / 4 E$.
The first term in Eq.~(\ref{probLBL}) is the dominant one, of order $\varepsilon^2$,
and gives the main sensitivity of long-baseline experiments
to the measurement of $\theta_{13}$.
The second term, of order $\varepsilon^3$,
is subdominant and gives the sensitivity
of long-baseline experiments to the
standard three-flavor CP-violating phase $\delta_{13}$.
The third term is proper of four-flavor mixing
and depends on the new mixing angles
$\theta_{14}$ and $\theta_{24}$,
and on the new CP-violation phase $\delta_{14}$.
Since also the third term is of order $\varepsilon^3$,
its effect in the determination of CP violation can be as large as the second term
and must be taken into account in the analysis of long-baseline data
in the case of 3+1 mixing.
The indication in favor of a large $\delta_{13}$ around $3\pi/4$
obtained from the data of the current long-baseline experiments T2K and NO$\nu$A
(taking into account the reactor constraints on $\theta_{13}$)
in the case of three-flavor mixing
persists in the 3+1 scheme~\cite{Palazzo:2015gja,Capozzi:2016vac},
but the precise determination of $\delta_{13}$ in the future dedicated experiments
DUNE and Hyper-Kamiokande
may be affected by the presence of $\delta_{14}$
\cite{deGouvea:2014aoa,Klop:2014ima,Berryman:2015nua,Gandhi:2015xza,Agarwalla:2016mrc,Agarwalla:2016xxa,Dutta:2016glq,Choubey:2017cba,Choubey:2017ppj,Agarwalla:2018zko,Gupta:2018qsv}.

%
%

\section{Direct neutrino mass experiments}
\label{sec:numass}

As described in Sec.~\ref{sub:Direct}, direct neutrino mass experiments are a complementary way of searching for light sterile neutrinos. These experiments are based on single $\beta$ decays, where an electron neutrino favour eigenstate is created, which is a superposition of mass eigenstates. Correspondingly, also the spectrum is given as a superposition of the spectra corresponding to each mass eigenstate $m_{\nu_i}$, weighted by their mixing amplitude $|U_{ei}|$ to the electron favour. As a result, a fourth mass eigentstate, with a mass significantly different to the three light mass eigenstates, will imprint itself as a kink-like signature in the beta decay spectrum, see Fig.~\ref{fig:kink}.

A challenge of direct neutrino mass experiments is that the signal rate close to the endpoint (where the signal of the neutrino mass and also the signal of a light sterile neutrino is maximal) is extremely low. The key requirements are thus 1) an ultra-strong radioactive source, 2) an extremely low background level, 3) a high energy resolution of about 1 eV, and 4) a precise understanding of the spectral shape.

Currently, the super-allowed decay of tritium with a half-life of  $T_{1/2}=12.3$~years and a kinematic endpoint of $E_0 = 18.6$~keV and the electron-capture decay of $^{136}$Ho (with $T_{1/2}=4500$~years and $E_0 = 2.8$~keV) are considered. In the following the working principles of the major experimental approaches: KATRIN~\cite{Angrik:2005ep}, Project-8~\cite{Asner:2014cwa}, ECHo~\cite{Gastaldo:2017edk} and Holmes~\cite{Faverzani:2016ajz}, and their sensitivity to light sterile neutrino will be discussed. Fig.~\ref{fig:sensitivity} summarizes the expected sensitivity of KATRIN, ECHo, and future tritium- and holmium-based experiments.

\onefig[htb]{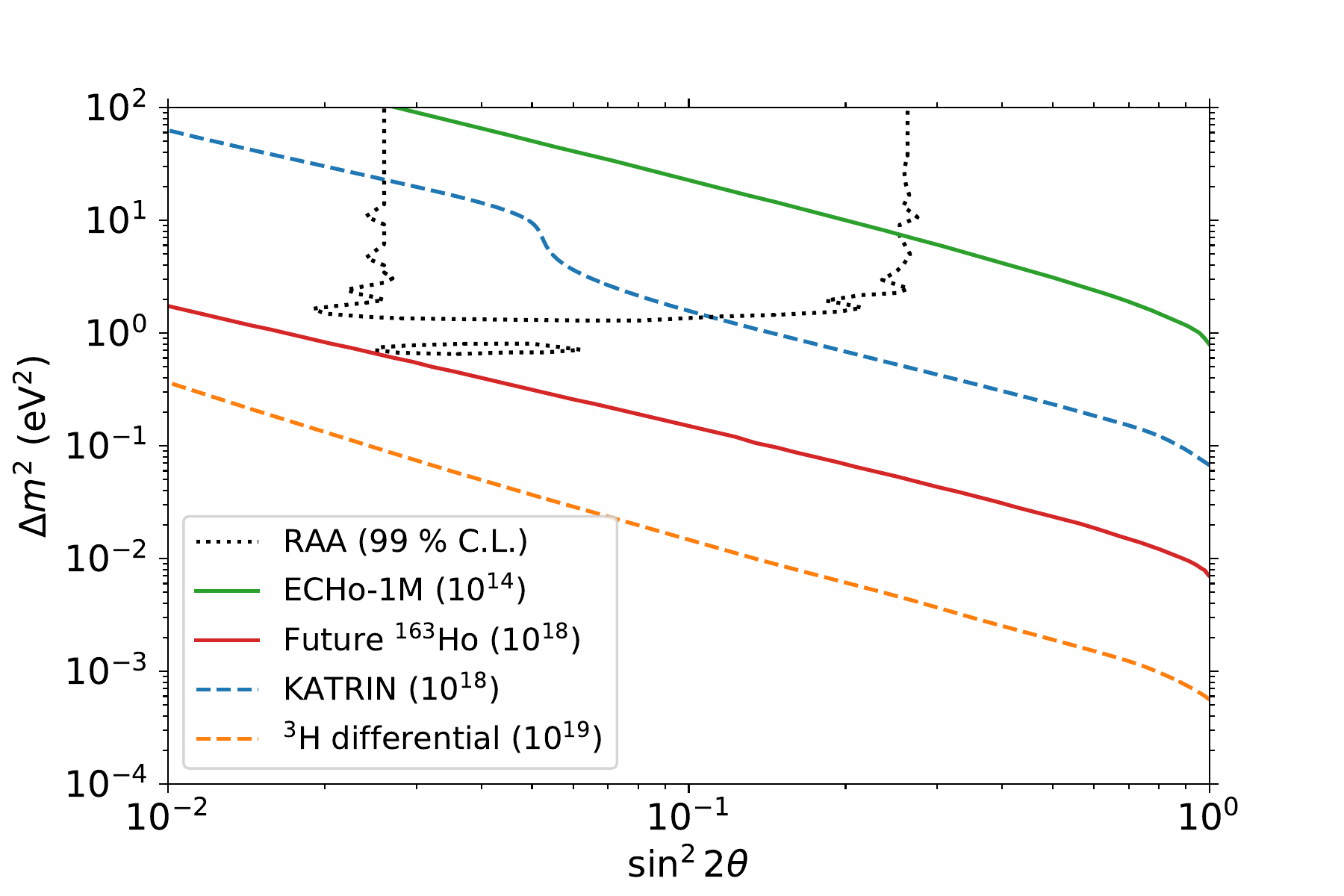}{Prospective sensitivity of tritium and holmium-based direct neutrino mass experiments to light sterile neutrino. In dashed blue the sensitivity of a 3-year KATRIN measurement is shown, based on the design parameters~\cite{Angrik:2005ep}. The small decrease in sensitivity at $\sin^2(2\theta)=5\cdot10^{-2}$ stems from the specific choice of the measurement time distribution assumed for this study. A future atomic tritium experiment with increased statistics ($10^{19}$ beta decays), and with a differential energy measurement with a resolution of 1~eV (Full-Width-Half-Maximum) could, from a statistical point-of-view, reach the sensitivity depicted with the yellow dashed line. The green solid line depicts the sensitivity the ECHo-1M could reach. A possible upgrade to higher statistics ($10^{18}$ decays) would increase the sensitivity accordingly (red line)~\cite{Gastaldo2016}. \label{fig:sensitivity}}

\subsection{KATRIN}
The Karlsruhe Tritium Neutrino (KATRIN) experiments is designed to measure the neutrino mass with a sensitivity of 200~meV (90\% CL)~\cite{Angrik:2005ep}. It performed its first commissioning run with tritium in 2018 and stared neutrino mass data taking in 2019. Its tritium source strength and spectroscopic quality allow to extend the physics program of KATRIN to also perform a competitive search for light sterile neutrinos without any hardware modification~\cite{Formaggio:2011jg, Esmaili:2012vg, Riis:2010zm}.

KATRIN combines a ultra-luminous gaseous molecular tritium source with the well-established magnetic adiabatic collimation and electrostatic (MAC-E) filter technology~\cite{Otten:2008zz,Lobashev:1985mu}. The 70\,m long beamline starts with the so-called windowless gaseous tritium source (WGTS) which contains about 30\,$\mu$g of tritium, providing an ultra-high and stable decay rate of $10^{11}$ decays/s. The WGTS beam tube is situated in a magnetic field, which is oriented in beam direction. All $\beta$-electrons that are emitted in the forward direction are guided along the field lines out of the WGTS and towards the spectrometers. The spectrometers work as MAC-E filters allowing only those electrons with enough kinetic energy to be transmitted. By counting the transmitted electrons as a function of the filter potential, the integral tritium spectrum is determined. The MAC-E Filter technology allows to combine high angular acceptance with ultra-sharp energy filtering.

A first tritium measurement campaign in 2018, demonstrated the integrity of the entire beamline, the stability of the system and a good understanding of the obtained tritium spectra. The first tritium data was also successfully used to perform a first sterile neutrino search in a mass range of 100--1,000\,eV. Based on a 14-day data set, a sensitivity to the mixing angle of $\sin^2(2\theta_{14}) < 10^{-2}$ could be reached (details of this preliminary result, will be subject of a different publication). 

Figure~\ref{fig:sensitivity} shows the sensitivity, KATRIN could reach with a 3-years data taking phase, based on the design parameters~\cite{Angrik:2005ep}. Recent studies have shown that an elevated background can strongly affect the KATRIN sensitivity. KATRIN successfully applied methods to reduce radon-induced backgrounds~\cite{G_rhardt_2018}, however, further measures are needed and are being explored to reach the design background rate of 10~mcps.

\subsection{Project-8}
Project-8 has demonstrated a new idea of a precise measurement of the $\beta$-decay electrons via Cyclotron Radiation Emission Spectroscopy (CRES)~\cite{Monreal:2009za, Asner:2014cwa, Esfahani:2017dmu}. This technique may allow to superseed the KATRIN sensitivity, based on the following advantages: 1) the scaling to larger tritium activity may be technically more feasible, 2) it is planned to make use of an atomic tritium source, which will overcome a limiting uncertainty in a molecular-based tritium-based experiment, 3) an essentially background-free setup may be feasible, 4) the technique provides access to the differential beta decay spectrum.

The general idea of this technique is to measure the coherent electromagnetic cyclotron radiation of the $\beta$-electron. As opposed to KATRIN, where the electron has to be extracted from the gaseous tritium source to measure its energy, here, the tritium source is transparent to the cyclotron radiation. The cyclotron frequency depends on the kinetic energy via the relativistic $\gamma$ factor. 

The technical realization of this approach consists of a magnetic trap inside of an antenna array or wave guide. The magnetic field determines the frequency range and radiated power of the $\beta$-electrons. For instance, for 18.6-keV electrons in a 1-T magnetic field the cyclotron frequency is 27.009~GHz and the radiated power is approximately 1.2~fW at a pitch angle of $90^{\circ}$. The density of tritium gas in the trap cell is limited, since scatterings of the electron with the background gas would lead to angle changes and hence a breaking of the storage condition. Consequently, the desired amount of tritium, the allowed tritium density, and the acceptance angle determine the size of the experiment. As compared to size of the KATRIN apparatus, the latter could in principle be smaller while reaching the same neutrino mass sensitivity.

Most importantly, concepts for an atomic tritium source are being developed by the Project-8 collaboration. This would allow to avoid the inevitable broadening of the beta-decay spectrum due to the rotational and vibrational excitation of the daughter molecule. 

Similarly to the KATRIN experiment, the Project-8 data obtained for a neutrino mass measurement can also be used to search for light sterile neutrinos. This measurement will be highly complementary to a MAC-E-Filter result, as the systematic uncertainties will be completely different. In Project-8 major systematics arise from inhomogeneities and inaccurate knowledge of the magnetic field. It will be a challenge to keep these uncertainties at an acceptable level, especially when scaling the experiment to a large-volume trap. The expected sensitivity of the final phase of Project-8 is expected to exceed the KATRIN sensitivity both for neutrino mass and sterile neutrinos. Figure~\ref{fig:sensitivity} shows the statistical sensitivity of a future atomic tritium experiment, with an energy resolution of 1~eV and a total statistics of $10^{19}$ decays.

\subsection{ECHo and Holmes}
Currently, two experiments explore the approach of using electron capture of ${}^{163}$Ho to probe the neutrino mass: ECHo~\cite{Gastaldo:2017edk, PhysRevLett.119.122501} and HOLMES~\cite{Faverzani:2016ajz, Giachero:2016xnn}. These experiments are complementary to tritium-based techniques both from a technical point-of-view and the fact that in this case an effective electron neutrino mass (as opposed to an effective electron-antineutrino mass) is studied.

The basic idea is to place the ${}^{163}$Ho source inside an absorber material with low heat capacity. X-rays and electrons emitted in the de-excitation of the ${}^{163}$Dy$^{*}$ daughter atom create phonons in the absorber material and cause a small temperature increase. This temperature change is detected by ultra-sensitive thermometers such as transition edge sensors (TES) or magnetic metallic calorimeters (MMC). 

The calorimetric concept avoids a number of systematic effects as compared to the MAC-E-filter technology. In particular energy losses due to scattering during the extraction of the electron from the gaseous tritium source are completely circumvented. Furthermore, the intrinsic energy broadening due to the final state distribution of molecular tritium is not present.  

However, the micro-calorimetric technique involves a different class of systematic effects and technical challenges. As opposed to the KATRIN experiment where only the electrons close to the endpoint are considered, in these experiments every single decay is detected. The total decay rate is typically twelve orders of magnitudes higher than the decay rate only in the last few eVs away from the endpoint. Hence, pile-up becomes a serious concern. To limit pile-up 1) a fast rise time is needed and 2) the source needs to be spread over a large number of detectors. To operate such a detector array in a cryogenic environment, however, a sophisticated multiplexed read-out technology is necessary. These topics are subject of active research and development of the ECHo and Holmes experiments.

Again, the neutrino mass data of a holmium-based experiment can be used to perform a search for light sterile neutrinos. As demonstrated in~\cite{Gastaldo2016} with a total statistics $10^{16}$ decays a sensitivity comparable to KATRIN can be achieved. Figure~\ref{fig:sensitivity} shows the sensitivity of the targeted ECHo-1M-stage with $10^{14}$ decays and the sensitivity of a possible future upgrade with $10^{18}$ decays, which would allow to cover the entire allowed parameter space of the reactor anomaly.

%
%

\section{Global status of active-sterile neutrino oscillations}
\label{sec:status}

The numerous experimental results obtained so far in the search
of light sterile neutrinos need to be analyzed in a consistent framework
that can take into account the positive indications and the constraints
following from the negative results.

In the following,
we discuss first the current status of
$\nua{e}$ disappearance in Section~\ref{sub:nuedis}
and then the results of global fits of
$\nua{\mu}\to\nua{e}$
appearance
and
$\nua{e}$ and $\nua{\mu}$ disappearance
in Section~\ref{sub:global}.

\subsection{Electron neutrino disappearance}
\label{sub:nuedis}

The study of short-baseline electron neutrino disappearance
due to active-sterile neutrino mixing
was originally motivated by the reactor and Gallium anomalies described in
Sections~\ref{sub:RAA} and \ref{sub:GA}, respectively.
However,
the discovery of the 5 MeV excess in the reactor neutrino spectra
of the
RENO \cite{RENO:2015ksa},
Double Chooz \cite{Abe:2014bwa},
and
Daya Bay \cite{An:2015nua}
experiments
with respect to the theoretical predictions based on the Huber-Muller fluxes \cite{Huber:2011wv,Mueller:2011nm}
raised strong doubts on the Huber-Muller fluxes and their estimated uncertainties
(see Refs.~\cite{Huber:2016fkt,Hayes:2016qnu}).
Since the reactor antineutrino anomaly \cite{Mention:2011rk}
follows from the comparison of the experimental reactor neutrino rates
with the theoretical predictions based on the Huber-Muller fluxes,
an increase of the uncertainties of the Huber-Muller fluxes
lowers the statistical significance of the anomaly.
Lacking a reliable indication in favor of the reactor antineutrino anomaly,
the Gallium neutrino anomaly cannot be considered as a sufficient indication in favor of
short-baseline electron neutrino disappearance,
because it is based on only four data points
with large uncertainties.
Therefore, at present, in the study of
short-baseline electron neutrino disappearance
we can rely with some confidence only on data that do not depend on an absolute rate measurements.
Indeed,
the new experiments are designed for the
investigation of the oscillations by measuring the neutrino interaction rate at different distances.
Comparing the rates or, better, the energy spectra measured at different distances
from a source gives model independent information
on neutrino oscillations or other distance-dependent phenomena.
At present, there are several reactor neutrino experiments
operating with this method, as described in Section~\ref{sub:reactorexp}.

\begin{figure}[!t]
\begin{minipage}[b]{0.49\linewidth}
\subfigure[]{\label{fig:GGLL-2018-PLB-782-13-f3a}
\includegraphics*[width=\linewidth]{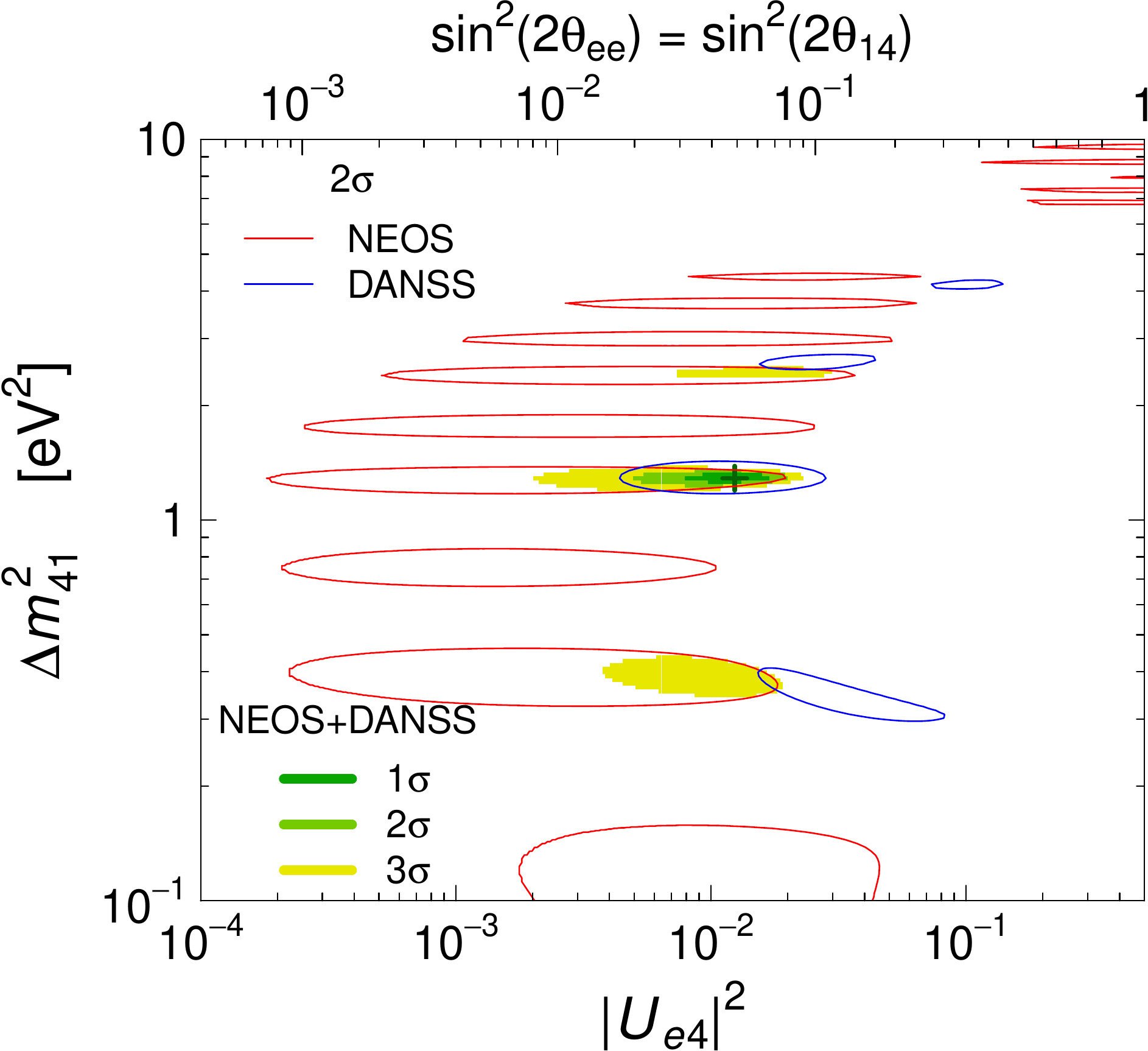}
}
\end{minipage}
\hfill
\begin{minipage}[b]{0.49\linewidth}
\subfigure[]{\label{fig:GGLL-2018-PLB-782-13-f3b}
\includegraphics*[width=\linewidth]{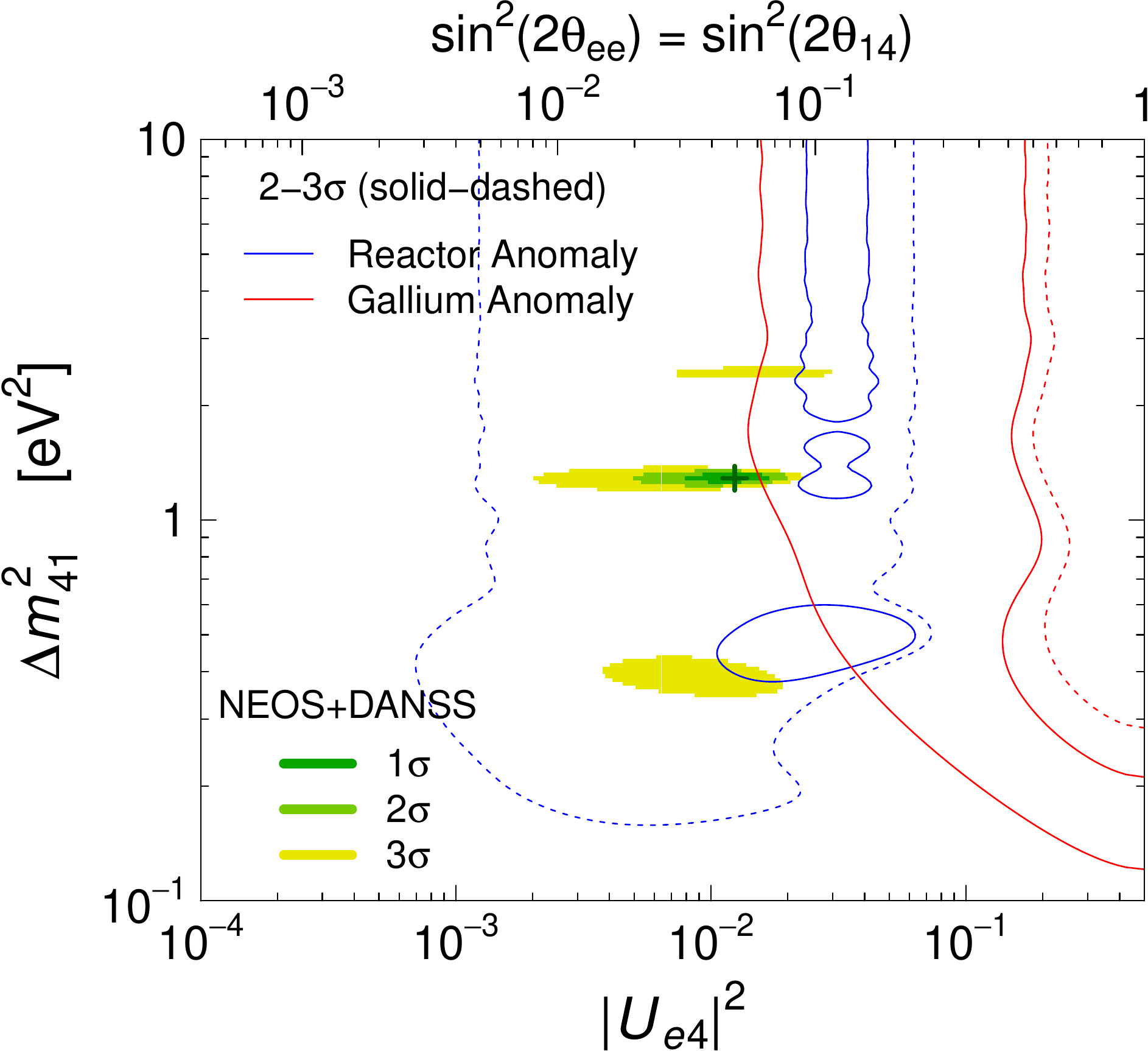}
}
\end{minipage}
\caption{ \label{fig:nuedis-GGLL}
Allowed regions in the
$|U_{e4}|^2$--$\Delta{m}^2_{41}$ plane obtained
in Ref.~\cite{Gariazzo:2018mwd}
from the combined analysis of the NEOS/Daya Bay and DANSS spectral ratios.
In both panels \subref{fig:GGLL-2018-PLB-782-13-f3a} and \subref{fig:GGLL-2018-PLB-782-13-f3b}
the shaded regions are allowed by the combined NEOS/Daya Bay and DANSS fit,
with the best-fit point indicated by a cross.
The red and blue lines in panel \subref{fig:GGLL-2018-PLB-782-13-f3a}
enclose, respectively, the NEOS/Daya Bay and DANSS $2\sigma$ allowed regions.
The blue and red lines in panel \subref{fig:GGLL-2018-PLB-782-13-f3b}
delimit, respectively, the reactor and Gallium anomaly allowed regions
at $2\sigma$ (solid) and $3\sigma$ (dashed; without a lower Gallium limit on $|U_{e4}|^2$).
}
\end{figure}

The currently most intriguing result is obtained from the combined fit of the
NEOS/Daya Bay spectral ratio \cite{Ko:2016owz}
and the
ratio of the spectra measured
at 10.7 and 12.7 meters from a reactor
in the DANSS \cite{Alekseev:2018efk}
experiment.
These experiments already achieved a sensitivity to very small values of $|U_{e4}|^2$,
of the order of $10^{-2}$,
for $\Delta{m}^2_{41} \sim 1 \, \text{eV}^2$.

Figure~\ref{fig:GGLL-2018-PLB-782-13-f3a}
shows the allowed regions in the
$|U_{e4}|^2$--$\Delta{m}^2_{41}$ plane
obtained in Ref.~\cite{Gariazzo:2018mwd}
from the NEOS/Daya Bay and DANSS spectral ratios.
There is a remarkable
overlap of the allowed regions of the two sets of data
for
\begin{equation}
|U_{e4}|^2 = 0.012 \pm 0.003
\quad
\text{and}
\quad
\Delta{m}^2_{41} = (1.29 \pm 0.03) \, \text{eV}^2
,
\label{ue4dm2bf}
\end{equation}
that determine the best-fit region of the combined analysis
(with two other regions at
$\Delta{m}^2_{41} \simeq 0.4 \, \text{eV}^2$
and
$\Delta{m}^2_{41} \simeq 2.5 \, \text{eV}^2$
that are allowed only at $3\sigma$).
The combined NEOS/Daya Bay and DANSS allowed region is confronted in
Fig~\ref{fig:GGLL-2018-PLB-782-13-f3b} \cite{Gariazzo:2018mwd}
with the regions preferred at 2 and 3 $\sigma$
by the reactor antineutrino anomaly
and the Gallium neutrino anomaly.
There is a tension at $2\sigma$
that indicates that the reactor and Gallium anomalies have been
somewhat overestimated.

\begin{figure}[!t]
\begin{minipage}[b]{0.49\linewidth}
\subfigure[]{\label{fig:DHKMMMS-2018-JHEP-1808-010-f2}
\includegraphics*[width=\linewidth]{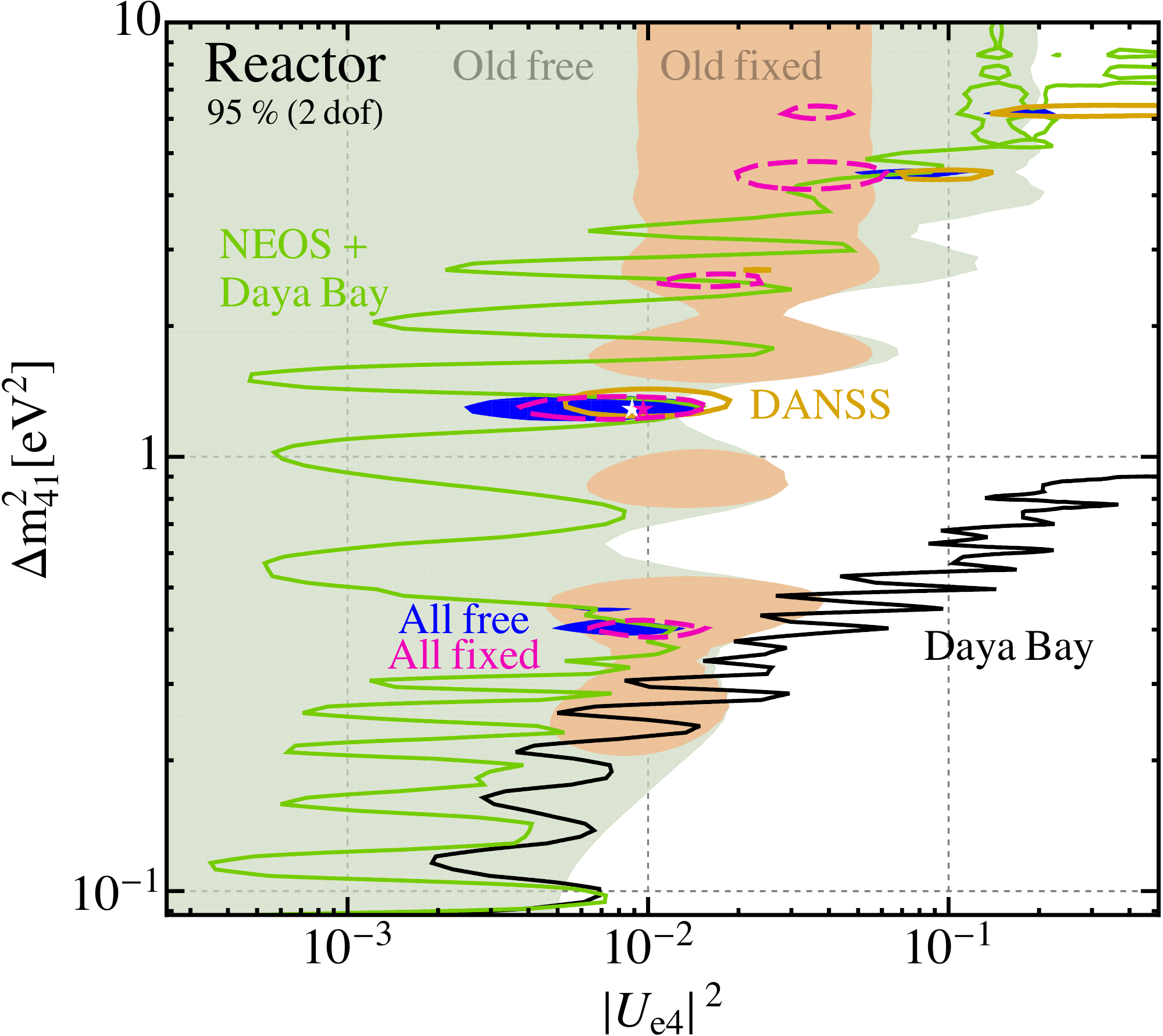}
}
\end{minipage}
\hfill
\begin{minipage}[b]{0.49\linewidth}
\subfigure[]{\label{fig:DHKMMMS-2018-JHEP-1808-010-f3}
\includegraphics*[width=\linewidth]{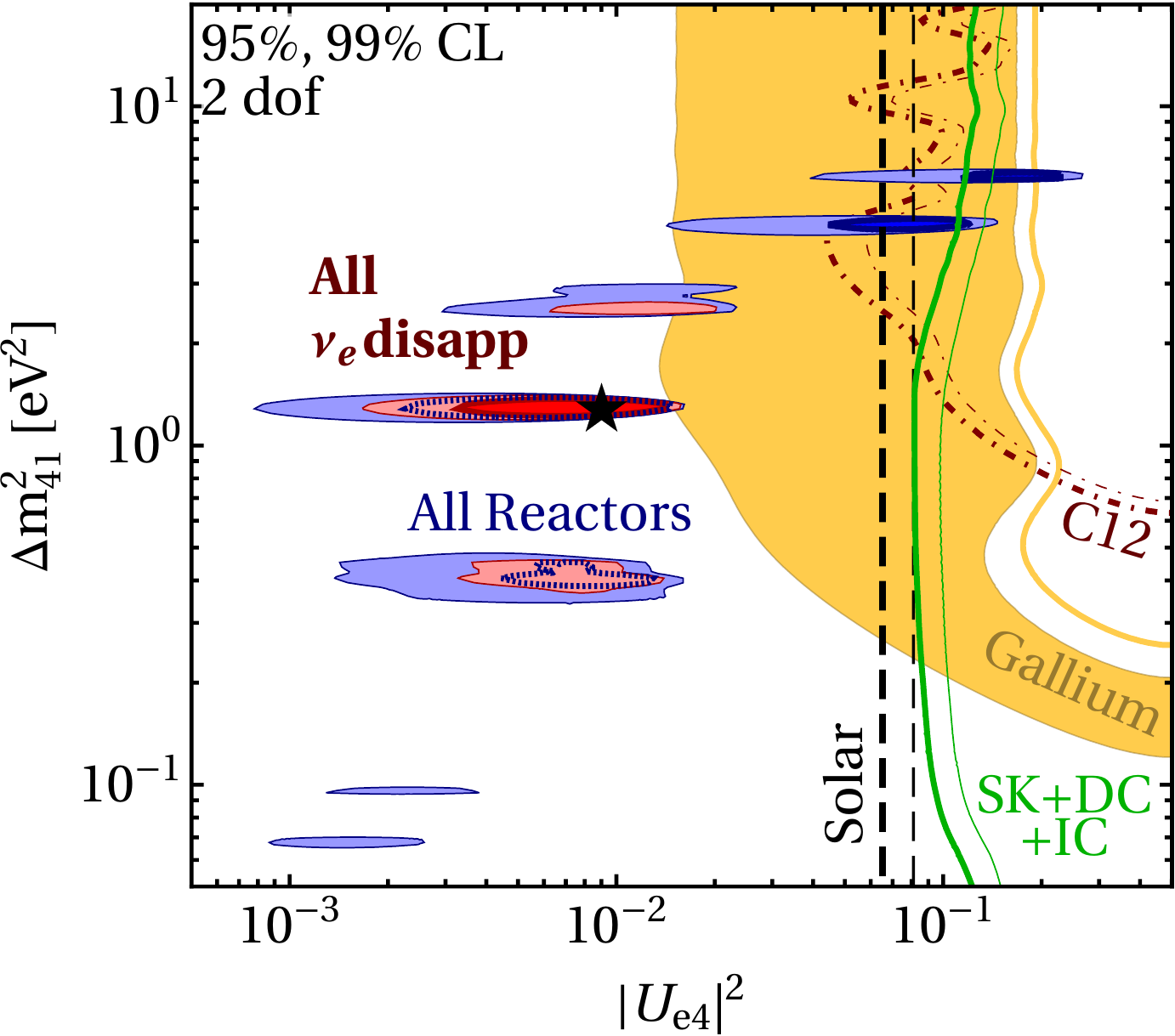}
}
\end{minipage}
\caption{ \label{fig:nuedis-DHKMMMS}
Allowed regions and exclusion curves in the
$|U_{e4}|^2$--$\Delta{m}^2_{41}$ plane obtained in Ref.~\cite{Dentler:2018sju}
from the analysis of different
$\nu_{e}$ and $\bar\nu_{e}$ disapperance datasets.
The allowed regions and exclusion curves in panel \subref{fig:DHKMMMS-2018-JHEP-1808-010-f2}
are at 95\% CL.
The blue shaded regions are allowed by the fit of all reactor data with free fluxes
(with the best-fit point indicated by a white star).
The magenta lines enclose the regions allowed by a fit of all reactor data
with the Huber-Mueller fluxes
(with the best-fit point indicated by a magenta star).
The light-shaded areas are allowed by the ``old'' reactor data
(i.e. without NEOS, Daya Bay and DANSS)
with fixed Huber-Mueller (light orange)
and free (light green) fluxes.
Also shown are the
Daya Bay \cite{An:2016luf} (black)
and
NEOS/Daya Bay \cite{Ko:2016owz} (green)
exclusion curves.
The allowed regions and exclusion curves in panel \subref{fig:DHKMMMS-2018-JHEP-1808-010-f3}
are at 95\% (dark shaded regions and thick curves)
and 99\% (light shaded regions and thin curves) CL.
The blue and red shaded regions are allowed, respectively,
by the combined fit of all reactor and all $\protect\nua{e}$ disappearance data
(with the best-fit point indicated by a black star).
Also shown are the solar exclusion curves (black dashed).
the Super-Kamiokande+DeepCore+IceCube (SK+DC+IC) exclusion curves (green solid),
and the $\nu_{e}$--${}^{12}\text{C}$ (C12)
scattering exclusion curves (dark red dash-dotted).
The analysis of Gallium data yielded the 95\% allowed yellow region and
the 99\% CL yellow exclusion curve.
}
\end{figure}

Figure~\ref{fig:DHKMMMS-2018-JHEP-1808-010-f2} \cite{Dentler:2018sju}
shows the regions in the $|U_{e4}|^2$--$\Delta{m}^2_{41}$ plane
that are allowed by the fit presented in Ref.~\cite{Dentler:2018sju}
of all reactor neutrino data,
including the NEOS/Daya Bay and DANSS spectral ratios,
with and without fixing the neutrino fluxes at the Huber-Mueller prediction.
The best-fit allowed region around
$|U_{e4}|^2 \simeq 0.01$
and
$\Delta{m}^2_{41} \simeq 1.3 \, \text{eV}^2$
is similar to that in Fig.~\ref{fig:nuedis-GGLL}
and it is almost independent of the assumption on the neutrino fluxes (free or fixed).
This happens because the NEOS/Daya Bay and DANSS spectral ratios
dominate the fit.
Moreover,
Fig.~\ref{fig:DHKMMMS-2018-JHEP-1808-010-f3} \cite{Dentler:2018sju}
shows that the best-fit region remains stable when the following constraints are
also taken into account:
the solar neutrino bound on $|U_{e4}|^2$
\cite{Giunti:2009xz,Palazzo:2011rj,Palazzo:2012yf,Giunti:2012tn,Kopp:2013vaa,Gariazzo:2017fdh,Dentler:2017tkw};
the ratio of the
KARMEN \cite{Armbruster:1998uk}
and
LSND \cite{Auerbach:2001hz}
$\nu_{e} + {}^{12}\text{C} \to {}^{12}\text{N}_{\text{g.s.}} + e^{-}$
scattering data at different distances from the source
\cite{Conrad:2011ce,Giunti:2011cp,Kopp:2013vaa};
the atmospheric neutrino constraint on $|U_{e4}|^2$
\cite{Maltoni:2007zf,Dentler:2018sju}
from the data of the
Super-Kamiokande \cite{Wendell:2014dka},
DeepCore \cite{Aartsen:2014yll}, and
IceCube \cite{TheIceCube:2016oqi}
experiments.

From Figs.~\ref{fig:nuedis-GGLL} and \ref{fig:nuedis-DHKMMMS}
(see also the discussions in Refs.~\cite{Gariazzo:2018mwd,Dentler:2018sju}),
we conclude that there is an intriguing model-independent
indication in favor of short-baseline $\nua{e}$ disapperance
due to active sterile mixing with the parameters in Eq.~(\ref{ue4dm2bf}).
Let us however emphasize that this indication depends crucially on the
agreement of the NEOS/Daya Bay and DANSS spectral ratios
and needs to be checked in other experiments,
as the ongoing
\textsc{Stereo} \cite{Almazan:2018wln},
PROSPECT \cite{Ashenfelter:2018iov},
SoLid \cite{Abreu:2018ajc}, and
Neutrino-4 \cite{Serebrov:2018vdw}
reactor experiments.

\begin{figure}[!t]
\begin{minipage}[b]{0.49\linewidth}
\subfigure[]{\label{fig:DHKMMMS-2018-JHEP-1808-010-f5}
\includegraphics*[width=\linewidth]{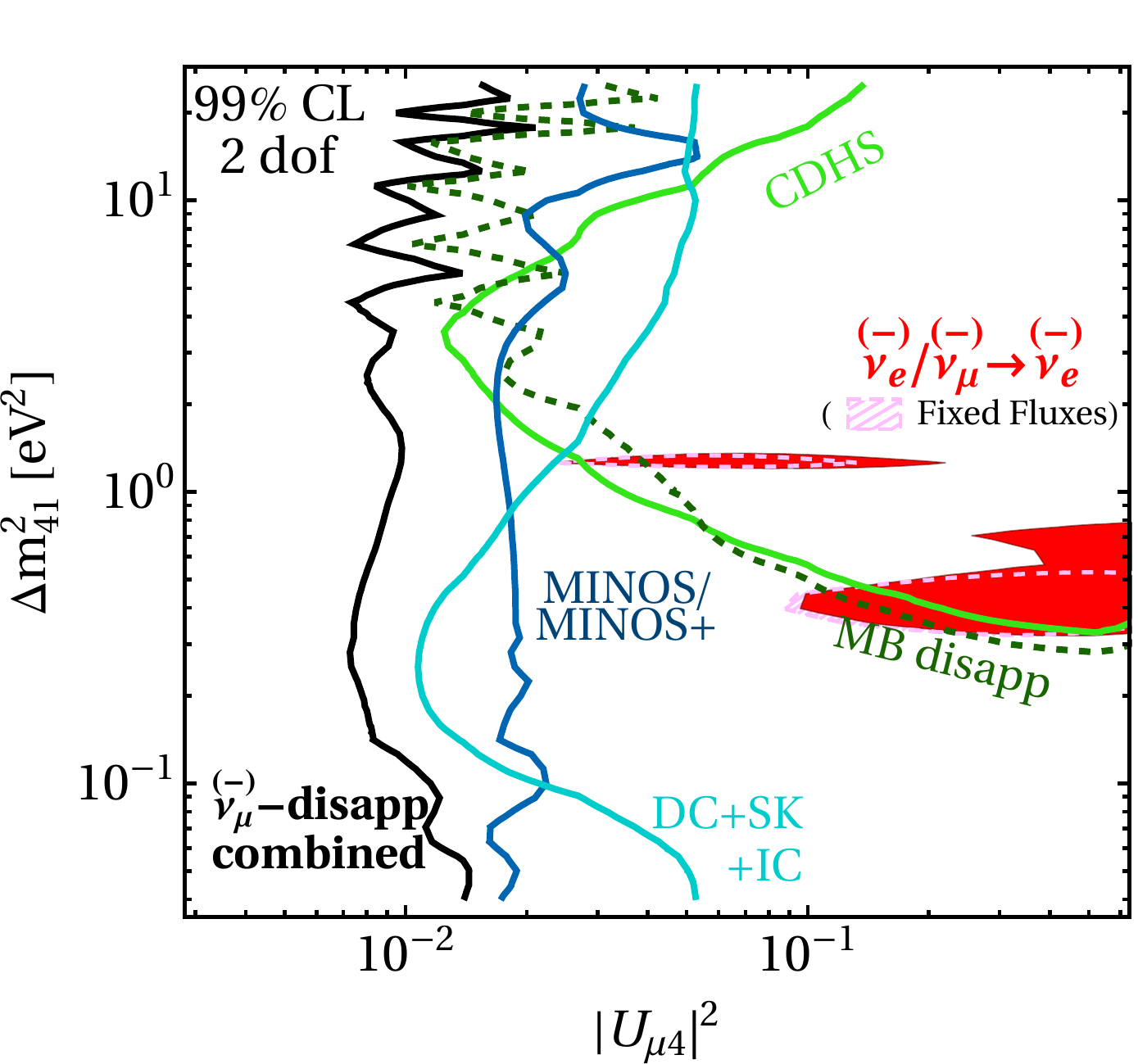}
}
\end{minipage}
\hfill
\begin{minipage}[b]{0.49\linewidth}
\subfigure[]{\label{fig:DHKMMMS-2018-JHEP-1808-010-f7}
\includegraphics*[width=\linewidth]{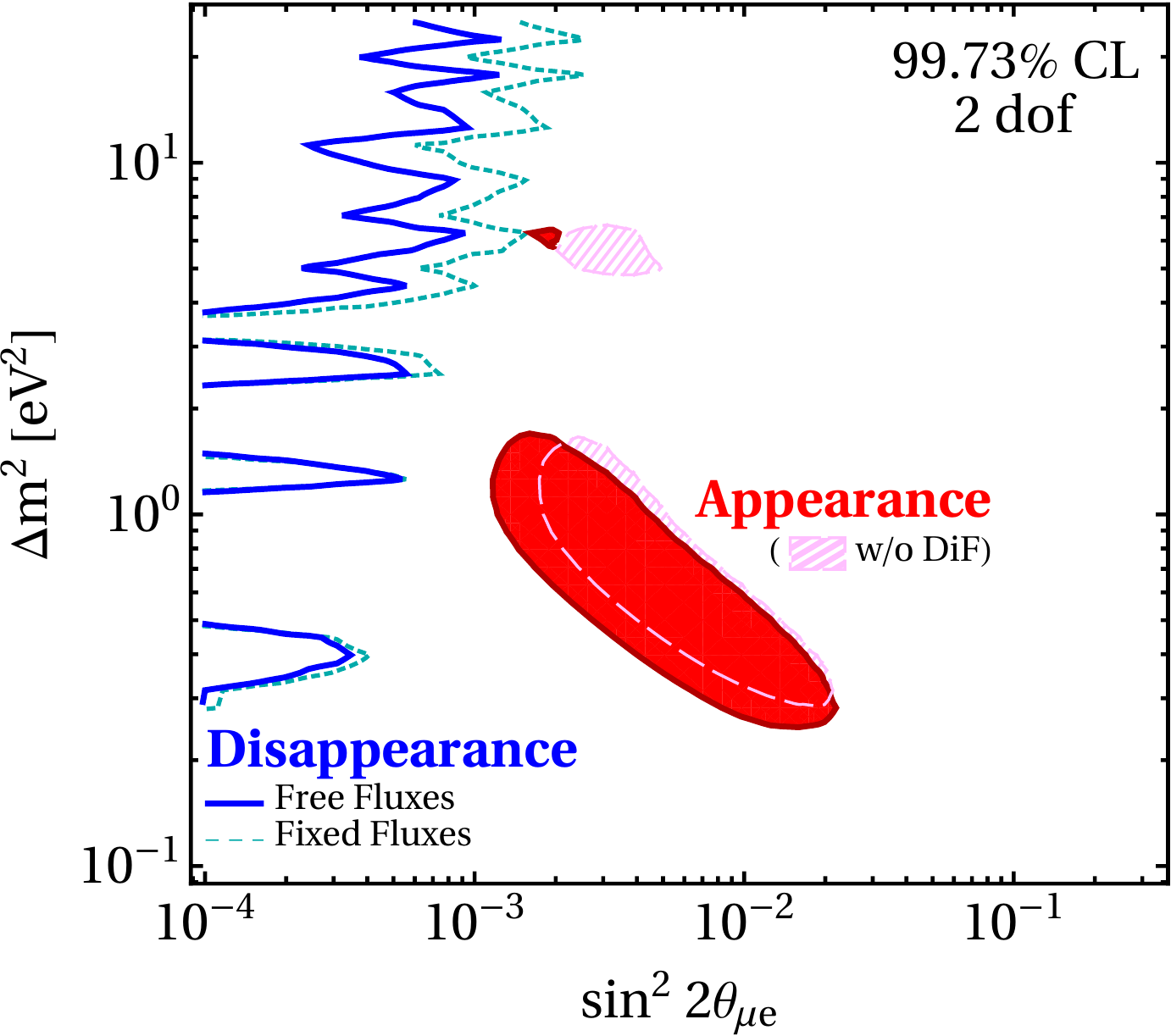}
}
\end{minipage}
\caption{ \label{fig:numdis-DHKMMMS}
Exclusion curves and allowed regions
obtained in Ref.~\cite{Dentler:2018sju}
from short-baseline $\protect\nua{\mu}$ disappearance and
$\protect\nua{\mu}\to\protect\nua{e}$ appearance data.
Panel \subref{fig:DHKMMMS-2018-JHEP-1808-010-f5} shows the 99\% CL exclusion curves in the
$|U_{\mu4}|^2$--$\Delta{m}^2_{41}$
plane obtained from the data of the
CDHS \protect\cite{Dydak:1983zq} (light green solid),
SciBooNE-MiniBooNE \protect\cite{Mahn:2011ea,Cheng:2012yy} (dark green dashed),
MINOS\&MINOS+ \cite{Adamson:2017uda} (blue solid), and
Super-Kamiokande+DeepCore+IceCube \cite{Wendell:2014dka,Aartsen:2014yll,TheIceCube:2016oqi}
(cyan solid)
experiments,
and the total combined exclusion curve (black solid).
Also shown are the allowed regions inferred from
the combination of $\protect\nua{e}$ disappearance
and
$\protect\nua{\mu}\to\protect\nua{e}$ appearance data
with free (red) and fixed Huber-Mueller (pink) reactor fluxes.
Panel \subref{fig:DHKMMMS-2018-JHEP-1808-010-f7} shows the
regions in the
$\sin^2(2\theta_{\mu e})$--$\Delta{m}^2_{41}$
plane
allowed at 99.73\% CL (i.e. $3\sigma$)
by $\protect\nua{\mu}\to\protect\nua{e}$ appearance data
with (red) and without (pink) LSND decay in flight (DIF) data \protect\cite{Aguilar:2001ty}.
Also shown are the combined $\protect\nua{e}$ and $\protect\nua{\mu}$ disappearance exclusion curve
with free (blue solid) and fixed (cyan dashed) reactor fluxes.
}
\end{figure}

\subsection{Global appearance and disappearance fit}
\label{sub:global}

Besides the indication of short-baseline $\nua{e}$ disapperance
discussed in the Section~\ref{sub:nuedis},
there are indications of short-baseline
$\nua{\mu}\to\nua{e}$ oscillations
found more than 20 years ago in the LSND experiment
\cite{Athanassopoulos:1995iw,Aguilar:2001ty}
and more recently in the MiniBooNE experiment
\cite{AguilarArevalo:2008rc,Aguilar-Arevalo:2013pmq,Aguilar-Arevalo:2018gpe}
(see Section~\ref{sec:lsndminiboone}).
In order to fit the short-baseline
$\nua{e}$ disapperance
and
$\nua{\mu}\to\nua{e}$ data
in the framework of active-sterile neutrino mixing,
one must take into account also the data of the experiments that searched for short-baseline
$\nua{\mu}$ disappearance,
on which there is currently no positive indication.
Actually, the negative results of short-baseline $\nua{\mu}$ disappearance searches
imply stringent bounds on $|U_{\mu4}|^2$
that generate a strong appearance-disappearance tension
\cite{Okada:1996kw,Bilenky:1996rw,Bilenky:1999ny,Grimus:2001mn,Maltoni:2002xd,Maltoni:2007zf,Kopp:2011qd,Giunti:2011gz,Giunti:2011hn,Giunti:2011cp,Conrad:2012qt,Archidiacono:2012ri,Archidiacono:2013xxa,Kopp:2013vaa,Giunti:2013aea,Gariazzo:2015rra,Gariazzo:2017fdh,Dentler:2018sju,Giunti:2019aiy}
due to the approximate relation
\begin{equation}
\sin^2(2\theta_{e\mu})
\simeq
\frac{1}{4}
\,
\sin^2(2\theta_{ee})
\,
\sin^2(2\theta_{\mu\mu})
\label{appdis}
\end{equation}
between the amplitude in Eq.~(\ref{eqn:Pemu})
of $\nua{\mu}\to\nua{e}$ oscillations
and the survival amplitudes of $\nua{e}$ and $\nua{\mu}$
in Eqs.~(\ref{see}) and (\ref{smm}), respectively\footnote{
Note that the appearance-disappearance tension cannot be alleviated
by considering more than one sterile neutrino,
because there are relations of the type (\ref{appdis})
for each additional sterile neutrino
\cite{Giunti:2015mwa}.
}.

Figure~\ref{fig:DHKMMMS-2018-JHEP-1808-010-f5}
\cite{Dentler:2018sju}
compares the exclusion curves in the
$|U_{\mu4}|^2$--$\Delta{m}^2_{41}$
plane with the allowed regions inferred from
the combination of $\nua{e}$ disappearance
and
$\nua{\mu}\to\nua{e}$ appearance data.
The appearance-disappearance tension is clear from the fact that these
allowed regions lie in the area that is excluded by the $\nua{\mu}$ disappearance experiments.

Figure~\ref{fig:DHKMMMS-2018-JHEP-1808-010-f7}
\cite{Dentler:2018sju}
shows the appearance-disappearance tension in the
$\sin^2(2\theta_{\mu e})$--$\Delta{m}^2_{41}$
plane,
where the region allowed by
$\nua{\mu}\to\nua{e}$ appearance data
lies in the area excluded by
the combined $\nua{e}$ and $\nua{\mu}$ disappearance data.

The appearance-disappearance tension has always been present in the analyses
of short-baseline neutrino oscillation data
\cite{Okada:1996kw,Bilenky:1996rw,Bilenky:1999ny,Grimus:2001mn,Maltoni:2002xd,Maltoni:2007zf,Kopp:2011qd,Giunti:2011gz,Giunti:2011hn,Giunti:2011cp,Conrad:2012qt,Archidiacono:2012ri,Archidiacono:2013xxa,Kopp:2013vaa,Giunti:2013aea,Gariazzo:2015rra,Gariazzo:2017fdh,Dentler:2018sju,Giunti:2019aiy}
with alternating strength since the discovery of the LSND anomaly in 1995
\cite{Athanassopoulos:1995iw}.
The tension was increased in 2016 \cite{Gariazzo:2017fdh} by the
NEOS \cite{Ko:2016owz} limits on $|U_{e4}|^2$
and by the MINOS \cite{MINOS:2016viw} and IceCube \cite{TheIceCube:2016oqi}
bounds on $|U_{\mu4}|^2$.
The confirmation of a small value of $|U_{e4}|^2$ by the DANSS experiment \cite{Alekseev:2016llm}
and the very stringent MINOS\&MINOS+ \cite{Adamson:2017uda}
limits on $|U_{\mu4}|^2$
led to a further dramatic increase of the tension in 2018 \cite{Gariazzo:2018mwd,Dentler:2018sju},
that is quantified by an appearance-disappearance parameter goodness-of-fit
\cite{Maltoni:2003cu}
smaller than $10^{-6}$
\cite{Dentler:2018sju,Giunti:2019aiy}.
This result disfavors the
neutrino oscillation explanation of the LSND and MiniBooNE anomalies.
However,
since it has been obtained with a combined analysis of many experimental data in
the specific framework of (3+1) active-sterile mixing,
it cannot be considered to be a definitive conclusion on the
LSND and MiniBooNE anomalies,
that need to be checked directly.
This will be done in the
Short-Baseline Neutrino (SBN) program \cite{Tufanli:2017mwt}
and in the
J-PARC Sterile Neutrino Search at J-PARC Spallation Neutron Source (JSNS$^2$)
experiment \cite{Ajimura:2017fld}.

%
%

\section{Cosmology}
\label{sec:cosmology}

The situation on the side of oscillation experiments is thus rather inconclusive, and other hints are worth exploring. We remember from Sec.~\ref{sec:cosmoeffects} that the existence of light sterile neutrinos normally implies the presence of a new population of relic particles in the universe, with consequences on cosmological observables such as the abundance of primordial elements, the spectrum of CMB anisotropies and the distribution of Large Scale Structures. We will now review the constraints on sterile neutrinos that can be inferred from cosmological observations.

\subsection{Bounds on the sterile neutrino density alone\label{sec:cosmo:density}}

A fit of BBN predictions to the measured abundance of primordial elements gives bounds on the density of relativistic relics when the Universe had a temperature $T \simeq 0.07$~MeV, and thus on the parameter $\Neff$ (see Sec.~\ref{sec:cosmoeffects:BBN}). The contribution of the nearly sterile state $\nu_4$, $\Delta N _4$, should be at least smaller than the total $N_\mathrm{eff}$. If active neutrinos thermalize in the early universe (as expected in the standard cosmological model), they contribute to $\Neff$ by 3.045~\cite{deSalas:2016ztq}, and we can further assume that $\Delta N_4 \leq (N_\mathrm{eff}-3.045)$, where the inequality accounts for the possibility of other light relics (e.g. light axions). 

BBN bounds on $N_\mathrm{eff}$ (and thus on $\Delta N_4$) are mostly sensitive to primordial helium abundance measurements, but one needs to combine helium and deuterium data to remove the degeneracy between the two parameters of the standard BBN model, $\omega_b$ and $N_\mathrm{eff}$. There are small controversies on the modelling of systematic errors in helium and deuterium measurements and on theoretical errors in BBN codes. For this work, we first computed bounds on $\Neff$ based on helium data from~\cite{Aver:2015iza} and deuterium data from~\cite{Cooke:2017cwo}, and then checked their dependence on various observational and theoretical uncertainties described in~\cite{Aghanim:2018eyx}. The results are always compatible with the conservative bounds 
\begin{equation}
    \Neff = 2.9 \pm 0.5~, \,\,\, \mathrm{(95\%CL,~Helium+Deuterium+BBN)}
\end{equation}
that also agree very well with~\cite{Pitrou:2018cgg}. Note that these bounds are marginalised over the baryon abundance, and are completely independent of any other cosmological parameter and of neutrino masses. They only assume the validity of the standard BBN model, with a negligible chemical potentials for the electron neutrino $|\mu_{\nu_e}| \ll T_\nu$ (this asumption will be released in some of the models discussed in Sec.~\ref{sec:cosmo:reconcile}). In conclusion, a conservative treatment of standard BBN and primordial abundance data tells us that a fully thermalized population of $\nu_4$ neutrinos is excluded at the 4$\sigma$ level.

CMB temperature and polarisation data from the Planck satellite give us a completely independent measurement of $\Neff$ in the framework of the minimal \LCDM{} cosmological model~\cite{Aghanim:2018eyx},
\begin{equation}
    \Neff = 2.92 \pm 0.37~, \,\,\, \mathrm{(\twosigma,~CMB+}\Lambda\mathrm{CDM)}
\end{equation}
excluding $\Neff=4$ at the 5.8$\sigma$ level. Finally, the combined CMB+BBN bounds presented in 
eq.~(77) of~\cite{Aghanim:2018eyx} raise the exclusion level to 7$\sigma$ or even 8$\sigma$.

One may think that CMB bounds on $\Neff$ are not as model-independent as BBN bounds, and could easily be evaded in extended cosmological model. This is however far from obvious. References~\cite{DiValentino:2015ola,DiValentino:2016hlg} fitted Planck data with many additional cosmological parameters at the same time, accounting for neutrino masses, a dark energy equation of state, a running of the primordial spectrum index, primordial tensor modes, etc. They still find $\Neff = 2.93^{+0.51}_{-0.48}$ (\twosigma) even with Planck 2015 data alone. The authors of~\cite{DiValentino:2016ikp} went even further. They tried to see whether $\Neff=4$ could be reconciled with cosmological data by paying the highest price: on top of floating the minimal \LCDM{} parameters plus ($\Neff$, $\sum m_{\nu,\mathrm{active}}$, $m_4$), they assign full freedom to the primordial spectrum of fluctuations (instead of the conventional power law assumption). Despite of this effort, their marginalized bound remains $\Neff<3.53$ (\twosigma, Planck 2015+BAO). 

It is worth noticing that direct measurement of the Hubble parameter from superNO$\nu$Ae luminosity are in significant tension with other cosmological data. If the tension is not caused by underestimated systematics in one of the data sets, the current standard cosmological model is ruled out at 4.4$\sigma$~\cite{Riess:2019cxk}. When this tension first emerged around 2011, it raised hopes that a higher value of $\Neff$ could reconcile the data sets, because $\Neff$ and $H_0$ have partially counter-acting effects on the time of radiation-to-matter equality and on some characteristics of the CMB temperature spectrum. Thus, $\Neff\simeq 4$ became slightly prefered over $\Neff\simeq 3$ at some point (see e.g.~\cite{Komatsu:2010fb}). However, all recent studies agree that such a simple solution does not work when more recent and precise CMB data are taken into account: it cannot solve the tension without raising other ones, at least when the \LCDM{} model is extended in a straightforward way with a free $\Neff$ plus active and/or sterile neutrino masses~\cite{Aghanim:2018eyx}. It remains nevertheless possible that more complicated extensions with more subtle physical ingredients do resolve the tension. In principle, they could be compatible with a high $\Neff$, and maybe with a population of sterile neutrinos. We will actually review one example of such models~\cite{Archidiacono:2016kkh} in Sec.~\ref{sec:cosmo:reconcile}. 

\subsection{Joint cosmological bounds on density and mass\label{sec:cosmo:mass}}

\begin{figure}[htbp]
\centering
\includegraphics[width=0.7\textwidth]{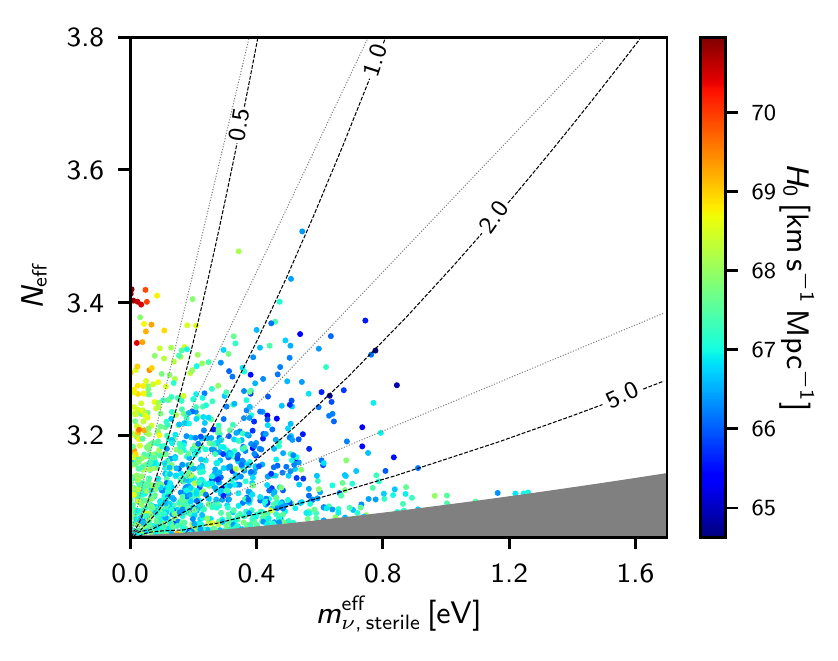}
\caption{Joint CMB bounds on $(\Neff, \, \meff)$ from Planck 2018 temperature, polarization and lensing data (figure taken from~\cite{Aghanim:2018eyx}).}
\label{fig:Planck}
\end{figure}

In Sec.~\ref{sec:cosmoeffects:LSS}, we defined the effective parameters $(\Neff, \, \meff)$. Fig.~\ref{fig:Planck} presents the joint bounds on these parameters obtained by the Planck collaboration when using only CMB data~\cite{Aghanim:2018eyx}. The density of points reflects the posterior probability on these two parameters (marginalized over all other model parameters). The black dashed lines correspond to fixed values of the particle mass for a model of little relevance to sterile neutrinos (namely, early decoupled thermal relics). However the thinner lines show the same fixed values of the mass for the Dodelson-Widrow model, $m_4 = \meff / \Delta N_4$. The vertical axis only extends up to $\Neff=3.8$ and does not include the case of thermalized sterile neutrinos with $\Delta N_4=1$ and $m_4=\meff$.

As expected, the upper bound on $\meff$ gets stronger when $\Neff$ increases, since the mass effect is weighted by $\Delta N_4$. No bound on the particle mass can be inferred from this analysis, since for an arbitrary small number density and small $\Delta N_4$ (i.e. $\Neff \longrightarrow 3$) some arbitrarily large physical masses are compatible with CMB data. Models with too high masses are irrelevant in the context of light sterile neutrino scenarios. The region close to the  $\Neff=3$ axis rather corresponds to models of warm or cold dark matter, whose mass is unconstrained by CMB data. Thus this region can be excluded when computing sterile neutrino parameter bounds. In the Planck analysis this is achieved by cutting the grey shaded region, which would correspond to a particle mass above a threshold of about 30~eV for the Dodelson-Widrow scenario. After removing this area, the individual constraints on each of the two parameters read
\begin{equation}
\Neff < 3.31 \, , \qquad \meff < 0.67\,{\rm eV}, \,\,\, \mathrm{(\twosigma,~CMB)}
\end{equation}
including Planck 2018 data from temperature, polarization and lensing~\cite{Aghanim:2018eyx}, as in Fig.~\ref{fig:Planck}.
This confirms that a sterile neutrino with a mass in the eV range can only be accommodated by cosmological data if $\Delta N_4<0.3$ (\twosigma), at least if its phase-space distribution is not dramatically different from a Dodelson-Widrow distribution. 

One additional interesting conclusion can be derived from Fig.~\ref{fig:Planck}, where the points have been colored according the value of the Hubble parameter in each allowed model. As expected, there is a trend to obtain larger $H_0$ values for larger $\Neff$. However, with a bound of $\Neff<3.3$, the high value preferred by direct measurement experiments ($H_0 \simeq 74$~km/s/Mpc~\cite{Riess:2019cxk}) is still out of reach. 

The bounds on $(\Neff, \, \meff)$ presented in~\cite{Aghanim:2018eyx} take into account the impact of sterile neutrinos mentioned in both Sec.~\ref{sec:cosmoeffects:CMB} (impact on CMB) and Sec.~\ref{sec:cosmoeffects:LSS} (impact on LSS), since they include CMB lensing information, which is a probe of the matter power spectrum in the recent universe, and thus of the formation of large scale structures. However, at the moment, the bounds on $\meff$ are dominated by information from primary CMB anisotropies, rather than probes of the small-scale matter power spectrum suppression due to the free streaming of massive $\nu_4$ neutrinos. Indeed, removing CMB lensing data and adding instead BAO data, which is blind to neutrino free-streaming, one gets approximately the same constraints~\cite{Aghanim:2018eyx}. The information contained in current galaxy and weak lensing surveys does not strengthen current bounds, but it will certainly do so in  the future (see Sec.~\ref{sec:cosmo:future}).

\subsection{Avenues for reconciling light sterile neutrinos with cosmology\label{sec:cosmo:reconcile}}

The results of the previous two sections have shown that in order to reconcile the (3+1) scenario with cosmology, one should reduce $\Neff$ with respect to 4.  This requires an incomplete thermalization of $\nu_4$ and/or of active neutrino.

Following~\cite{Foot:1995bm,Chu:2006ua}, a large primordial lepton asymmetry of the order of $L\simeq 10^{-2}$ would be sufficient for blocking active-to-sterile neutrino conversion in the early universe by effectively suppressing the mixing angle in the primeval plasma. Such an asymmetry would be very large compared to the one in the baryon sector, but a priori, still small enough to comply with BBN and CMB constraints on the chemical potential of active neutrinos~\cite{Mangano:2011ip,Castorina:2012md}. The authors of~\cite{Hannestad:2012ky} find that this mechanism can be efficient enough to reduce the density of $\nu_4$ neutrinos down to $\Delta N_4 \simeq 0.1$ when assuming $(\Delta m_{41}^2,\sin^2(2\theta_{14})) \simeq (1,0.1)$. However, reference~\cite{Saviano:2013ktj} revisited this model taking into account the possibility of different primordial asymmetries in each (active and sterile) flavor, and studying their evolution with time due to neutrino oscillations in the early universe. They stress that when the primordial asymmetries are sufficient to prevent sterile neutrino thermalization, the chemical potential of electron neutrinos at the BBN epoch has to be significantly different from zero, in such a way that primordial abundances could be in tension with observations. However this issue is still not precisely settled.

One can also question the fact that active neutrinos are fully thermalized and contribute to $\Neff = 3.045$ as in the Standard Model. To avoid this, one may assume that reheating after inflation takes place at an extremely low temperature $T_\mathrm{RH}$, of the same order of magnitude as the neutrino decoupling temperature. Then, neutrinos can never reach thermal equilibrium with photons. This solution has been proposed in the context of light sterile neutrinos in~\cite{Gelmini:2004ah,Yaguna:2007wi}. For instance, with $T_\mathrm{RH} \simeq 3$~MeV, each neutrino would be produced at the level of $\Delta \Neff \sim \frac{3}{4}$~\cite{deSalas:2015glj}, leaving room for four species summing up to $\Neff \sim 3$. It is also possible to reduce the density of active and sterile neutrinos relative to that of photons by invoking some entropy production taking place in the small lapse of time between neutrino decoupling and BBN~\cite{Ho:2012br}. In both cases, $\Neff$ can be adjusted to about three, but should not get much smaller in order to comply with the lower bound from BBN and CMB. Then, the problem is that in these models, active-sterile neutrino oscillations with $\sin^2(2\theta_{14}) \simeq 0.1$ should be efficient and lead to an equal share of $\Neff \simeq 3$ between the four neutrino states, and thus to $\Delta N_4 \simeq \frac{3}{4}$. Strictly speaking, the constraints on $(\Neff, \, \meff)$ seen in Sec.~\ref{sec:cosmo:mass} are not directly applicable here because they assume $N_\mathrm{eff, active}=3.045$. To derive cosmological bounds on these models, one would need a dedicated analysis with at least three free cosmological parameters related to the neutrino sector: ($N_\mathrm{eff, active}$, $N_\mathrm{eff, sterile}$, $\meff$). However the results can be roughly approximated by those of Sec.~\ref{sec:cosmo:mass} with $\Neff=N_\mathrm{eff, active} + N_\mathrm{eff, sterile}$ and $\meff=m_4/\Delta N_4$. Then models with $\Neff \simeq 3$, $\Delta N_4 \simeq \frac{3}{4}$ and $m_4\simeq 1$~eV would give roughly $\meff \sim \frac{3}{4}$~eV, in 2.5$\sigma$ tension with CMB constraints. We see that this category of models is only marginally consistent with current cosmological bounds.

Finally, $\nu_4$ neutrinos could be coupled through ``secret interactions'' to some (vector or scalar) boson belonging to a dark sector. Through finite temperature effects, the interaction generates an effective potential that suppresses active-sterile mixing in the early universe and may achieve $\Delta N_4<0.3$~\cite{Hannestad:2013ana,Dasgupta:2013zpn,Archidiacono:2014nda,Saviano:2014esa,Chu:2015ipa}.

At low temperature, the effective potential becomes irrelevant and active-sterile oscillations are efficient again (at least for the large mixing angle values suggested by anomalies in the oscillation data, $\sin^2(2\theta_{14}) \sim 0.1$). This could be problematic, especially in the case of interactions between a massive gauge boson and the sterile neutrino flavor state
\cite{Mirizzi:2014ama,Cherry:2016jol,Forastieri:2017oma,Chu:2018gxk}.
In this scenario, at neutrino decoupling,  $\Neff$ is still close to 3 and distributed equally among active neutrinos. Then, if active-sterile neutrino oscillations become efficient, $\Neff$ gets redistributed almost equally between the {\it four} mass eigenstates. In absence of secret interactions, this would take place with a constant $\Neff \simeq 3$. However, the efficient interactions mediated by the gauge boson would force this redistribution to take place in thermal equilibrium between all neutrino mass states, since they all contain at least a small sterile component~\cite{Mirizzi:2014ama}. Then entropy conservation would lead to an overall reduction of the temperature of the neutrino sector and bring $\Neff$ close to 2.7, which is still compatible with BBN bounds. With  $\Delta N_4\simeq 2.7/4 = 0.675$, the effect of a mass $m_4\sim {\cal O}(1\,\mathrm{eV})$ on CMB and LSS observables is marginally compatible with CMB bounds: approximating the impact of this specific model in terms of the parametrization of Sec.~\ref{sec:cosmo:mass}, one gets $\meff \simeq 0.675$~eV for $m_4\simeq 1$~eV, close to the 2$\sigma$ CMB upper bound. However, in this model, the relic neutrinos are not free-streaming like in \LCDM: due to secret interactions, they behave at least partially as a relativistic fluid. CMB data disfavor the combined effect of $\Neff<3$ and of self-interacting relativistic relics at a significant level~\cite{Mirizzi:2014ama,Cherry:2016jol,Forastieri:2017oma,Chu:2018gxk}.

The scenario of~\cite{Archidiacono:2014nda,Archidiacono:2015oma,Archidiacono:2016kkh} 
assumes instead a non-standard interaction between the mainly sterile mass state $\nu_4$ and a pseudoscalar field in a dark sector with a mass $m_\phi \ll m_4$. Assuming a pseudoscalar rather than a scalar boson allows to evade fifth force bounds. Like in the previous model, an effective potential suppresses active-sterile oscillations in the early universe, helping to pass BBN tests. There is a range in which the interaction is strong enough for such effects to take place and weak enough to avoid bounds from superNO$\nu$Ae energy loss~\cite{Farzan:2002wx}. Interestingly, this model differs from the one with heavy gauge boson mediators at the level of the late time cosmological evolution. Indeed, a significant population of light pseudoscalars is produced in the late universe, when the secret interaction is efficient. At times relevant for CMB and LSS, $\Neff$ remains typically in the range from 3 to 3.5, and accounts for two populations of relativistic relics beyond photons: first, free-streaming active neutrinos, and second, a self-coupled fluid of $\nu_4$ and $\phi$ particles. Initially, at temperatures $T>m_4$, the second fluid is composed of both relativistic $\nu_4$ neutrinos and relativistic pseudoscalars. Then $\nu_4$ neutrinos annihilate into lighter pseudoscalar particles, and one is left with a fluid of very light relics $\phi$ with a mass $m_\phi$ fulfilling bounds on $\meff$. The fact that part of the relativistic species are self-interacting at CMB times is not in contradiction with Planck data, because of a partial cancellation between the impact on CMB observables of a high $\Neff>3$ and of self interactions (due to modified ``neutrino drag'' and ``gravity boost'' effects, see Sec~\ref{sec:cosmoeffects:CMB}). This cancellation works better when $H_0$ is also high. Thus this model could help to resolve the tension between CMB data and direct measurements of $H_0$~\cite{Archidiacono:2016kkh}. The dedicated cosmological analysis presented in the last reference shows that this scenario is compatible with recent cosmological data. The collisional nature of the extra relativistic degrees of freedom during the CMB epoch offers an opportunity to test it specifically with more precise CMB experiments. 

Reference~\cite{Chu:2018gxk} gives some hints of other types of secret interactions, based on even more specific assumptions, that may also evade BBN, CMB and LSS bounds.

\subsection{Sensitivity of future cosmological data\label{sec:cosmo:future}}

Spectacular improvements on the sensitivity to the mass and density of a fourth neutrino mass state can be expected from the next generation of CMB and LSS experiments (see e.g.~\cite{Giusarma:2011ex,DiValentino:2016foa}). Forecasts on these parameters have been performed in the framework of a \LCDM{} cosmology with nine free parameters: the usual six parameters of the minimal \LCDM{} model, plus the sum of active neutrino masses $\sum m_{\nu,\mathrm{active}}$, and the parameters ($\Neff$, $\meff$) defined as in Sec.~\ref{sec:cosmoeffects:LSS} and~\ref{sec:cosmo:mass}. The current bounds reported in Sec.~\ref{sec:cosmo:mass} neglected the effect of $\sum m_{\nu,\mathrm{active}}$, but this approximation would be inappropriate at the level of precision of future experiments. In sensitivity forecasts, final bounds on $\Neff$ and $\meff$ are marginalised over all other parameters including $\sum m_{\nu,\mathrm{active}}$.

Combining Planck CMB data with future LSS data will not change significantly the sensitivity to $\Neff$ and $\Delta N_4$, but will vastly improve constraints on $\meff$, thanks to a direct sensitivity to the free-streaming effect of $\nu_4$ neutrinos. According to~\cite{DiValentino:2016foa}, galaxy and weak lensing surveys from the DESI~\cite{Levi:2013gra} and Euclid~\cite{Amendola:2016saw} collaborations should have a sensitivity $\sigma(\meff)\simeq 0.06$~eV.

Some even more spectacular progress can be expected from future CMB experiments. An hypothetical satellite mission like CORE-M5 would bring the sensitivity down to
\begin{equation}
\sigma(\Delta N_4)\sim 0.05 \, , \qquad \sigma(\meff)\sim 0.04 \, \mathrm{eV} \, ,
\end{equation}
using only temperature and polarisation information~\cite{DiValentino:2016foa} (thanks to CMB lensing extraction, one could obtain even better numbers). The CORE-M5 satellite project was not accepted by ESA in 2017, but these sensitivities are representative of what could be expected in general from a next generation of CMB satellite like PICO~\cite{Sutin:2018onu} or CORE, covering both large and small scales. Very similar numbers could be obtained by combining a satellite like LiteBird~\cite{Suzuki:2018cuy} covering large scales with a ground-based experiment like CMB-S4~\cite{Abazajian:2016yjj} covering small scales.

With such numbers, and assuming the existence of a fourth neutrino mass state with $m_4 \sim {\cal O}(1\,\mathrm{eV})$, one could accurately test most of the models summarised in Sec.~\ref{sec:cosmo:reconcile}. For instance, scenarios with an (ad hoc) low reheating temperature or entropy production could be confirmed or ruled out, since they predict $\meff \sim {\cal O}(1\,\mathrm{eV})$; secret interactions mediated by gauge bosons would be probed for the same reason, and also because they tend to achieve $\Neff < 3$. Future CMB data will also have an enhanced sensitivity to the assumption that all or part of relativistic relics could be self-interacting~\cite{DiValentino:2016foa}. This should provide a strong test of models with secret interactions between sterile neutrinos and a light pseudoscalar.

All these bounds assume a plain \LCDM{} cosmology - and thus, implicitly, that the current tension with direct measurements of $H_0$ will disappear after taking into account some yet unknown systematic effects. If this is not the case, and if the solution to this problem really requires new physics (either related to sterile neutrinos or totally different), the previous forecasts might be inapplicable, but future prospects will be equally exciting.

%
%

\section{Conclusions}
\label{sec:conclusions}

Since eV-mass sterile neutrinos have been pointed out as a consistent explanation of the short-baseline anomalies detailed in Sec.~\ref{sec:hints}, considerable effort has been taken to either disprove or consolidate this hypothesis, exploring not only a puzzling consistency between the anomalous results but also a possible window to physics beyond the Standard Model. 

The anomalies have triggered a wave of experimental activity to search for a more unambiguous signature for the existence of sterile neutrinos: informed by the global best fit values of the original anomalies, an excellent testing ground is the exploration of the shape regime (Sec.~\ref{sec:signatures}) of short-baseline oscillations where sterile neutrinos are expected to leave a distinctive pattern at an $L/E$ ratio corresponding to $\Delta m^2_{41}\geq 1\un{eV^2}$.

Since 2011, a variety of new results has led to greater experimental sensitivity for both the $\protect\nua{e}\to\protect\nua{e}$ and $\protect\nua{\mu}\to\protect\nua{\mu}$ disappearance channels: reactor antineutrino experiments (Sec.~\ref{sub:reactorexp}) have been able to rule out a large fraction of the parameter space originally favoured by reactor and gallium anomalies, although current results point towards a new best-fit region at similar $\Delta m^2_{41}$ but smaller amplitude. Also the limits on the $\nu_\mu\to\nu_\mu$ disappearance amplitude have grown significantly tighter (Sec.~\ref{sub:atmospherics} and Sec.~\ref{sub:lblacc}). Contrariwise, recent results from MiniBooNE even increase the significance of the anomalous $\nu_\mu\to\nu_e$ appearance signature (Sec.~\ref{sec:lsndminiboone}). Thus, it seems that the current inconsistency between (negative) disappearance and (positive) appearance data  will only be resolved when the reactor searches will have reached their full sensitivity and the Fermilab SBN program will provide an independent and more sensitive test of the appearance anomaly (Sec.~\ref{sub:sblacc}). A disappearance search based on radioactive $\nu_e$ sources would provide a cross-check of the gallium anomaly and CPT invariance (Sec.~\ref{sub:sourceexp}).

Direct neutrino mass experiments offer an interesting alternative handle to investigate the light sterile neutrino problem (Sec.~\ref{sec:numass}). KATRIN is sensitive to the characteristic kink that eV-scale sterile neutrinos would leave in the tritium decay spectrum close to the endpoint. KATRIN started data taking in 2019, and hence first relevant results can be expected in a few years from now. For a value of $m_4\approx2\un{eV}$, the expected sensitivity of KATRIN corresponds to a matrix element of $|U_{e4}|^2=0.1$ (90\,\%\,CL), while for larger masses the sensitivity increases significantly. Hence, the KATRIN data will provide information relevant to the best-fit values of the original anomaly. Future tritium- and holmium-based experiments, such as extensions of KATRIN, Project-8, ECHo, and Holmes, are planned to surpass this sensitivity, being able to test the current best-fit values for $|U_{e4}|^2$. However, more than a decade will pass before these results become available.

Meanwhile, the joint bounds on the number and combined mass of light neutrinos that are derived from observations of the Cosmic Microwave Background (CMB), Large Scale Structures (LSS) and Big Bang Nucleosynthesis (BBN) strongly disfavour an additional light neutrino state on the eV-mass scale (Sec.~\ref{sec:cosmology}). It should be noted, though, that the restrictions posed by this standard interpretation of cosmological data can be somewhat alleviated if additional physics mechanisms are added, e.g.~by avoiding the thermalization of the sterile neutrino state in the early Universe (Sec.~\ref{sec:cosmo:reconcile}). 

While recent results from reactor and atmospheric neutrino searches have considerably tightened the limits on the active-sterile mixing parameters, the parameter space preferred by the original anomalies is not yet fully tested.  However, a much more complete picture is expected to emerge only a few years from now: the running reactor/source experiments and the SBN program at FNAL are very likely to deliver a definite answer on the occurrence of active-sterile oscillations in the 1-eV$^2$ range. Absolute mass experiments and cosmological observations will provide a complementary set of constraints, limiting the mass and number of light neutrino species. Therefore, a definite conclusion on the existence of an eV-scale sterile neutrino seems well within range!


\section*{Acknowledgements}
JL would like to acknowledge very useful exchanges with Maria Archidiacono and Stefano Garriazzo. Fermilab is managed by Fermi Research Alliance, LLC (FRA), acting under Contract No. DE-AC02-07CH11359.

\bibliographystyle{ieeetr}
\bibliography{main}

\end{document}